 \documentclass[final,5p,times,twocolumn,authoryear]{elsarticle}

\usepackage{amssymb}
\usepackage{lipsum}
\usepackage{url}
\usepackage{subcaption}
\usepackage{amsmath}

\journal{Nuclear Physics A}

\begin{document}
\begin{frontmatter}



\title{Commissioning and Performance of the \emph{Trinity} Demonstrator}


\author[a]{Mahdi Bagheri}
\author[a]{Adam Barletta}
\author[a]{Jordan Bogdan}
\author[b]{Anthony M. Brown}
\author[a]{Luigi Cedeno}
\author[a,cor1]{Mariia Fedkevych}
\author[a]{Srikar Gadamsetty}
\author[a]{Eliza Gazda}
\author[c]{Jamie Holder}
\author[d]{Eleanor Judd}
\author[e]{Dave Kieda}
\author[f]{Evgeny Kuznetsov}
\author[a]{Nolan Lew}
\author[a]{Oscar Romero Matamala}
\author[a]{Arnav Menon}
\author[a]{A.~Nepomuk Otte}
\author[a]{Mathew Potts}
\author[e]{Wayne Springer}
\author[a]{Sofia Stepanoff}
\author[a]{Ace Wilcox}
\author[a]{Angelina Zhang}
\affiliation[a]{
organization={Georgia Institute of Technology, School of Physics, Center for Relativistic Astrophysics},
                addressline={837 State Street NW}, 
                city={Atlanta},
                state={GA},
                postcode={30332-0430}, 
                country={U.S.A.}}

\affiliation[b]{
organization={Department of Physics and Centre for Advanced Instrumentation, University of Durham},
                addressline={South Road}, 
                city={Durham},
                state={},
                postcode={DH1 3LE}, 
                country={UK}}

\affiliation[c]{
organization={University of Delaware, Department of Physics and Astronomy and the Bartol Research Institute},
city={Newark},
state={DE},
postcode={19716},
country={U.S.A.}}

\affiliation[f]{
organization={University of Alabama in Huntsville, Center for Space Plasma and Aeronomic Research},
addressline={NSSTC, CSPAR, 320 Sparkman Drive},
city={Huntsville},
state={AL},
postcode={35805},
country={U.S.A.}}

\affiliation[d]{
organization={University of California at Berkeley, Space Sciences Laboratory},
addressline={7 Gauss Way},
city={Berkeley},
state={CA},
postcode={94720},
country={U.S.A}}

\affiliation[e]{
organization={University of Utah, Department of Physics and Astronomy},
addressline={115 S 1400 E \#201},
city={Salt Lake City},
state={UT},
postcode={84112},
country={U.S.A}}

\cortext[cor1]{Corresponding Author, mfedkevych3@gatech.edu}

\begin{abstract}
The \emph{Trinity} Demonstrator is a proof of concept prototype for the \emph{Trinity} Neutrino Observatory, which is sensitive to astrophysical neutrinos above PeV energies. The Demonstrator is a one-square meter class imaging atmospheric Cherenkov telescope deployed on Frisco Peak, Utah, and remotely operated. The light-collection surface is equipped with 77 mirror facets with 15\,cm diameter, and its $3.87^\circ\times3.87^\circ$ field of view is instrumented with a 256-pixel camera yielding a $0.24^\circ$ angular resolution. The camera signals are digitized with a 100\,MS/s, and 12-bit resolution switched capacitor array readout. We discuss the Demonstrator's design, the telescope's deployment on Frisco Peak, and its commissioning.
\end{abstract}




\end{frontmatter}




\section{Introduction}
\label{sec:Intro}

The neutrino sky above PeV energies is very much unexplored. Below PeV energies, IceCube's detection of the galactic plane \citep{galacticplaneIceCube2023}, the extragalactic diffuse neutrino flux \citep{astrophysical2013}, and potentially the first two high-energy neutrino sources, the Seyfert galaxy NGC 1068 and the blazar TXS 0506+056 \citep{NGC2022,txs0506}, are first-hand examples demonstrating the impact of high-energy neutrino astrophysics on time-domain and multi-messenger astrophysics.

It is commonly accepted that IceCube only sees the metaphorical tip of the iceberg. Instruments that reach above PeV energies can thus be expected to open an entirely new window on the Universe, the very-high-energy (VHE, $>1$\,PeV) and ultra-high-energy (UHE, $>1$\,EeV) neutrino bands. VHE and UHE neutrinos will provide us with a new and complementary way of studying accretion and jet formation around supermassive black holes, the acceleration, propagation, and composition of ultra-high energy cosmic rays, and a pathway to search for new physics beyond the standard model of particle physics to only name some important topics \citep{snowmass}.

The sensitivity limits of IceCube and Pierre Auger Observatory constrain the fluxes of VHE/UHE neutrinos to extremely low values \citep{IceCube_Aartsen_2018, Auger_Aab_2019, IceCube_meier2023searchextremelyhighenergy} even though KM3NET has recently detected a neutrino with 120\,PeV \citep{Aiello2025}. The low neutrino fluxes translate into requirements for new instruments to monitor areas of thousands of square kilometers or hundreds of cubic kilometers of ice or water. These vast areas and volumes must be instrumented with sufficiently fine granularity to detect VHE and UHE neutrinos. Among the proposed techniques are radio arrays deployed in-ice \citep{gen2_Aartsen_2021, RNO_Aguilar_2021}, on the surface \citep{GRAND_lvarez_Mu_iz_2019, BEACON_Southall_2023}, or on sub-orbital balloons \citep{PUEO_Abarr_2021}, in-ice RADAR \citep{RTE_Prohira_2020}, particle detectors deployed in valleys \citep{tambo_thompson2023tambosearchingtauneutrinos}, and finally imaging air-shower Cherenkov telescopes pointing at the horizon \citep{chant_Neronov_2017,poemma_Olinto_2021,brown2021trinityimagingaircherenkov}. 

The proposed \emph{Trinity} PeV-neutrino Observatory is a system of Cherenkov telescopes deployed on several mountaintops. The telescopes detect air showers that result from tau neutrinos that enter the Earth under an angle of less than ten degrees. At such small angles, the length of the trajectory inside the Earth is just right for VHE/UHE tau neutrinos to undergo a charged current interaction and the resulting tau to decay in the atmosphere once the tau has emerged from the ground \citep{Fargion2001}. The decay of the tau in the atmosphere initiates a particle shower causing the emission of Cherenkov light. 

The mirrors of a \emph{Trinity} telescope collect Cherenkov light from the air shower and project an image of the air shower onto the telescope's focal surface where a silicon-photomultiplier camera records the image. By analyzing the image one can then reconstruct the neutrino's energy and arrival direction. The technique to image air showers was originally developed for VHE gamma-ray astronomy, where it has been successfully used for the last thirty years \citep{Hillas1985, whipplecrab_nuimeprn12618}. 

Once completed, the \emph{Trinity} Observatory will comprise eighteen 60-degree wide-angle Cherenkov telescopes deployed at different sites to maximize \emph{Trinity}'s acceptance for diffuse neutrino fluxes. \textit{Trinity}'s core energy range is between 1\,PeV and 1\,EeV \citep{Trinity_Wang_2021}. We develop \emph{Trinity} in three stages. The first stage is the \emph{Trinity} Demonstrator, which we report here. The second stage will be \emph{Trinity} One, the first \emph{Trinity} telescope. In the final stage, once \emph{Trinity} One is operational, we anticipate scaling to the full 18-telescope \emph{Trinity} Observatory. 

The \emph{Trinity} Demonstrator's goal is to establish the \emph{Trinity} concept by validating and demonstrating that projected sensitivities for \emph{Trinity} can be met. This includes understanding how different types of background events impact the sensitivity and how to efficiently suppress them in the analysis. Finally, the Demonstrator points toward the direction where NGC 1068 and TXS 0506+056 cross the horizon, with the goal of constraining VHE/UHE neutrino emission from both objects.

\section{Description of the Demonstrator Telescope}

\begin{figure}[!htb]
\centering
\includegraphics[angle=0, width=.9\columnwidth]{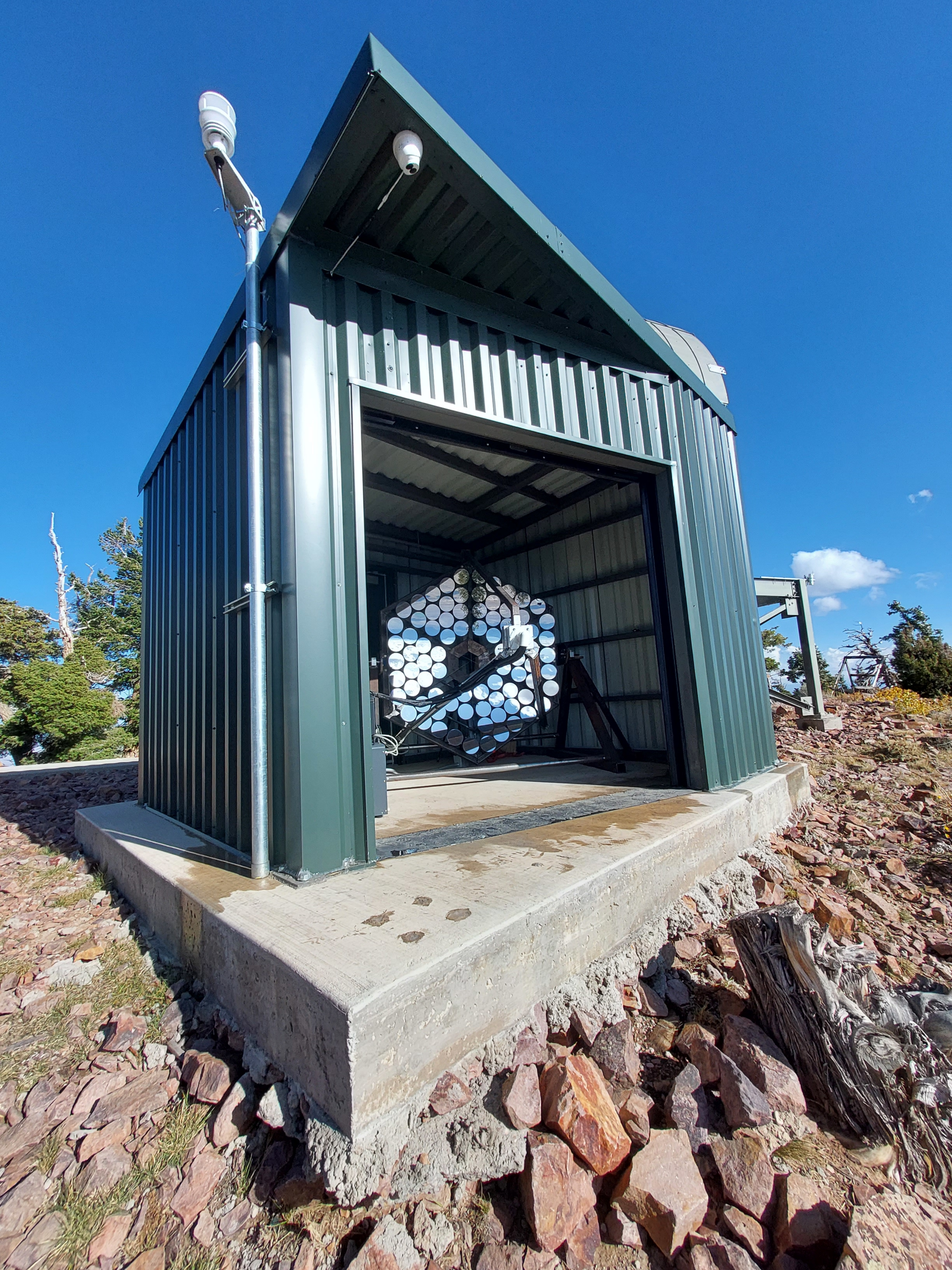}
\caption{The \emph{Trinity} Demonstrator inside its building on Frisco Peak, Utah.}
\label{fig:Demonstrator}
\end{figure}

The \emph{Trinity} Demonstrator is deployed on Frisco Peak, Utah, 2930~m above mean sea level (see Figure \ref{fig:Demonstrator}). Compared to the future \emph{Trinity} telescopes, the \emph{Trinity} Demonstrator's light-collection area is about ten times smaller and has a different optical design. Yet, the Demonstrator shares key parameters with it, such as the angular resolution, sampling speed of the readout system, and nearly identical vertical field of view. In sharing these, the Demonstrator can prove the viability of the \emph{Trinity} concept.

In the following description of the Demonstrator we distinguish between the telescope structure, which includes the optical support structure with its mirrors and the camera housing, the movable cabinet on the floor next to the telescope with the data acquisition and the power supplies, the camera, and auxilliary systems.

\subsection{The Optics}\label{sec:optics}

\begin{figure}[!htb]
\centering
\includegraphics[angle=0,width=1.\columnwidth]{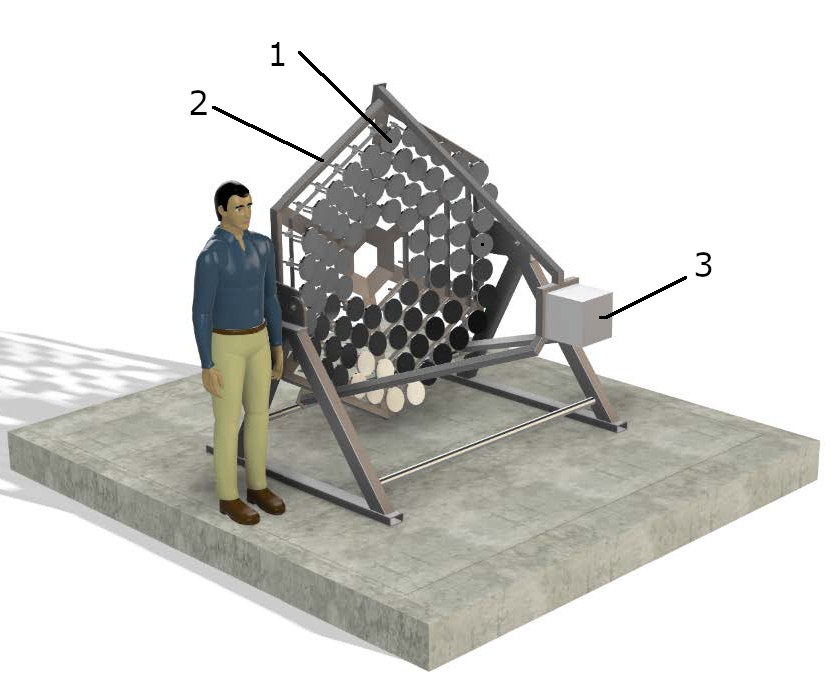}
\caption{CAD drawing of the \emph{Trinity Demonstrator}: 1 - mirror facet; 2 - optical support structure; 3 - camera housing.}
\label{fig:CADDemonstrator}
\end{figure}

Figure \ref{fig:CADDemonstrator} shows a CAD drawing of the telescope with its main components. The optical support structure (OSS) supports the 84 mirror facets and positions them on the surface of a sphere with a radius of 1.48\,m, equal to the focal length of the mirror facets. \emph{i.e.} a Davies-Cotton optics \citep{Davies1957}. The camera focal plane is positioned to be at the center of the sphere.

\begin{figure}[!tb]
\centering
\includegraphics[angle=0,width=0.9\columnwidth]{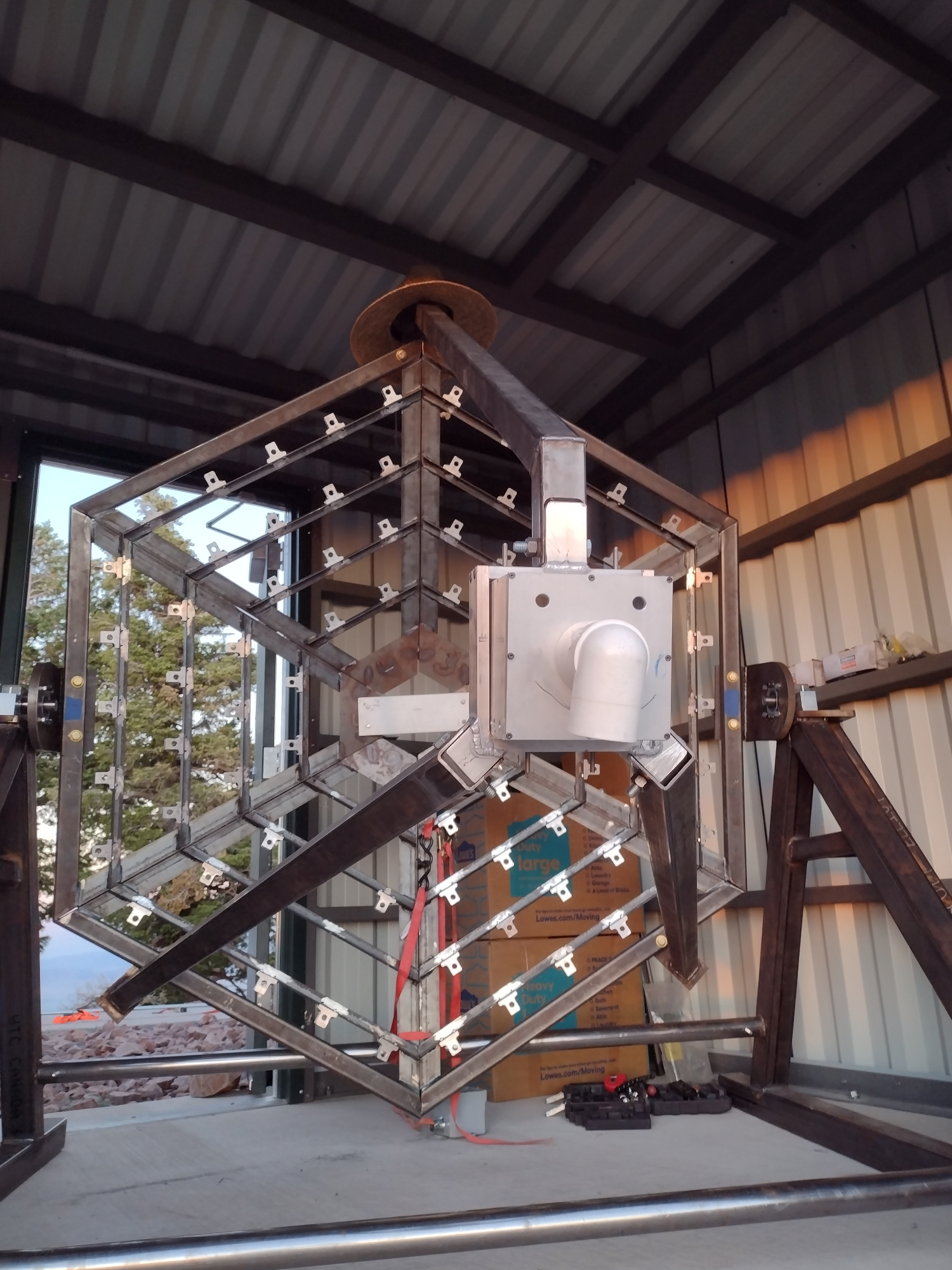}
\caption{The optical support structure of the \emph{Demonstrator} after installation on Frisco Peak.}
\label{fig:OSS}
\end{figure}

Figure \ref{fig:OSS} shows the mechanical structure of the Demonstrator installed on Frisco Peak before the installation of the mirrors. The OSS consists of six triangles or sectors, forming a hexagon when assembled. Each sector can hold up to 14 mirrors.

The OSS is installed into its support structure, which consists of two side supports interconnected with two horizontal supports at the bottom. Ball bearings on each side support allow the OSS to freely rotate up and down. The OSS support structure is bolted into the foundation of the Demonstrator building, with the telescope's optical axis permanently aligned at an azimuth angle of $280^\circ\pm0.06^\circ$. The \emph{Demonstrator} structure was designed to be assembled in the field by at most three persons. On-site assembly of the OSS took about three hours. 

\begin{figure}[!tb]
\centering
\includegraphics[angle=0,width=0.9\columnwidth]{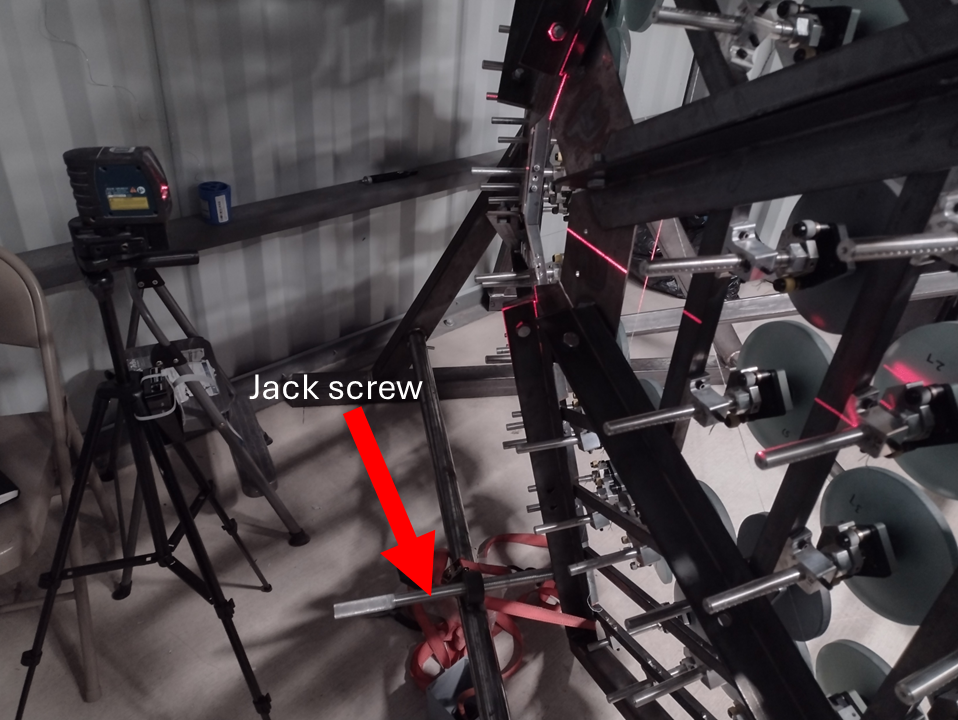}
\caption{The Jack screw for adjusting the pointing of the telescope in elevation, indicated by red arrow}
\label{fig:jackscrew}
\end{figure}

Because the center of gravity of the fully populated OSS is forward of the hinges, the OSS naturally wants to tilt down. A jack screw that is installed into a threaded block mounted on the center of the rear interconnecting support counters the downward motion and is used to adjust the pointing of the telescope in elevation  (see Figure \ref{fig:jackscrew}).  In its normal observing position, the center of the telescope camera points $1.56^\circ$ below the horizon.

The mirror facets are concave spherical surface mirrors with a glass substrate and protected aluminum coating. Each facet is round and has a 15\,cm diameter. The mirrors were repurposed from a previous experiment, and their properties and history are not well known. At present, the OSS is equipped with 77 facets with a combined light-collection area of 1.36\,m$^2$. 

\begin{figure}[!tb]
\centering
\includegraphics[angle=0,width=1\columnwidth]{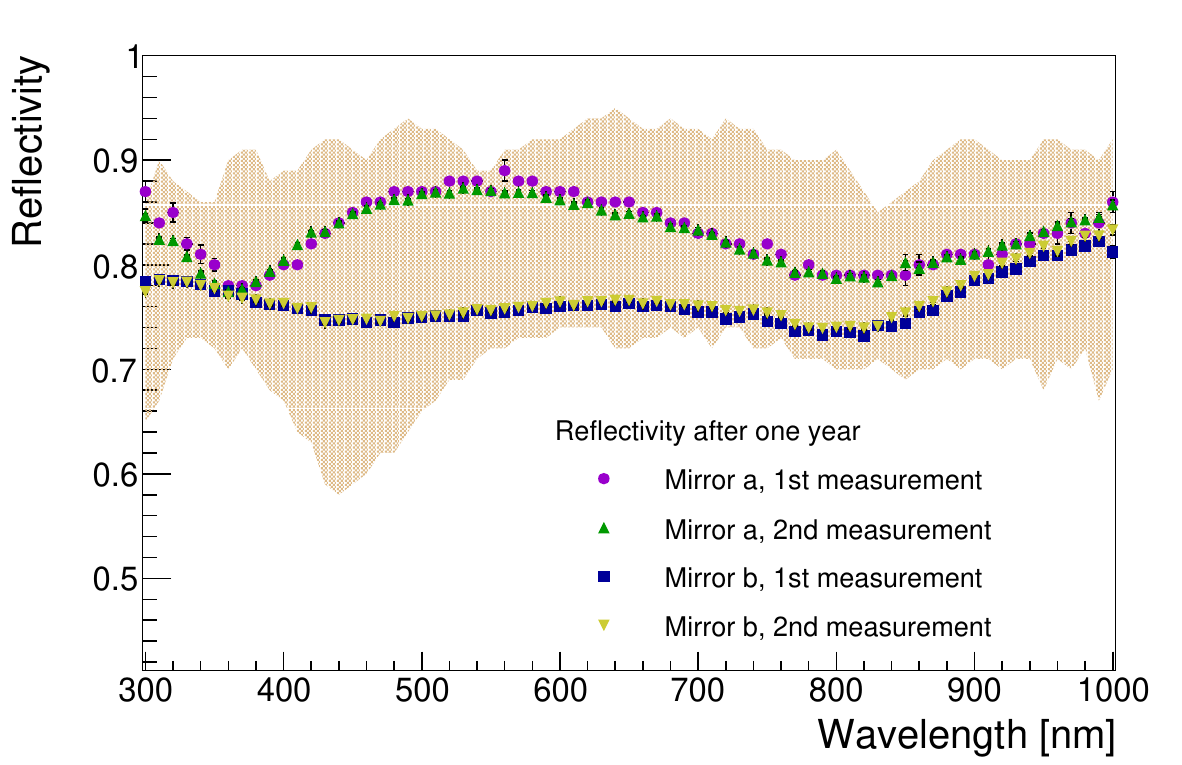}
\caption{Shaded band: Reflectivity of ten mirrors measured before telescope deployment. Data points: Reflectivity measurements of two mirrors after one year of telescope operation. The reflectivity of each mirror was measured twice at different points of their surfaces.}
\label{fig:reflectivity}
\end{figure}

The shaded band in Figure \ref{fig:reflectivity} shows the range of mirror reflectivities of 10 facets measured before the telescope was deployed. The reflectivity is relatively flat between 300\,nm and 1,000\,nm with an average of about 80\%. The four reflectivity curves in the same figure are measurements of two mirror facets recorded after one year of operation. The measurement of each mirror was repeated at a different location on the mirror surface and agree with one another. The reflectivity of both mirrors is fully contained within the band of the pre-deployment reflectivity, which is why we conclude that the mirrors did not significantly degrade during the first year of operation. We attribute the different wavelength dependencies to different batches of mirror production.

We measured the reflectivities with a Xenon arc lamp coupled to a monochromator. Light of a selected wavelength was first recorded by directly illuminating a photodiode and then by reflecting the light first off the mirror under test with an angle of incidence of about $10^\circ$. During the measurement, the intensity of the Xenon lamp was independently monitored with a second photodiode. Care was taken to ensure the light spot was fully contained in both photodiodes. 

\begin{figure}[!tb]
\centering
\begin{subfigure}[b]{1.\columnwidth}
\includegraphics[angle=0,width=1.\columnwidth]{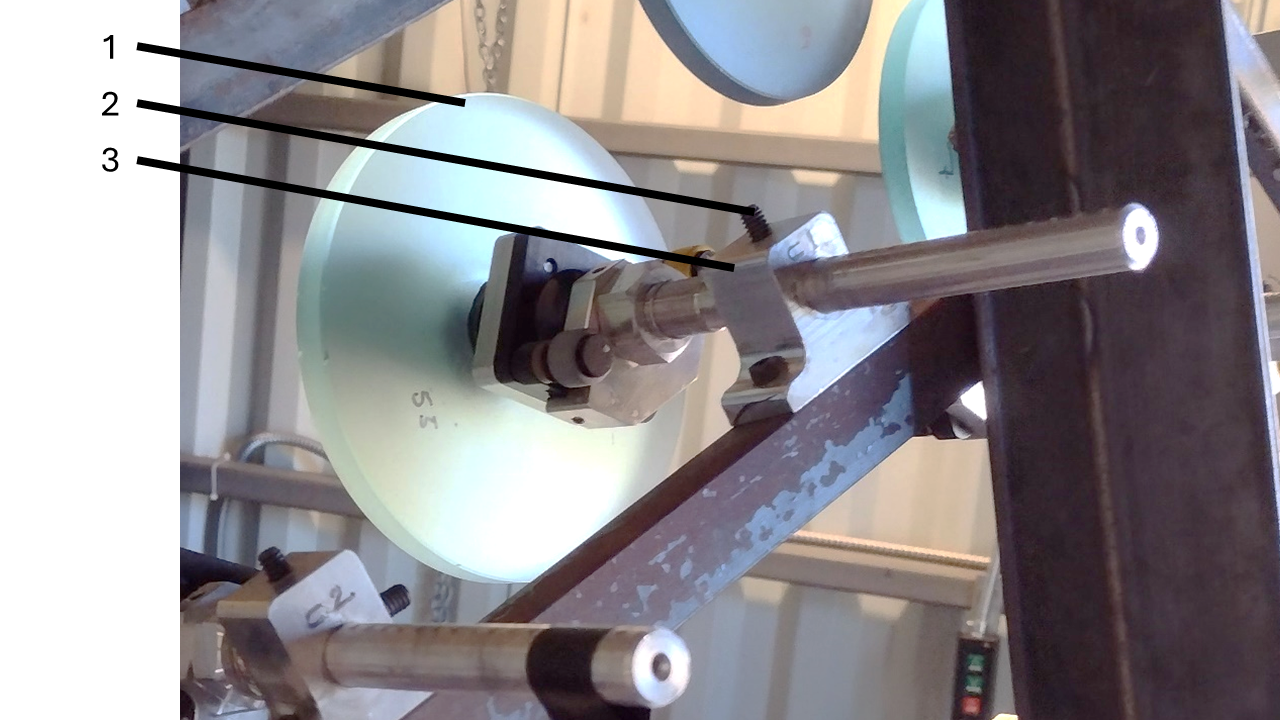}
\caption{Mirror mount and adjustment: 1 - mirror facet; 2 - set screw to fix axial mirror position after coarse alignment; 3 - mirror mount interface block on the OSS}
 \label{fig:mirrormount}
\end{subfigure}

\begin{subfigure}[b]{1.\columnwidth}
\centering
\includegraphics[angle=0,width=0.6\columnwidth]{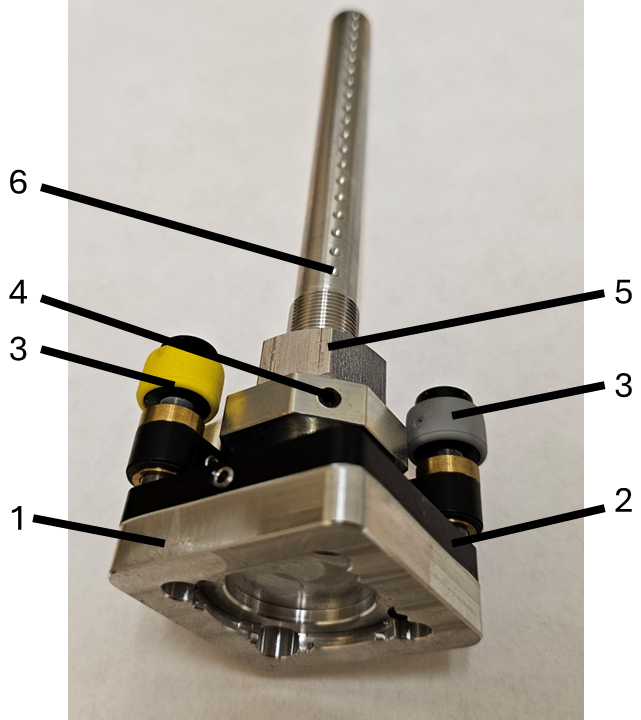}
\caption{Mirror mount without mirror: 1 - milled aluminum receptacle; 2 - tip/tilt IM100.T2a optics mount; 3 - tip/tilt alignment thumb screws; 4 - set screw to lock fine adjustment; 5 - nut to fine adjust the distance of the mirror to the focal plane; 6 - mounting rod with 1/4-inch marks for coarse adjustment}
 \label{fig:mirrormountalone}
\end{subfigure}
\caption{Mirror mount with and without mirror.}
\end{figure}

Figure \ref{fig:mirrormount} shows how a mirror is installed in the OSS, and the mirror mount on its own is shown in Figure \ref{fig:mirrormountalone}. To interface the mirror facet with its mount, a Quad-Lock\footnote{\url{https://www.quadlockcase.com/collections/shop-mounts/products/quad-lock-st-adhesive-mount-twin-pack?variant=179958462}} is glued to the back of a mirror facet with Epoxy Adhesive DP190 from 3M. The quad-lock attaches to an aluminum milled receptacle that is mounted on a IM100.T2a optics mount from SISKIYOU, which is used to adjust the tip/tilt of the facet (see section \ref{sec:mirroralignment}). To position the mirror at the right distance from the camera, i.e.\ one 1.48\,m focal length, the rod of the mount can be moved axially in 1/4-inch steps. For fine-tuning, the tip/tilt assembly can be moved along the rod with the help of a fine-threaded nut. Once the mirror is at the right distance, it is tightened in place with two set screws.

\subsection{The Camera and Readout System}\label{sec:CameraNReadout}

\begin{figure}[!htb]
\centering
\includegraphics[angle=0,width=0.9\columnwidth]{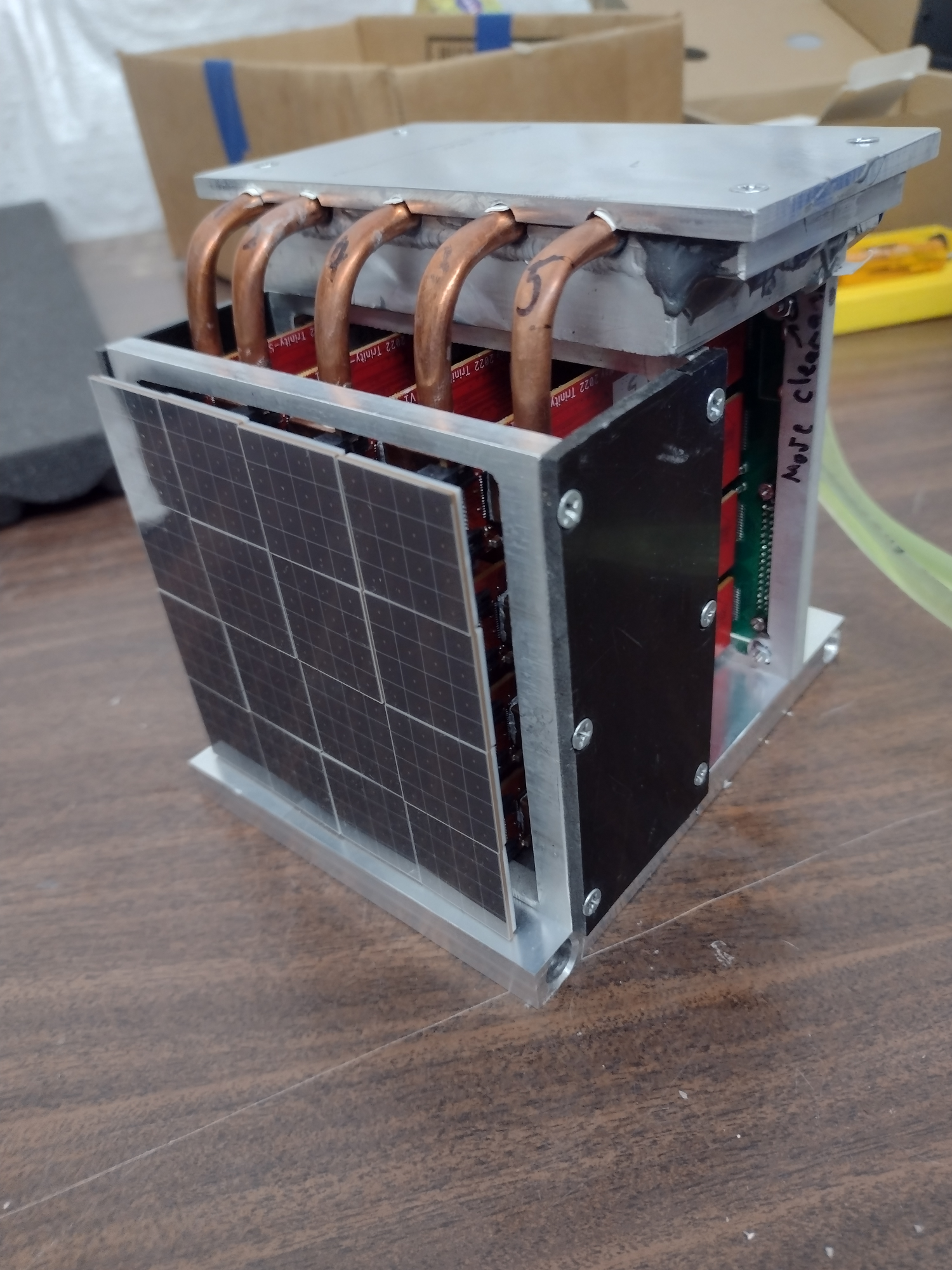}
\caption{The \emph{Demonstrator} camera. The front of the camera consists of 16 SiPM matrices. The copper tubes are heatpipes, which are part of the camera cooling system. }
\label{fig:camera}
\end{figure}

\begin{figure}[!htb]
\centering
\includegraphics[trim = 80 430 120 90, clip,angle=0,width=0.9\columnwidth]{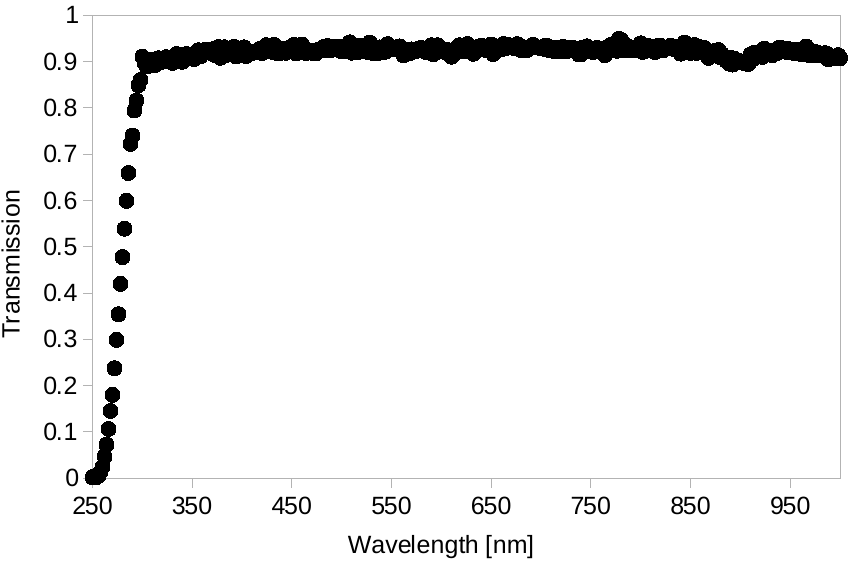}
\caption{Transmission of the 6\,mm thick entrance window of the camera.}
\label{fig:windowtransmission}
\end{figure}

The focal plane of the Demonstrator is instrumented with the 256-pixel silicon photomultiplier (SiPM) camera shown in Figure \ref{fig:camera}. Installed in the Demonstrator, the camera observes a $3.87^\circ\times3.87^\circ$ region and is protected with a 6\,mm thick Clear UVT Acrylic Plastic window from EMCO Industrial Plastics. Figure \ref{fig:windowtransmission} shows the transmission of the window under normal incidence, which is 93\%\,$\pm$1\% above 350\,nm.

\begin{figure}[!htb]
\centering
\includegraphics[angle=0,width=0.9\columnwidth]{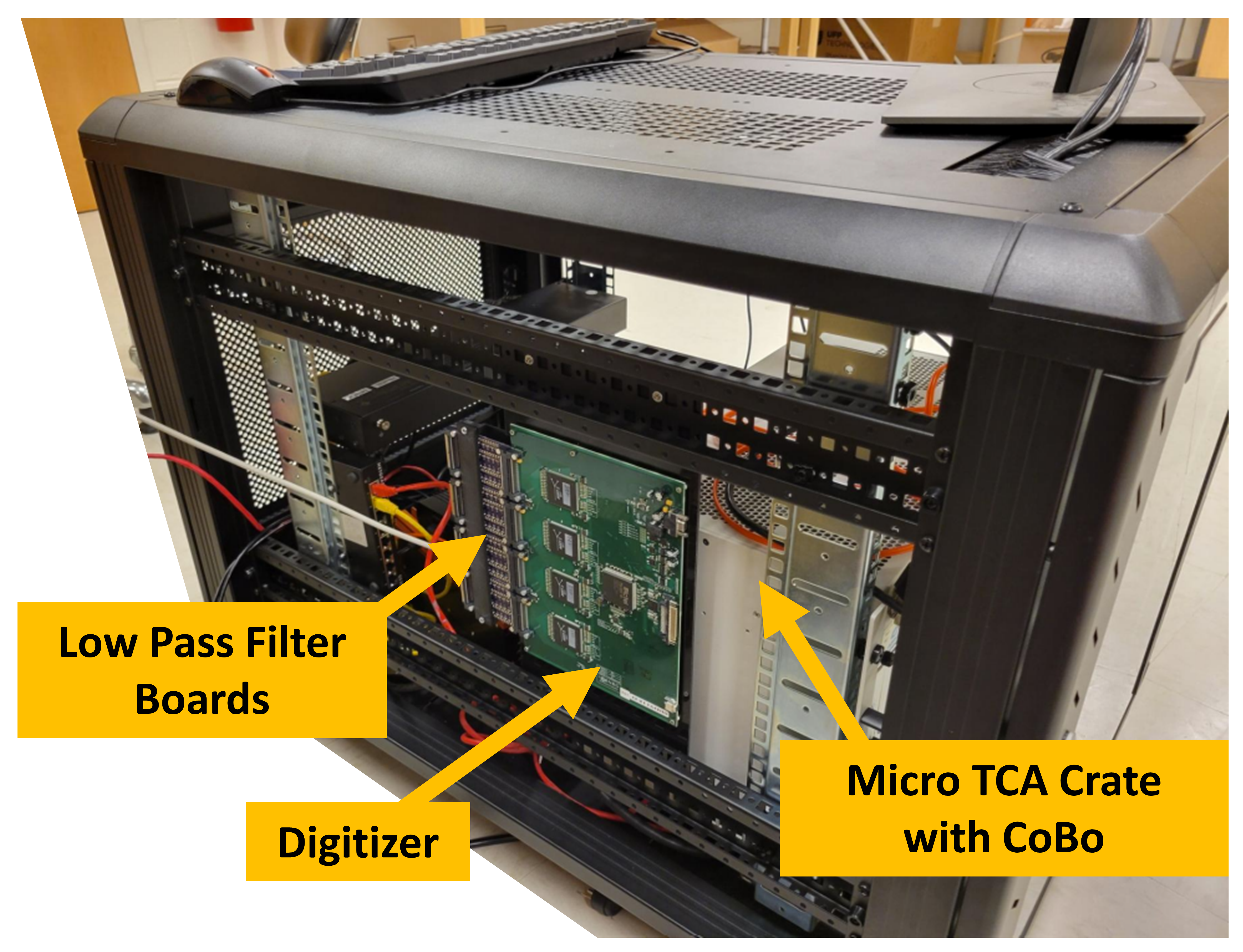}
\caption{Cabinet with the data acquisition. The Digitizer system consists of the AsAd board with 4 AGET chips and a Micro-TCA crate that hosts the CoBo module, which receives the digital data stream and forwards the data to the data acquisition computer in the cabinet. At the input of the digitizer, low-pass filter boards slow the SiPM signals to avoid a non-linear response in the AGET. The power supply board for the camera is installed on the other side of the cabinet and not visible. Figure from \citep{BAGHERI2025169999}}
\label{fig:cabinet}
\end{figure}

The SiPMs used in the camera are S14161-6050HS-04 devices from Hamamatsu. Each SiPM has a size of $6.25$\,mm\,$\times\,6.25$\,mm and consists of 14331 Geiger-mode cells. The S14161-6050HS-04 groups 16 SiPMs into a $4\times4$ matrix and is attached to a front-end readout board \citep{BAGHERI2025169999} where the SiPM signals are amplified and shaped with eMUSIC ASICS \citep{Gomez2016}. The eMUSICs also trim the bias voltage of each SiPM, which is used to flatfield the camera response (see Section \ref{HVFF}). The amplified signals are routed via the camera backplane and through Samtec micro-coaxial cables into a cabinet (see Figure \ref{fig:cabinet}) next to the telescope where the signals are digitized by a 256-channel AGET based digitizer \citep{Pollacco2018}. 

The AGET samples the SiPM signals at a rate of 100\,Megasamples per seconds and stores the analog amplitudes in a 512-cell ring-capacitor sampler. When the readout is triggered the signals stored in the ring sampler are digitized with 12-bit resolution and saved on the computer that is integrated into the cabinet. The cabinet also houses the power supplies for the SiPM bias voltage and camera electronics and a network switch.

Before deployment, we bench-tested the camera, readout, and trigger in the laboratory. Detailed results are given in \citep{BAGHERI2025169999}; we, therefore, restrict ourselves to a summary of the main characteristics. During data taking, we operate the SiPMs about 5\,V above the breakdown voltage of 39\,V, which yields a $>50$\% peak photon-detection efficiency at 470\,nm with a broad tail extending out to 1000\,nm. After shaping the SiPM signals with the eMUSIC and the low-pass filter in front of the digitizer, the full-width at half maximum of the digitized SiPM signals is 30\,ns. The dynamic range of the readout is linear up to 125 detected photoelectrons, which is half of the dynamic range in EUSO-SPB2 \citep{Gazda2023} because we operate the SiPMs at twice the gain, as we detail below. The average electrical crosstalk between camera pixels is 5\% but can be as high as 20\% in some combination of pixels \citep{BAGHERI2025169999}. When the readout is triggered, we record the full $5.12\,\mu$s traces of all camera pixels, which takes 1.44\,ms. Combined with an average trigger rate of less than 10 events per second, the effective dead time of the Demonstrator is below 1.4\%.

The trigger is implemented in a Mesa Electronics’ 7I80HD-25 Field Programmable Gate Array (FPGA) Ethernet Anything I/O card, which receives the discriminator outputs of the 32 eMUSICs of the camera. Inside an eMUSIC, each of its eight input channels connects to a discriminator with an adjustable threshold. The eight discriminator outputs are then combined with a logic OR, and the OR's output is then sent from the eMUSIC to the Mesa FPGA. The FPGA samples the trigger inputs and issues a readout command to the AGET system as soon as it senses a signal on any of the trigger inputs. Thus, the readout triggers when the signal amplitude in one camera pixel goes above the discriminator threshold \citep{BAGHERI2025169999}.

\subsection{Camera Calibration Flasher}

\begin{figure}[!htb]
\centering
\includegraphics[trim = 1500 1000 1300 1000, clip,angle=0,width=0.9\columnwidth]{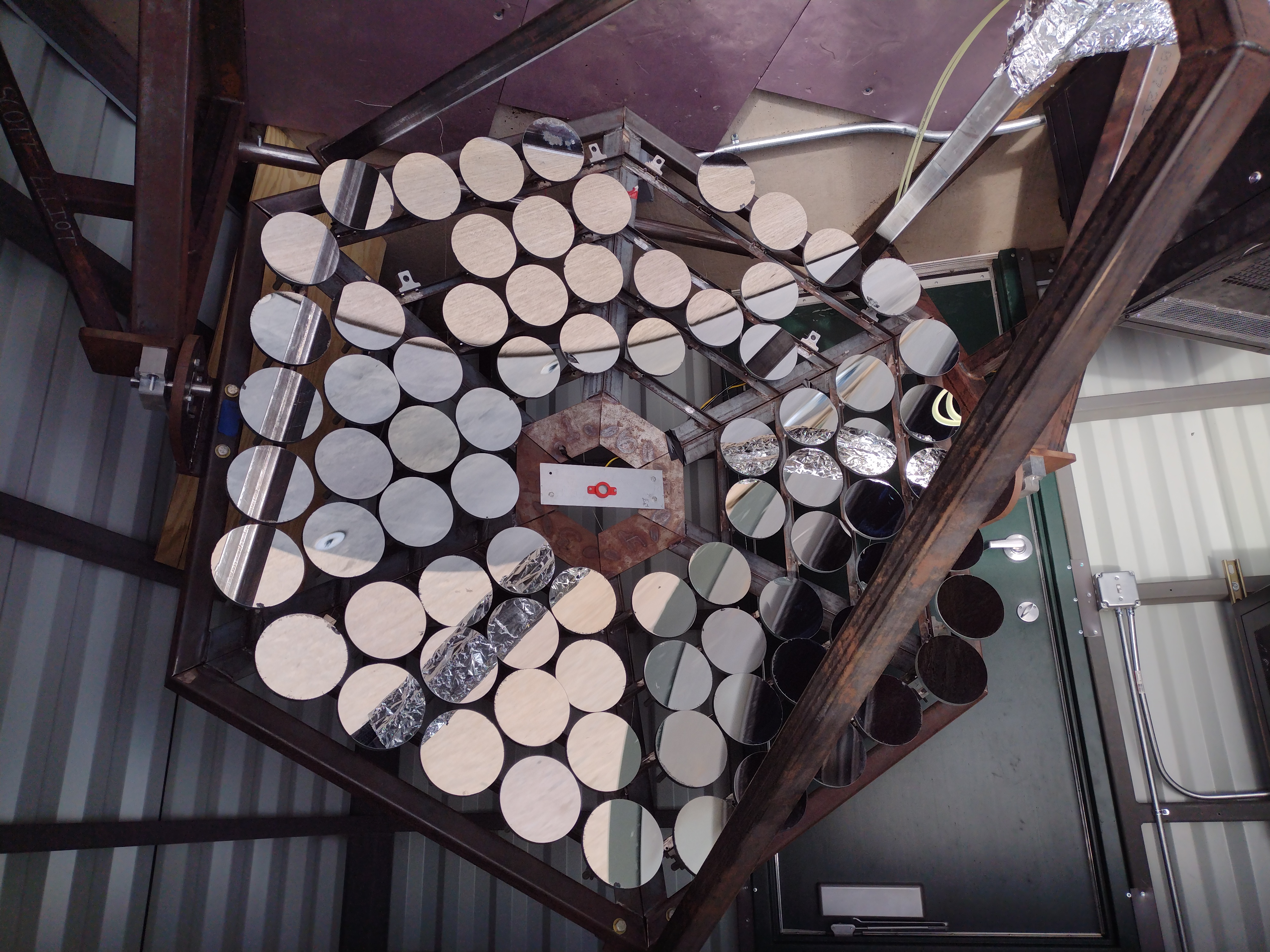}
\caption{Calibration flasher in the center of the OSS with a diffuser installed in its front.}
\label{fig:flasherFront}
\end{figure}

During observing, the stability of the camera response is monitored with a pulsed LED light source installed in the center of the OSS (see Figure \ref{fig:flasherFront}). The Trinity calibration system is derived from the flasher system built for the Small-Sized-Telescopes of the Cherenkov Telescope Array (CTA) \citep{Brown2015ICRC}, which evolved into an Uncrewed-Aerial-Vehicle based calibration system to cross-calibrate Imaging-Air-Cherenkov-Telescope arrays such as CTA \citep{Brown2018,Brown2022}.

\begin{figure}[!htb]
\centering
\includegraphics[angle=0,width=0.9\columnwidth]{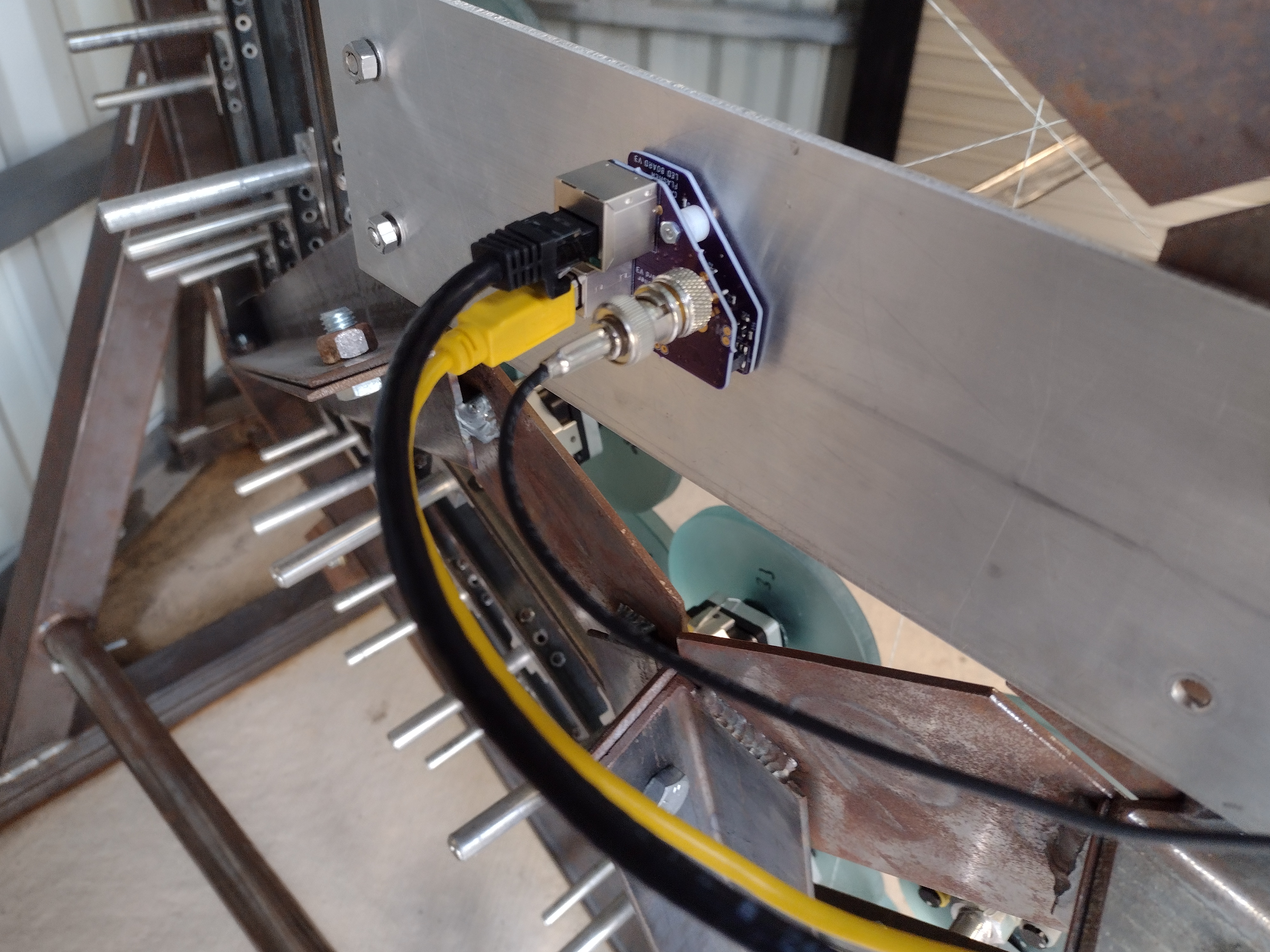}
\caption{Calibration flasher viewed from the back of the telescope.}
\label{fig:flasher}
\end{figure}

The Trinity flasher consists of a single UV LED, which emits light at a wavelength of 400 nm, within a 15 degrees opening angle. The intensity and duration of the light pulse is set by a microcontroller-controlled circuit and a series of resistors and photo relays \footnote{For nightly calibration, we use 6~ns long pulses.} (see Figure~\ref{fig:flasher}). Placed in front of the LED is a UV-fused silica ground glass diffuser DG10-220-MD to achieve uniform illumination of the camera. The calibration flasher is triggered with 1 Hz throughout observations.  We assessed the uniformity of the flasher intensity across the camera by moving a standalone SiPM across the focal plane, and found that the flasher intensity is uniform  within the 1\% uncertainty of our measurement. 

The flasher is triggered with a Transistor-Transistor Logic (TTL) signal from the telescope's trigger board and is powered via USB by the camera computer. The USB connection also allows in-situ programming of the flasher.

\subsection{Supporting Infrastructure}
\label{sec:Aux}
The telescope operation is supported by several auxiliary systems, which are described in this section.

\begin{figure}[!htb]
\centering
\includegraphics[angle=0,width=0.9\columnwidth]{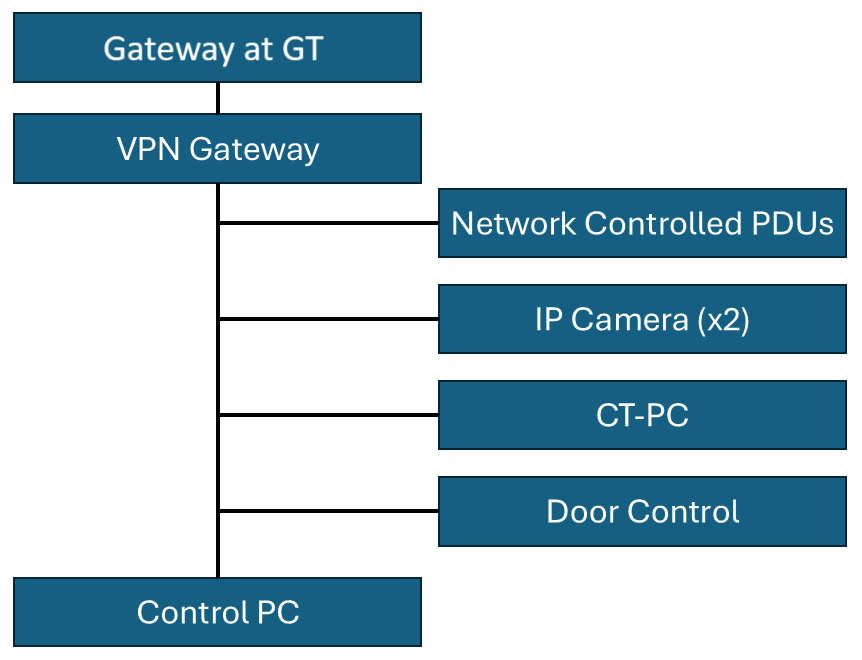}
\caption{Block diagram of the onsite network.}
\label{fig:NetworkBlockDia}
\end{figure}

\paragraph{Telescope Building} The Demonstrator is housed in an 8\,ft tall building with an $11 \times 12$ square foot footprint (see Figure \ref{fig:Demonstrator}). The sides and roof of the building are covered with corrugated steel sidings. The roof can be partially removed to observe cosmic-ray air showers. Utilities connecting to the building are one network cable and one 120\,V power line. The building is accessible through a door, and integrated into the west-facing side of the building is a roll-up door, which is opened during observing.
\paragraph{Network}\label{sec:Network}
The on-site network is accessible through a Palo Alto PA-440 VPN gateway only from a gateway machine at Georgia Tech.  Figure \ref{fig:NetworkBlockDia}, shows a block diagram of the devices connected to the on-site network. The CT-PC controls the telescope's camera and readout and stores the data taken with the telescope. From there, the data is transferred daily to the computing cluster at Georgia Tech. The Control PC manages the data from the weather station, two web cameras, and the horizon camera. It also controls the roll-up door of the building and automatically closes it in the presence of adverse weather conditions or if the sun is about to rise. Two Cyperpower PDU41001 network-controlled power distribution units (PDUs) with an SSH/HTTP server allow the cycling of the power for any device onsite, including the computers, cameras, weather station, heating mat, chiller, and the individual components of the telescope.

\paragraph{Roll-up door} The roll-up door in front of the telescope is a commercial product from Overhead Door. Opening and closing is controlled by an ethernet relay controller, uSwitch, from uHave control. The uSwitch has two relays. One relay controls the direction the shutter moves and the second relay engages the shutter motor. These relays can be controlled via a web page interface or through TCP commands sent to the uSwitch in a Python program. The uSwitch is powered with a 12\,V power supply connected to one of the PDU outlets.

\paragraph{Weather Station}
Attached to the north-west corner of the building is a Maximet 550 weather station from Gill Instruments (see Figure \ref{fig:Demonstrator}). It provides real-time wind speed and direction, atmospheric pressure, relative humidity, and temperature information. 

Data from the Maximet 550 is transmitted to the control computer via a serial connection and the weather information is sampled once per second and saved into an ASCII file. Additionally, the weather information is uploaded every 10 minutes to Windy.com\footnote{\url{https://www.windy.com/station/pws-f058b862?31.185,-113.291,5}}, an online public weather service. The weather station is powered by a 12\,V power supply connected to a PDU outlet.

\begin{figure}[!htb]
\centering
\begin{subfigure}[b]{1.\columnwidth}
\center{\includegraphics[angle=0,width=0.9\columnwidth]{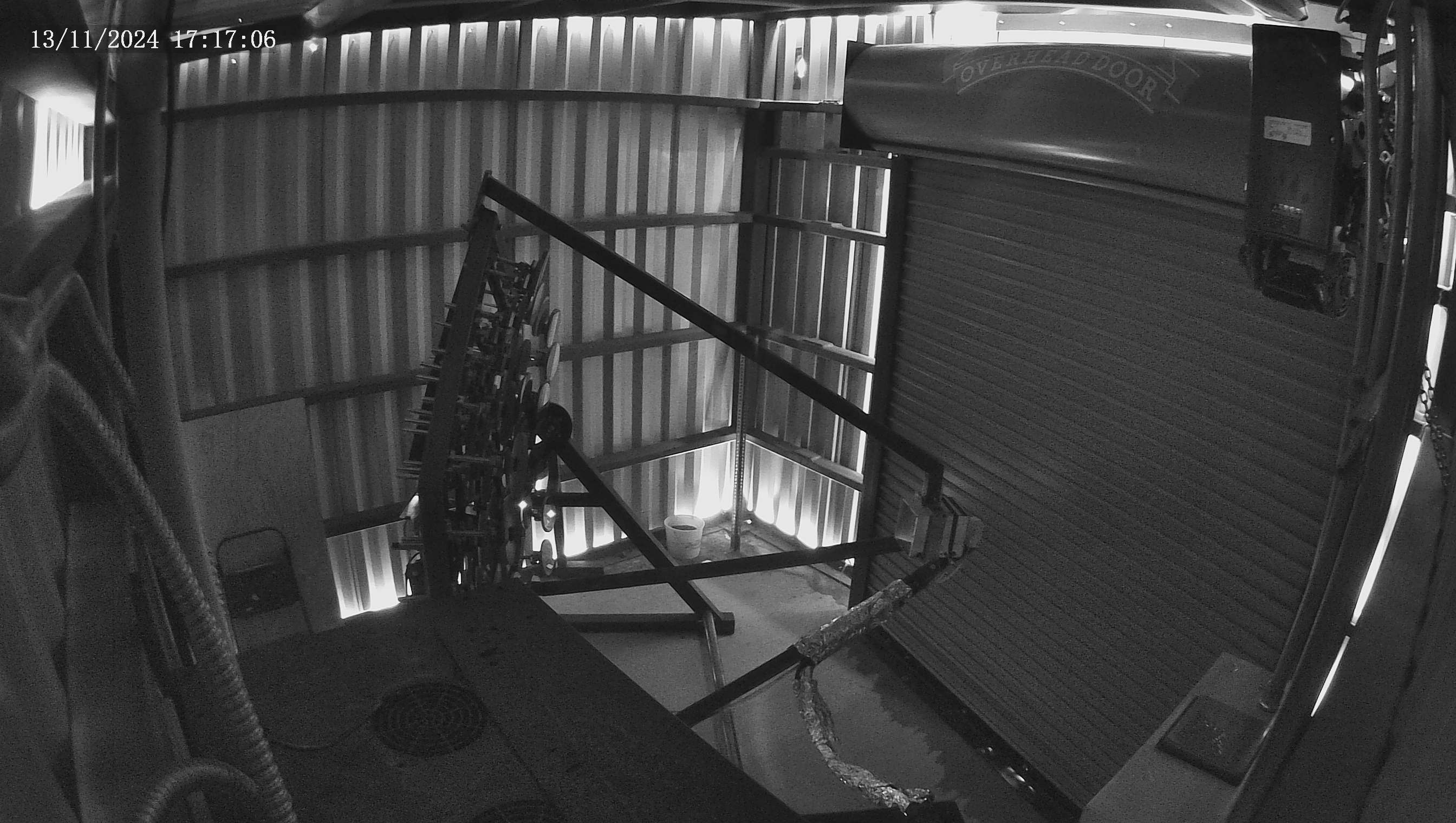}}
\caption{}
\label{fig:cameras_a}
\end{subfigure}
\vfill
\begin{subfigure}[b]{1.\linewidth}
\center{\includegraphics[angle=0,width=0.9\columnwidth]{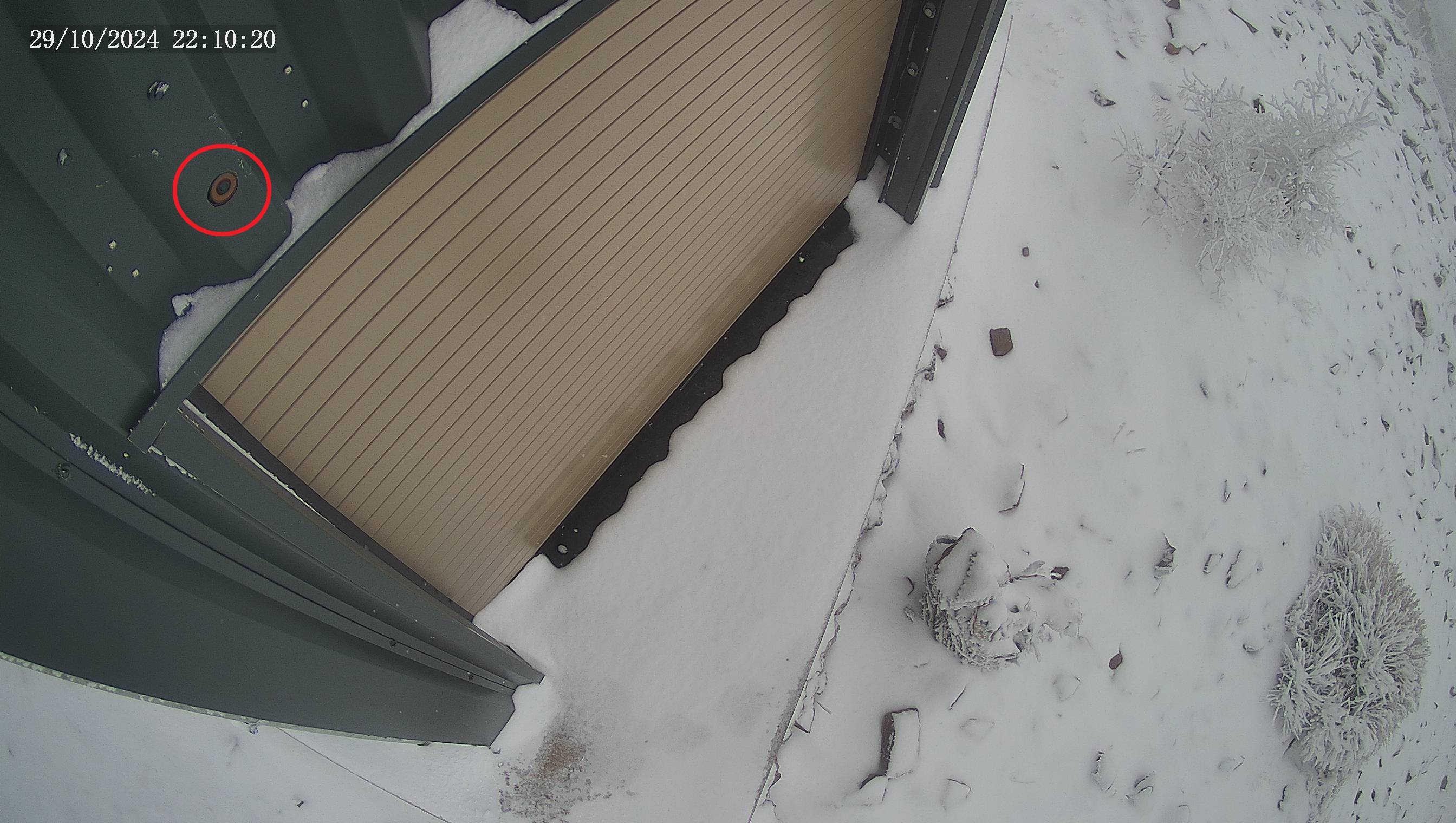}}
\caption{}
\label{fig:cameras_b}
\end{subfigure}
\caption{Pictures taken with the SV-VP4-N cameras installed (a) inside the building and (b) outside the building. The picture also shows the horizon camera installed encircled in red above the roll-up door in the upper left corner of the picture.}
\label{fig:cameras}
\end{figure}

\paragraph{Monitoring cameras}
There are three cameras installed on site. Two of them are internet protocol (IP) cameras, SV-VP4-N, manufactured by SureVision and powered over ethernet. The imaging sensor of the SV-VP4-N is a Starlight CMOS chip that excels in low-light conditions and has high IR sensitivity for night vision. Figure \ref{fig:cameras_a} shows a snapshot taken with the camera installed inside the building, and Figure \ref{fig:cameras_b} shows a snapshot taken with the camera installed on the outside of the building pointed at the roll-up door. Images of each camera are taken every minute and saved on the control computer. In addition, the indoor camera is programmed to start recording additional images every time it detects motion.

\begin{figure}[!htb]
\centering
\includegraphics[angle=0,width=0.9\columnwidth]{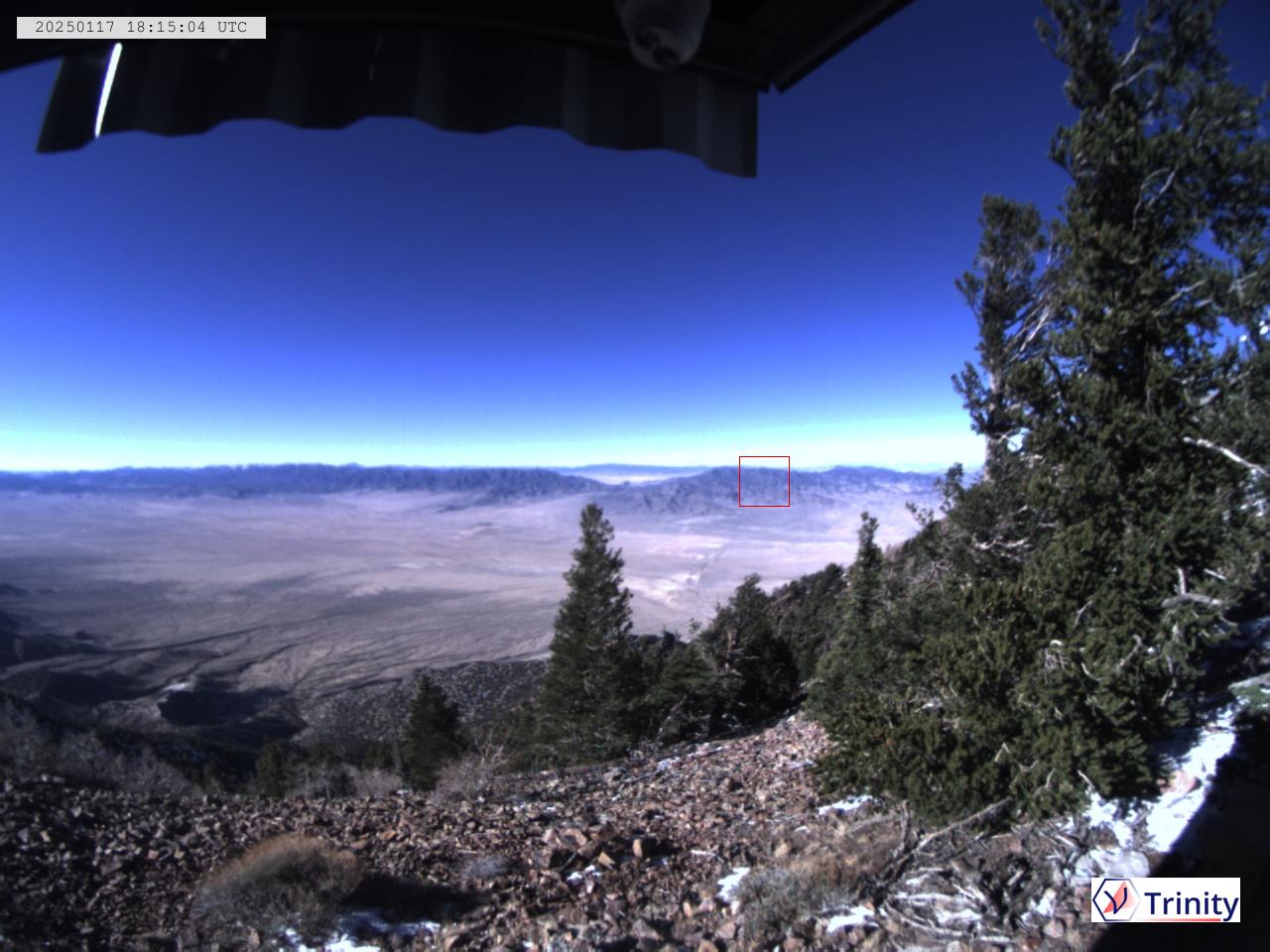}
\caption{Picture taken with the horizon camera. The red square shows roughly the pointing direction and field of view of the Demonstrator \label{fig:horizoncam}}
\end{figure}

The third camera is a planetary astronomy camera ZWO ASI120MC-S with a  $150^\circ$ field of view 28-mm lens for all-sky viewing. The imaging sensor is a 1/3'' CMOS AR0130CS with 1280$\times$960 pixels. The camera connects with USB 3.0 to the control computer. It is installed above the roll-up door and points at the horizon, which is why we call it the horizon camera (see Figure \ref{fig:cameras_b}). Figure \ref{fig:horizoncam} shows a snapshot taken with the horizon camera. 

The horizon camera allows us to assess the condition of the sky while observing. Pictures are taken every 5 minutes and saved to the control computer. 

The pictures of all three cameras are uploaded to the public \textit{Trinity} website to provide a live view of the site. Additionally, the snapshots are processed into daily timelapse videos and uploaded to YouTube\footnote{\url{https://www.youtube.com/@demonstratorcams}}

\paragraph{Heating Mat}

During winter, snow can build up in front of the roll-up door, preventing it from properly closing. To avoid build-up, we installed a $13\times365$\,cm$^2$ Tempurtech Heated Roof mat in front of the door. The 120\,W mat is connected to a PDU outlet with an auto-reset GFCI extension cord. Figure \ref{fig:cameras_b} shows the mat in action during a recent snow event.

\paragraph{Camera Cooling}

The camera cooling system is described in \citep{BAGHERI2025169999}. In summary, the front-end chips are thermally coupled to heat pipes, which transfer the heat into a liquid-cooled cold plate on top of the camera (see Figure \ref{fig:camera}).
The cold plate is connected to a CW-5200 Industrial Chiller by Omtech, which is set at a fixed temperature of $0^\circ$C. The chiller is positioned on the floor next to the readout cabinet. 

\section{Commissioning}
\subsection{Mirror Alignment}\label{sec:mirroralignment}

\begin{figure}[!htb]
\centering
\includegraphics[angle=0,width=0.9\columnwidth]{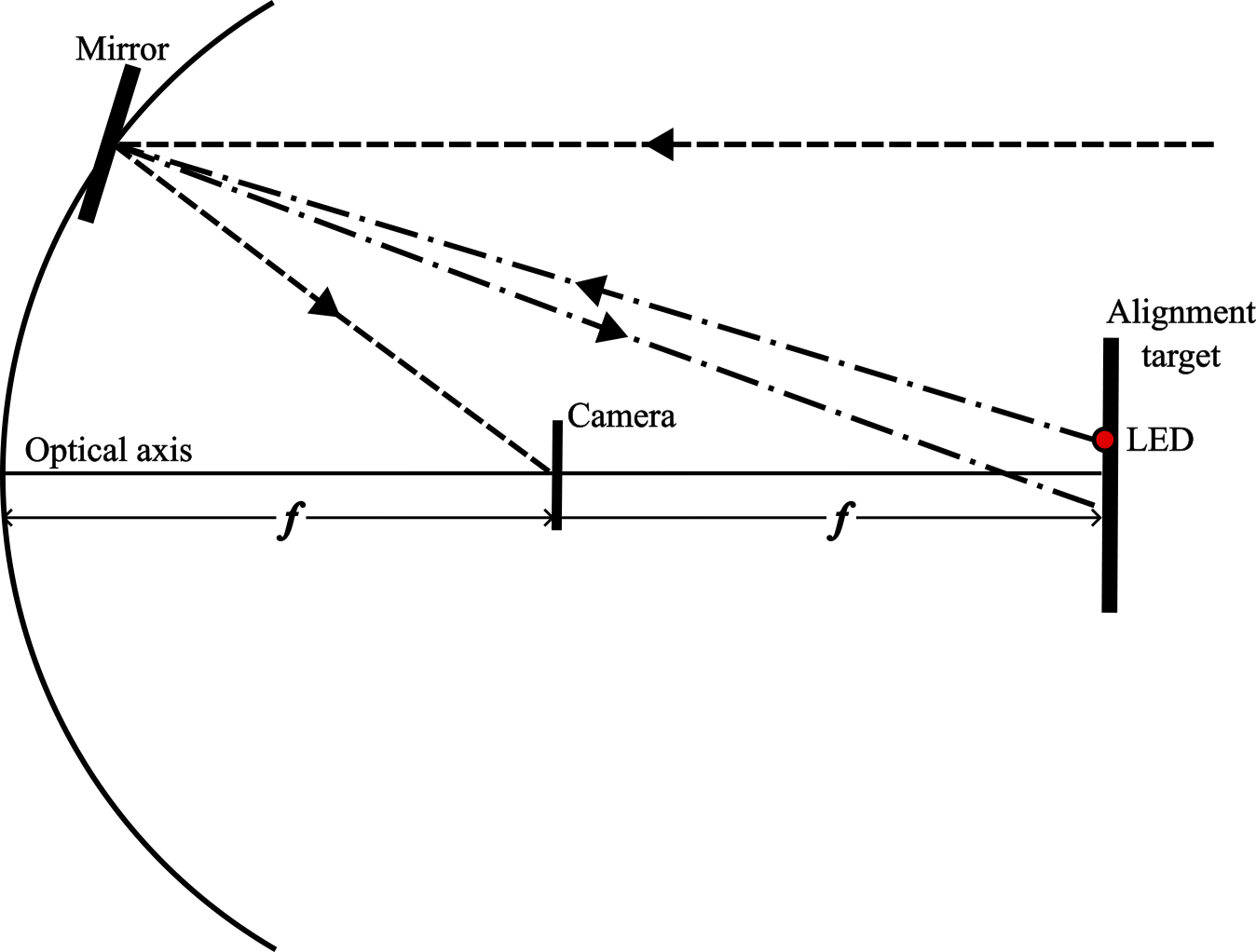}
\caption{Sketch of the alignment procedure. An LED is offset from the optical axis in the Alignment target and imaged back onto the target. The dot-shaded line represents one ray reflecting off one mirror facet. The dashed line represents a ray from a light source at infinity on the optical axis.}
\label{fig:alignmentSketch}
\end{figure}

\begin{figure}[!htb]
\centering
\includegraphics[angle=0,width=0.9\columnwidth]{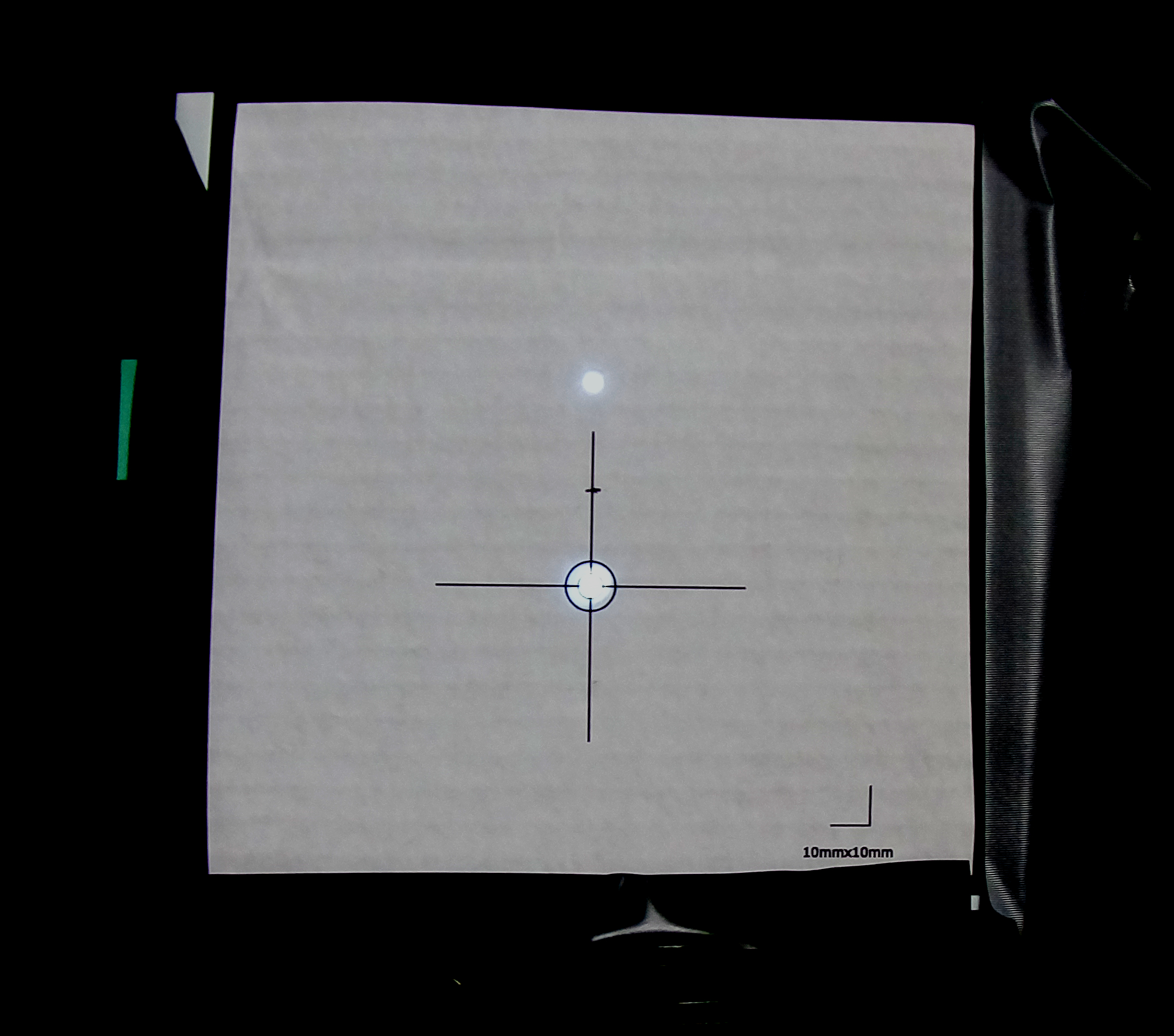}
\caption{Image of the mirror-alignment screen at the 2f point. The light spot on top is the LED behind the pinhole. Centered in the crosshair at the bottom is the image of the LED projected by all the mirror facets onto the screen after a complete alignment of all mirror facets.}
\label{fig:alignment}
\end{figure}

For aligning the mirror facets, we employ the method used by the Whipple Collaboration for their 10\,m Davies-Cotton telescope \citep{Kildea2007}. In this approach, all mirrors are aligned by referencing an image at twice the focal length, the 2f point. (see Figure \ref{fig:alignmentSketch}). We implemented the method by placing a screen along the optical axis at the 2f point or 2.96\,m from the dish center. The screen has a pinhole with an LED about 1 inch above the optical axis (see Figure \ref{fig:alignment}). According to geometrical optics, the fully aligned telescope must project an image of the LED at the same distance from the optical axis but on the opposite side of the optical axis. We accordingly aligned the reflected image of each mirror facet with the tip/tilt adjustment knobs of the mirror mounts. The LED image of the fully aligned Demonstrator is shown in Figure \ref{fig:alignment}. The mirror alignment takes two people to complete in two hours. 

\begin{figure}[!htb]
\centering
\includegraphics[trim = 20 5 35 15, clip,angle=0,width=0.9\columnwidth]{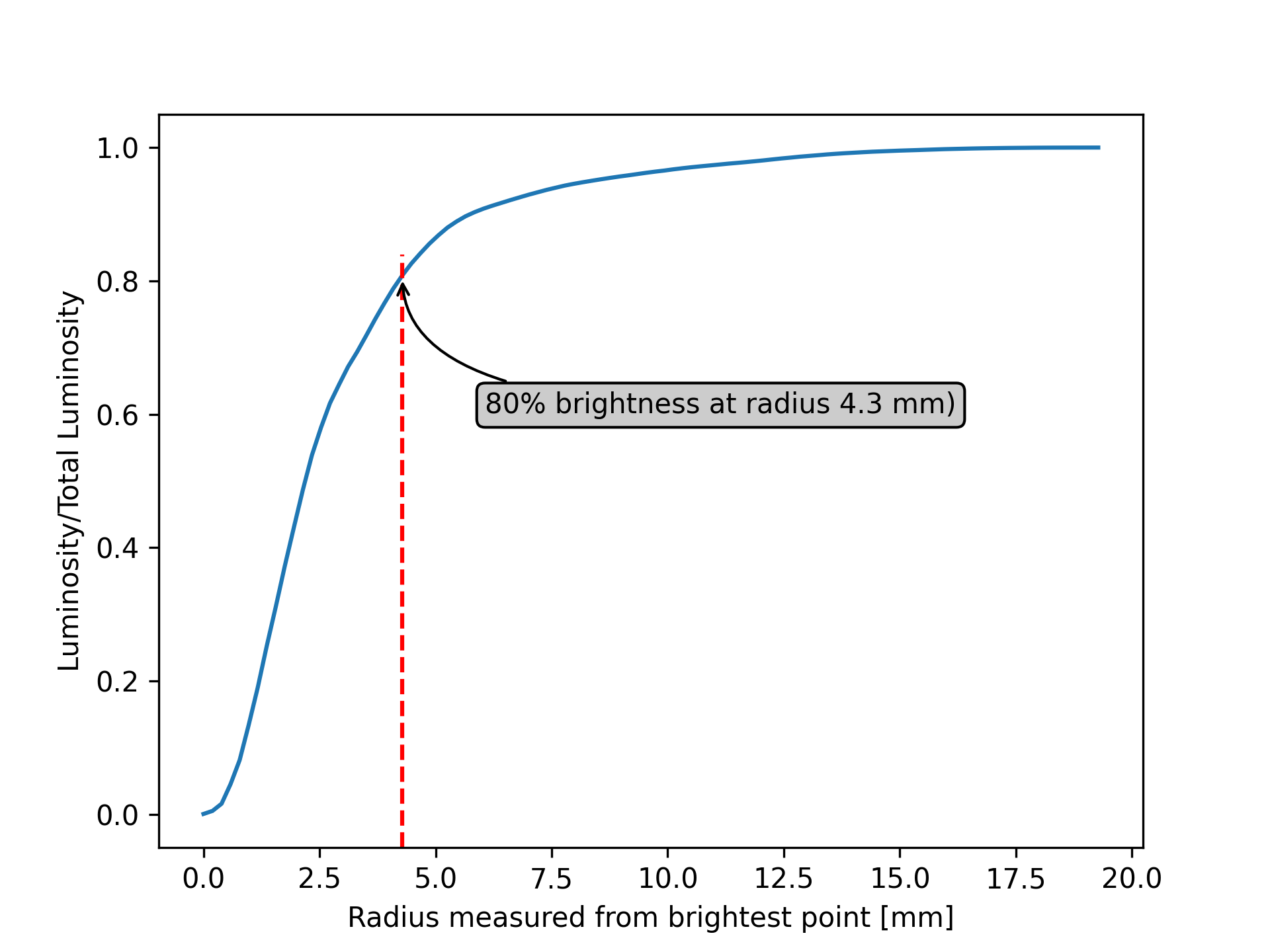}
\caption{Radial brightness distribution of an image at the 2f point after mirror alignment}
\label{fig:psf}
\end{figure}

We quantified the optical point spread function of the telescope by taking two pictures of the screen with a Raspberry Pi HQ Camera and an Arducam $2.8-12~\text{mm}$ Varifocal Lens; one picture with the LED on and one with the LED off for background subtraction. Figure \ref{fig:psf} shows the background-subtracted radial brightness distribution at the screen with an $80\%$ containment radius of  $4.3~\text{mm}$. However, the brightness distribution at the focal plane produced by a point source at infinity is expected to be narrower. From simulations tuned to our 2f-point alignment measurements, we conclude that the image of a point source at infinity on-axis has a $2.1$\,mm 80\% containment radius in the focal plane, or 1/3 of the size of a SiPM.

\begin{figure}[!htb]
\centering
\includegraphics[trim = 0 0 0 110, clip, angle=180,width=1.\columnwidth]{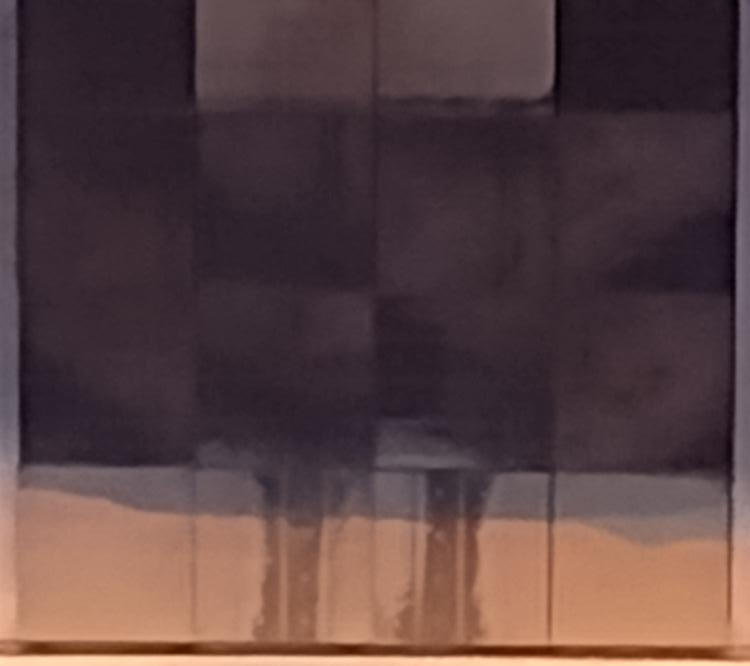}
\caption{Image of the first row of SiPM matrices (4x16 SiPM camera pixels) taken after sunset. Imaged onto the SiPM camera by the telescope optics is the mountain range in the west at which the Demonstrator points. The ridge is clearly resolved, demonstrating the optical performance of the Demonstrator. The dark vertical lines in the middle are reflection artifacts.}
\label{fig:horizon}
\end{figure}

The good optical performance of the Demonstrator is also demonstrated with Figure \ref{fig:horizon}. The picture shows the part of the SiPM camera that is pointed at the horizon and was taken shortly after sunset. The dark horizontal structure is the mountain range to the west, which the telescope optics images onto the SiPM camera. A perfect optics would image a sharp transition from dark to bright, as described by a step function, and any deviation from a step function is due to the telescope's PSF. Because the vertical scale of the image spans four rows of SiPMs and the transition happens within a fraction of the size of a SiPM, we conclude that the PSF across the entire lateral field of view must be well within one SiPM.

\subsection{Camera Flat-Fielding and Gain Calibration\label{sec:Characterization}}

\subsubsection{Flat-Fielding}\label{HVFF}

Each SiPM has a slightly different response due to differences in gain, breakdown voltage, and photon detection efficiency. Furthermore, the gain varies between readout channels. The result is an inhomogeneous response of the camera to light if all SiPMs are biased at the nominal 44\,V. To arrive at a homogeneous response, we adjust the bias of each SiPM with the following flat-fielding procedure.

We start by recording 3,000 flashes with the camera calibration flasher at a 42\,V SiPM bias voltage and another set of 3,000 flashes at a 44\,V. From the flashes recorded at 44\,V, we then determine the relative response of each SiPM to the camera median. To find the SiPM bias that minimizes the spread in the relative response, we make use of the linear dependence of the SiPM gain on its bias \citep{BAGHERI2025169998}, and that the linear response is fully characterized by the 42\,V and 44\,V data sets. With the linear response of each SiPM, we then calculate updated bias values that bring the response of each SiPM in line with the camera median. Finally, the changes in bias voltage are programmed into the eMUSIC bias trim voltage.

\begin{figure}[!htb]
\begin{subfigure}{1.\columnwidth}
\center{\includegraphics[width=0.9\columnwidth]
{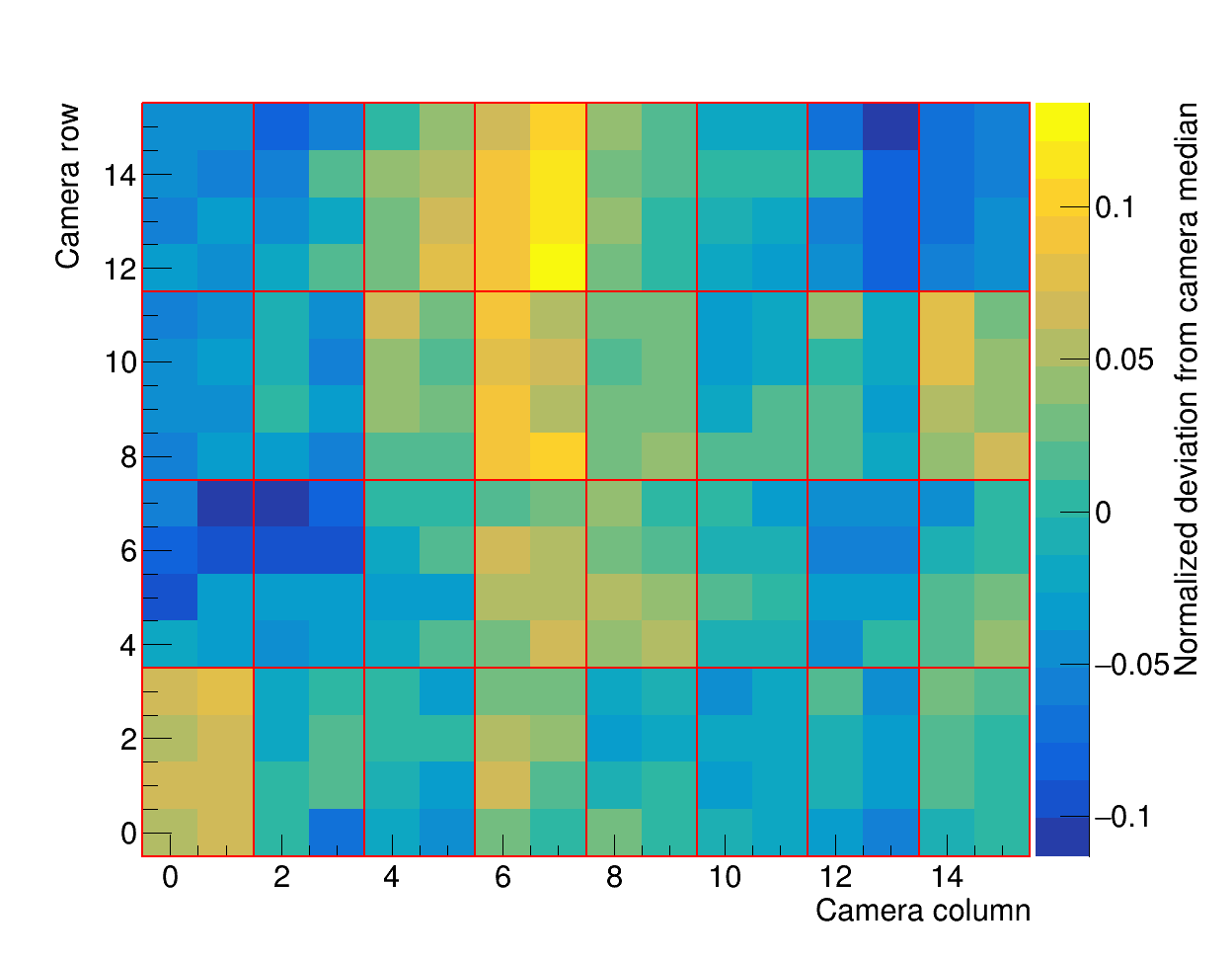}}
\caption{Relative response of each camera pixel to the camera median.}
\label{fig:dist_FF_a}
\end{subfigure}
\vfill
\begin{subfigure}{1.\columnwidth}
\center{\includegraphics[width=0.9\columnwidth]{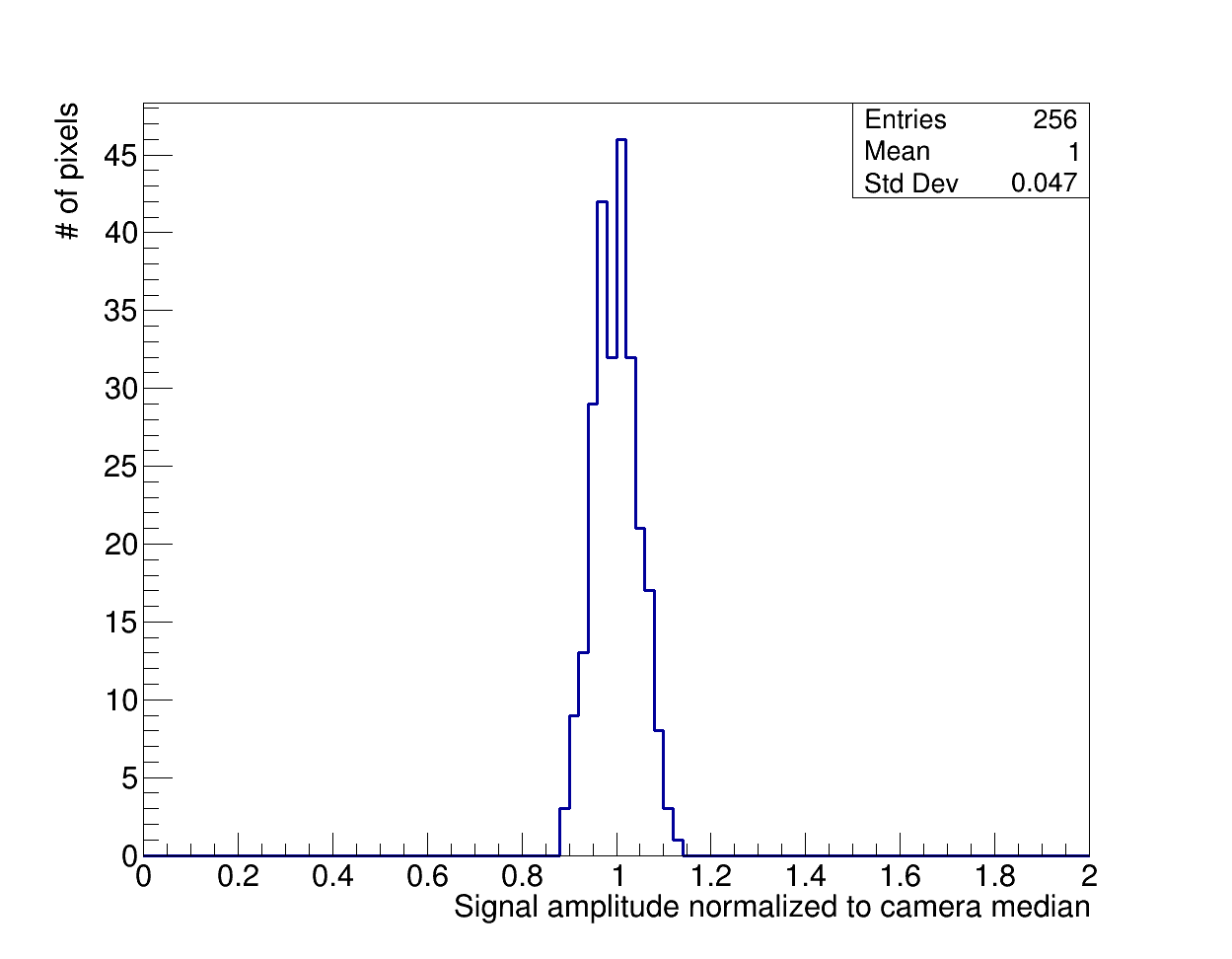}}
\caption{Distribution of the relative pixel responses to the camera median.}
\label{fig:dist_FF_b}
\end{subfigure}
\caption{Relative response of the flat-fielded SiPM camera to calibration flashes normalized to the camera median.}
\label{fig:dist_FF}
\end{figure}

After flat-fielding, the camera's response is uniform within a 5\% standard deviation (see Figure \ref{fig:dist_FF}). The SiPM bias remains fixed during observing, which causes the standard deviation to vary between 5\% and 10\% due to the non-uniform response of the SiPMs to changes in, e.g., background light intensity and temperature.

\subsubsection{Gain Calibration}

Essential for the event reconstruction is the calibration of the camera signals in units of photoelectrons, \emph{i.e.}, the signal amplitude one detected photon produces in the readout.

While SiPMs produce single photoelectron signals, the low bandwidth of the Demonstrator readout spreads the SiPM signals so much in time that the dark count signals significantly overlap preventing the identification of single photoelectron signals and a direct measurement of the calibration factor. Instead, we calibrate the intensity of the calibration flasher and infer the gain calibration factor from the flasher signals recorded with the Demonstrator readout. 

\begin{figure}[!htb]
\centering
\includegraphics[angle=0,width=0.9\columnwidth]{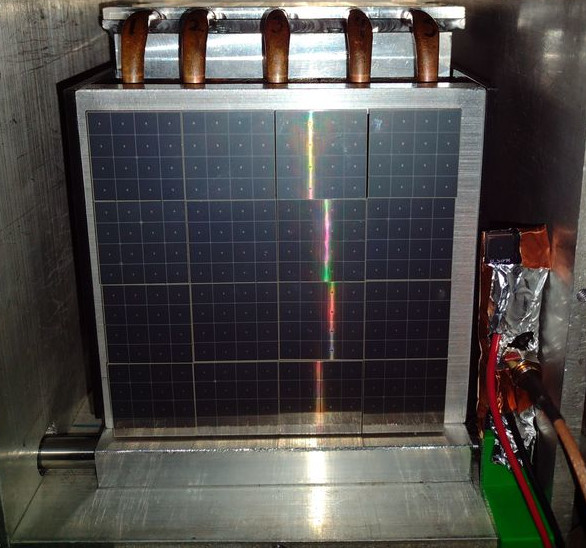}
\caption{Setup to calibrate the camera signals in photoelectrons. To the right of the camera is the calibration S14160-6050HS SiPM, which can slide across the camera to test the uniformity of the calibration flasher. \label{fig:GainCalibSetup}}
\end{figure}

To calibrate the flasher intensity in average number of photons detected by an SiPM per flash, we used a separate S14160-6050HS SiPM, the same SiPM type used in the camera. Figure \ref{fig:GainCalibSetup} shows that calibration SiPM installed in the camera. The signals of the calibration SiPM are amplified with a Mini-Circuits ZFL-500LN+ amplifier and then shortened with a passive RC-high-pass filter before being displayed on a 500\,MHz bandwidth TekTronix TDS 3054C oscilloscope. 

The calibration SiPM and the camera SiPMs are biased at the same 5\,V overvoltage and, therefore, a flash from the calibration flasher generates about the same number of photoelectrons in the calibration SiPM as in each camera SiPM.

\begin{figure}[!htb]
\begin{subfigure}{1.\linewidth}
\center{\includegraphics[width=1.0\linewidth]
{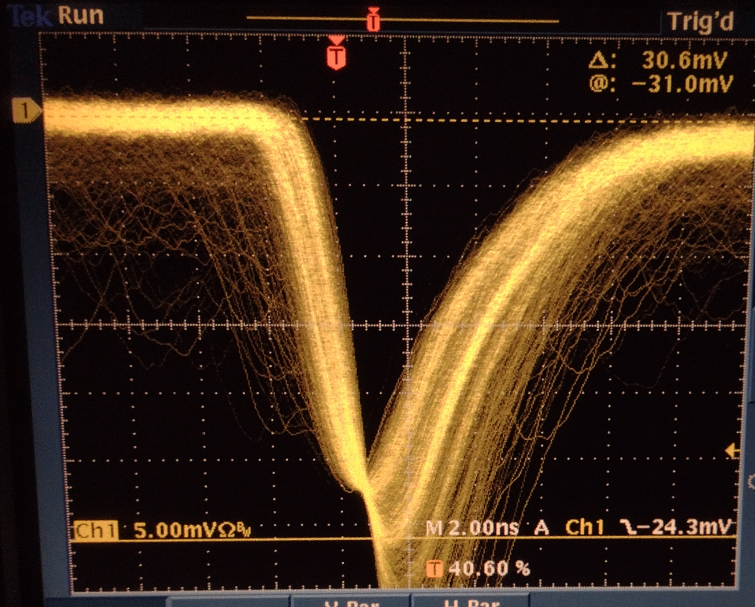}}
\caption{}
\label{fig:gain_calibration_a}
\end{subfigure}
\vfill
\begin{subfigure}{1.\linewidth}
\center{\includegraphics[width=1\linewidth]{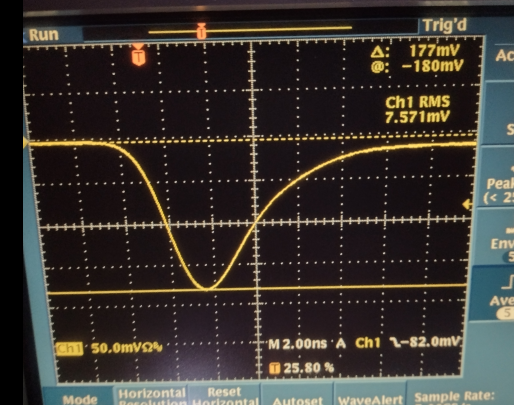}}
\caption{}
\label{fig:gain_calibration_b}
\end{subfigure}
\caption{ (a) Optical crosstalk from dark count signals of the calibration SiPM. The oscilloscope trigger is set slightly below 6 photoelectrons, and the horizontal yellow line is centered at the average 7 photoelectron signal. (b) Averaged signal (256 traces) of the calibration flasher recorded with the the calibration SiPM.}
\label{fig:gain_calibration}
\end{figure}

Figure \ref{fig:gain_calibration_a} shows optical crosstalk dark-count pulses of the calibration SiPM. The bands corresponding to the pile-up of 6, 7, and 8 simultaneously fired SiPM cells are clearly identifiable. From the average signal of 7 photoelectrons (PEs) at 30.7\,mV (horizontal solid yellow line), we determine a $4.37\pm0.01$mV/PE conversion factor for the calibration SiPM. 

We then flashed the calibration SiPM with the calibration flasher, obtaining an average amplitude of $177\pm1$\,mV (see Figure \ref{fig:gain_calibration_b}). Divided by the conversion factor, we obtain $40.5\pm0.2$ PEs as the average detected number of photons in the calibration SiPM and thus also in each camera SiPM. Because the calibration flasher signals are recorded by the camera SiPMs with a median amplitude of 975\,DC (DC = digital counts) in the camera readout, we finally arrive at a calibration factor of $24.1\pm0.1$ DC/PE per camera pixel. The calibration measurement was done at a temperature of $12.7\pm0.1^\circ$C. Changes of the calibration factor due to temperature are taken care off during the data analysis.

In bench testing, we measured that the camera response is linear up to 2,400 DC and becomes nonlinear at larger signals due to the AGET digitizer (see \cite{BAGHERI2025169999}). With a $24.1\pm0.1$ DC/PE calibration factor, the dynamic range of one channel in the readout is, therefore, $\sim$100 photoelectrons. The dynamic range is adequate for the Demonstrator with its small light-collection area. In the most extreme and most unlikely case that an air-shower develops along the telescope's optical axis, the Cherenkov signal from a 1 PeV neutrino-induced air shower saturates the readout if it is 25\,km or closer and a 100 PeV neutrino signal would saturate the readout if it is closer than 100\,km. However, in the majority of cases, the air shower will develop at an angle to the optical axis and at much larger distances. Therefore, the signal spreads out over several pixels and a much smaller signal will be detected per pixel \citep{Otte2019d}.

\subsection{Trigger Calibration and Flatfielding}\label{tFF}

To achieve a uniform trigger response across the camera, we had to flatfield the trigger thresholds of the camera pixels. We accomplished that by flashing the camera with the calibration flasher at a rate of 100\,counts per second (cps) and scanning the discriminator thresholds, recording the pixel trigger rate at each threshold setting. 

\begin{figure}[!htb]
\begin{subfigure}{1.\linewidth}
\center{\includegraphics[width=1\linewidth]
{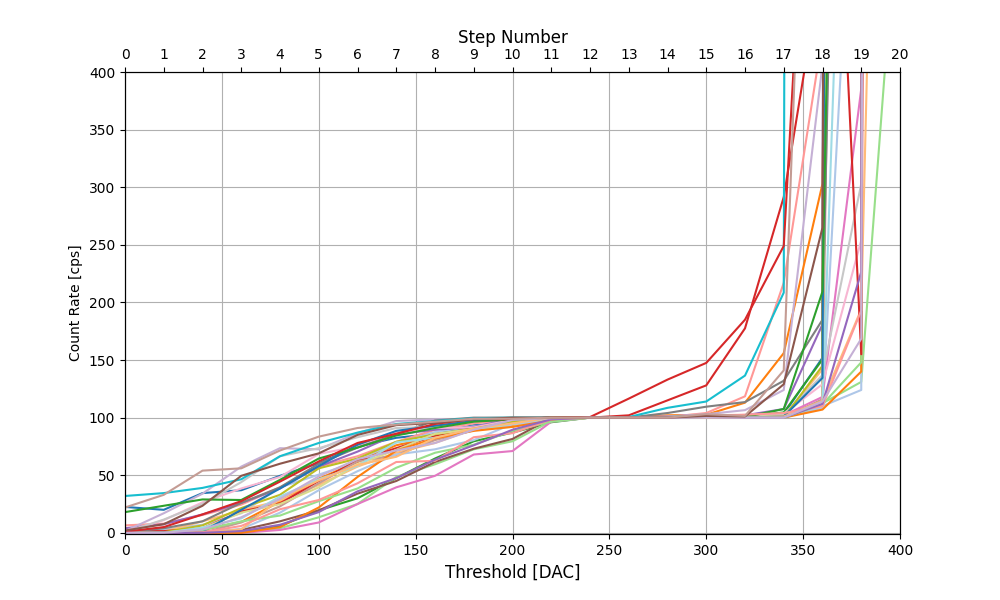}}
\caption{Uncalibrated trigger rate scan curves for 32 camera pixels.}
\label{fig:trigger_FF_a}
\end{subfigure}
\vfill
\begin{subfigure}{1.\linewidth}
\center{\includegraphics[width=1\linewidth]{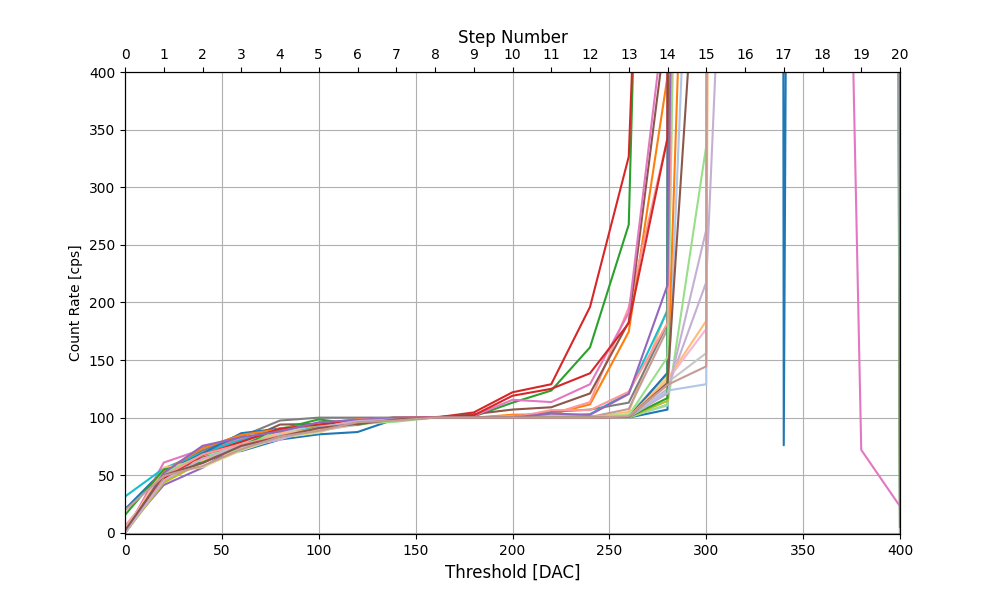}}
\caption{Calibrated trigger rate scan curve for the same 32 camera pixels.}
\label{fig:trigger_FF_b}
\end{subfigure}
\caption{Example trigger rate scans of 32 camera pixels before and after discriminator threshold calibration. The trigger flat-fielding procedure is described in section~\ref{tFF}.}
\label{fig:trigger_FF}
\end{figure}

Figure \ref{fig:trigger_FF_a} shows as an example the trigger-rate scans of 32 camera pixels. The trigger threshold changes from high to low with increasing digital-to-analog converter (DAC) value. The trigger rate in each scan changes from zero to a plateau at 100\,Hz, which is when a pixel triggers on all flasher signals. Decreasing the discriminator threshold further (higher DAC), the discriminator eventually triggers in the noise, resulting in a sharp increase of the trigger rate at about 350\,DAC.

The trigger rate scans vary from pixel to pixel because the discriminator thresholds are not calibrated. We flatfielded the trigger response by determining for each pixel the DAC value for which the discriminator triggers on 50\% of the flasher signals. We subtracted the camera median DAC value from the 50\% trigger-rate threshold setting and added the difference to the DAC setting of the corresponding pixel. Figure \ref{fig:trigger_FF_b} shows the thus calibrated trigger rate scans for 32 pixels. The trigger flatfielding reduces the dispersion in the trigger response from $31.0$\,DAC to $4.0$\,DAC.

In the final step, we calibrated the trigger threshold in photoelectrons using a) the known intensity of the flasher of 48\,PE  per flash, b) the camera averaged threshold values where the efficiency triggering on flasher signals reaches 50\% (22\,DAC), and c) the threshold value where the trigger rates diverge due to noise, \emph{i.e.} 242\,DAC = 0\,PE.

For observing, we found that a trigger threshold of 150\,DAC corresponding to 20\,PE provides stable operations during dark and moonless nights with a trigger rate of less than 0.5\,Hz.

\section{Performance Measurements}
\subsection{Direct Muon Hits in the Camera SiPMs}

One source of background events for the Demonstrator is direct muon hits in the camera's SiPMs. When a muon traverses a SiPM, its ionization trail fires one or more SiPM cells due to a combination of the diffusing charge cloud from the ionization and optical crosstalk. In normal data taking, these events are recorded if the number of fired SiPM cells is above the trigger threshold of 20\,PE.

\begin{figure}[!htb]
\begin{subfigure}{1.\columnwidth}
\centering{\includegraphics[trim={3cm 5cm 5cm 1cm},width=0.8\columnwidth]
{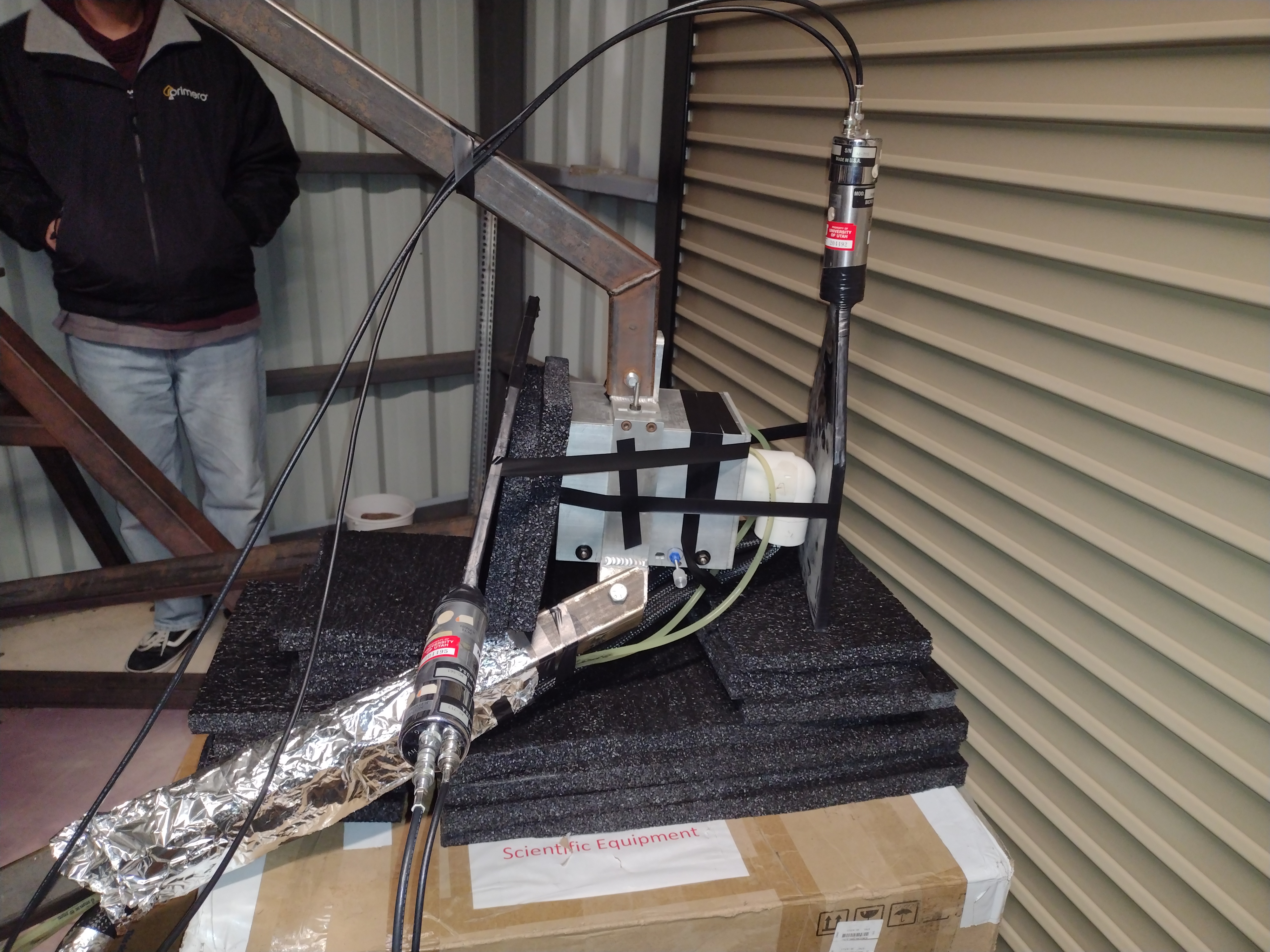}}
\vspace{5pt}
\caption{Horizontal muon paddle configuration.}
\end{subfigure}
\begin{subfigure}{1.\columnwidth}
\centering{\includegraphics[width=0.8\columnwidth]{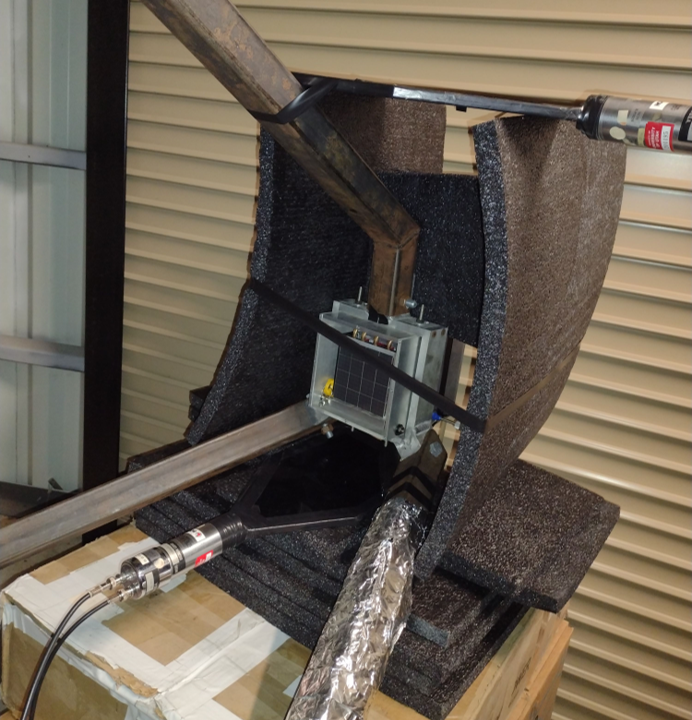}}
\caption{Vertical muon paddle configuration.}
\end{subfigure}
\caption{The two muon paddle configurations used to characterize muon signals in the camera SiPMs.}
\label{fig:muonpaddles}
\end{figure}

We identified muon signals in the camera by installing two plastic scintillator detectors (muon paddles) above and below the camera (vertical configuration) and in a second measurement at the front and back of the camera (horizontal configuration), see Figure \ref{fig:muonpaddles}. In the measurement, a coincidence logic built from NIM modules triggers the Demonstrator readout when a muon traverses both paddles. In the vertical configuration, the coincidence selects muons originating from air showers above the Demonstrator, and in the horizontal configuration, it selects muons originating from air showers developing at the horizon. 

\begin{figure}[!htb]
\centering
\includegraphics[trim={1cm 11cm 0.5cm 0.2cm},clip,angle=0,width=1.\columnwidth]{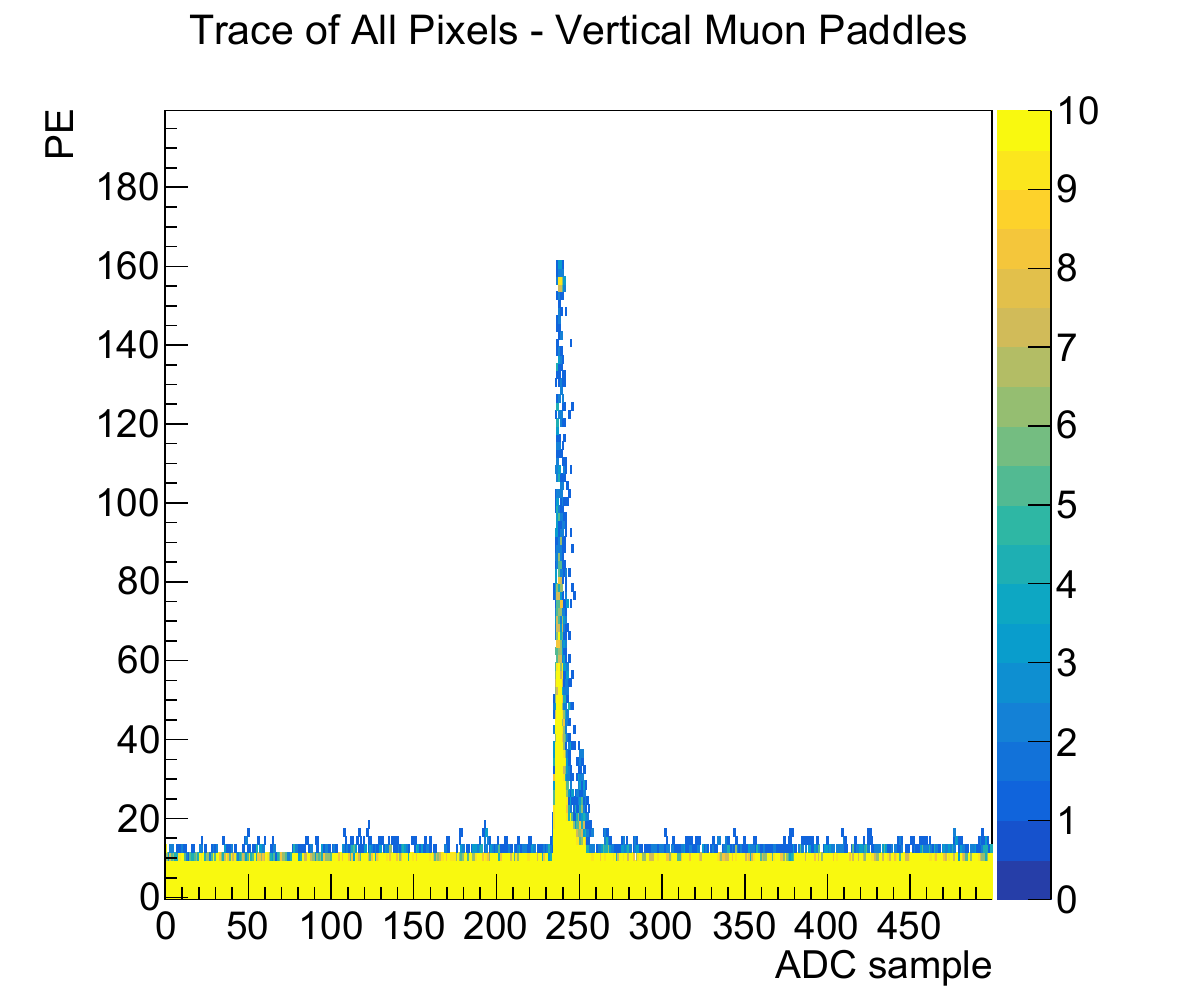}
\caption{Overlaid traces of all camera SiPMs of events triggered by the muon-paddle coincidence.}
\label{fig:muontraces}
\end{figure}

\begin{figure}[!htb]
\begin{subfigure}{1.\columnwidth}
\center{\includegraphics[trim=1.5cm 6.5cm 0.5cm 2cm,clip,width=\columnwidth]
{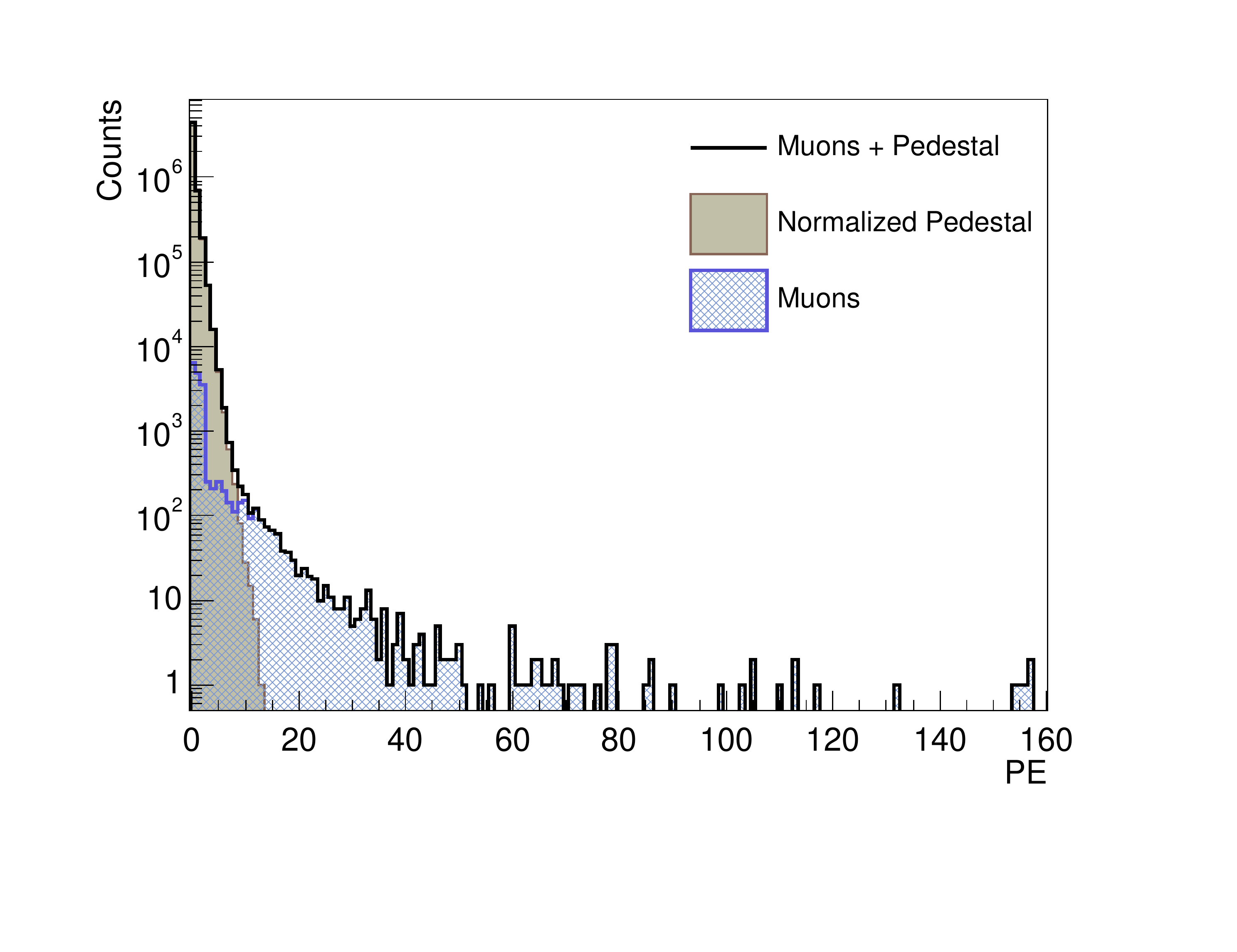}}
\caption{horizontal muon paddle configuration.}
\end{subfigure}
\vfill
\begin{subfigure}{1.\columnwidth}
\center{\includegraphics[trim={1.5cm 6.5cm 0.5cm 2cm},clip,width=\columnwidth]{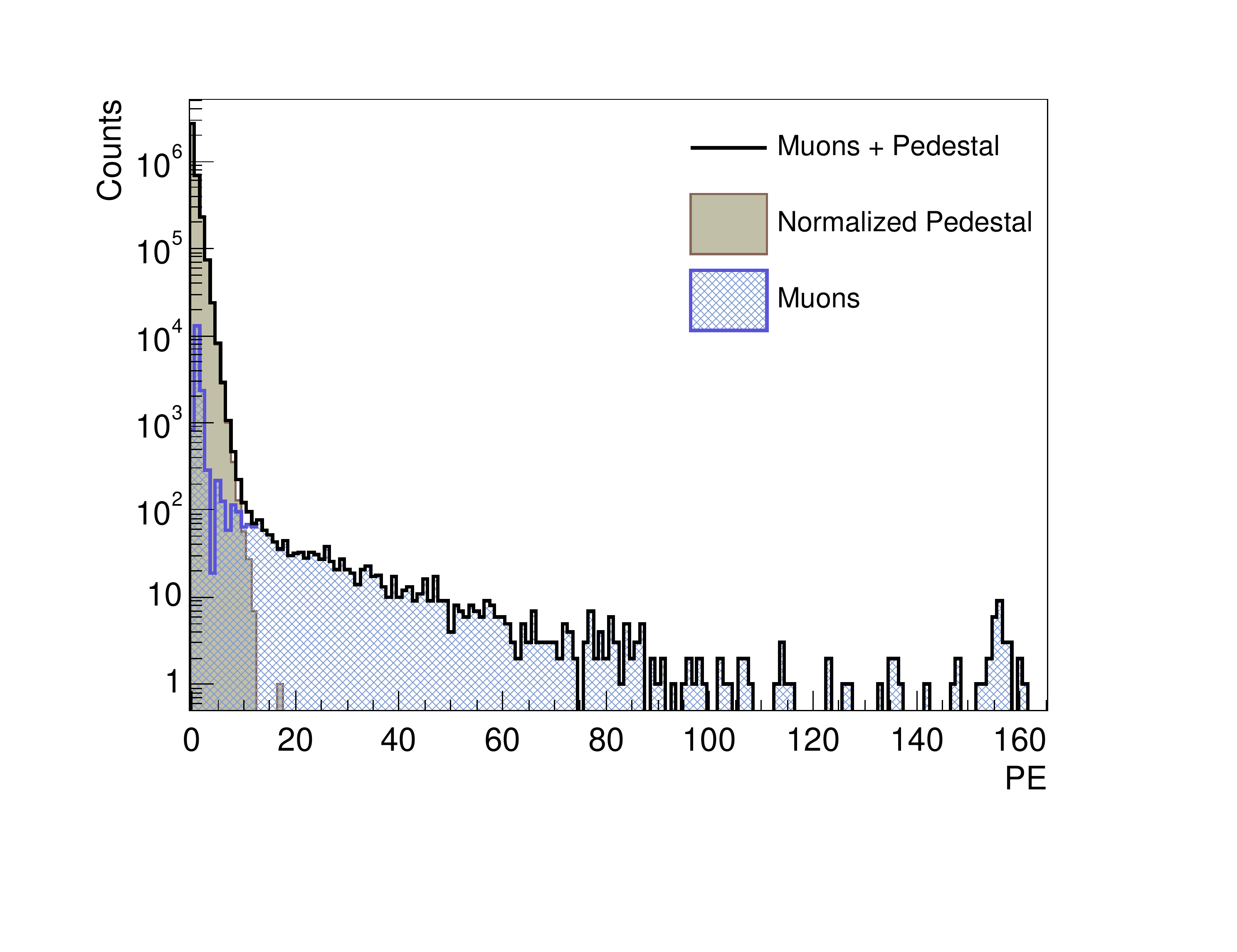}}
\caption{Vertical muon paddle configuration.}
\end{subfigure}
\caption{Muon signal distribution after pedestal subtraction for both muon-paddle configurations.}
\label{fig:musigdistr}
\end{figure}

We recorded coincidence events for 120 minutes in the vertical and 420 minutes in the horizontal. Figure \ref{fig:muontraces} shows the overlaid traces of all SiPMs for all coincidence events in the vertical configuration. Clearly identifiable are the muon signals between ADC samples 230 and 250. We processed the events by extracting the maximal signal amplitude for each SiPM in a window centered on the average muon signal arrival time. The distributions of the extracted signals are shown in Figure \ref{fig:musigdistr} labeled \emph{Muons + Pedestal}. To remove the noise-only events, we extracted a pedestal distribution from an earlier time window in the traces, normalized it to the first five bins in the \emph{Muon + Pedestal} distribution, and then subtracted it from the \emph{Muon + Pedestal} distribution. The corresponding pedestal subtracted distributions, labeled \emph{Muons} are shown in the same figure.

The signal distributions are characterized by a power law that falls less steeply in the vertical configuration than in the horizontal configuration. This can be explained by the longer average trajectory a muon leaves in a SiPM in the vertical configuration, whereas in the horizontal configuration, the selected muons hit the SiPMs head-on, and the trajectory is considerably shorter, resulting in less ionization and, on average, smaller signals than in the vertical configuration.

\begin{figure}[!htb]
\centering
\includegraphics[trim={0.2cm 10cm 9.9cm 0.75cm},clip,angle=0,width=1.\columnwidth]{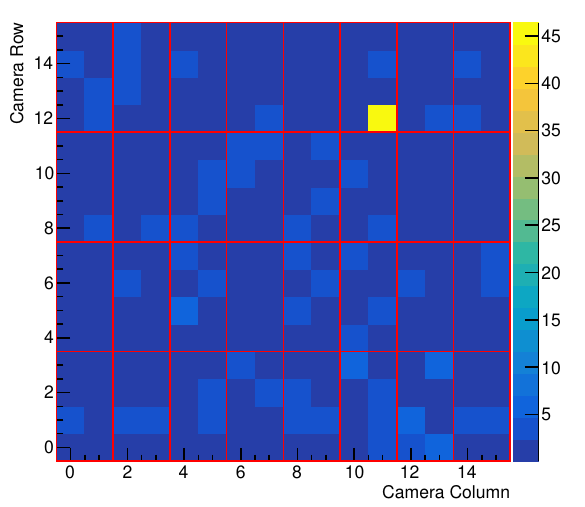}
\caption{One recorded muon event. The color scale gives the number of recorded photoelectrons.}
\label{fig:muonevent}
\end{figure}

It is evident from the distributions that muons can produce signals that are above the trigger threshold of 20\,PE and are thus recorded during normal data taking and need to be discarded in the data analysis. Fortunately, the vast majority of muon signals are confined to individual SiPMs, which results in just one camera pixel light up in the recorded event (see Figure \ref{fig:muonevent}), whereas the expected air-shower images spread over several camera pixels. Muon events are thus discarded at the image-cleaning stage of the analysis, which requires several adjacent camera pixels with a signal to be considered a reconstructable event. Analyzing the muon data with such a standard image-cleaning algorithm used in the Cherenkov community, no event survived.

\subsection{Cosmic Ray Air-Showers}

\begin{figure}[!htb]
\centering
\includegraphics[width=0.9\columnwidth]{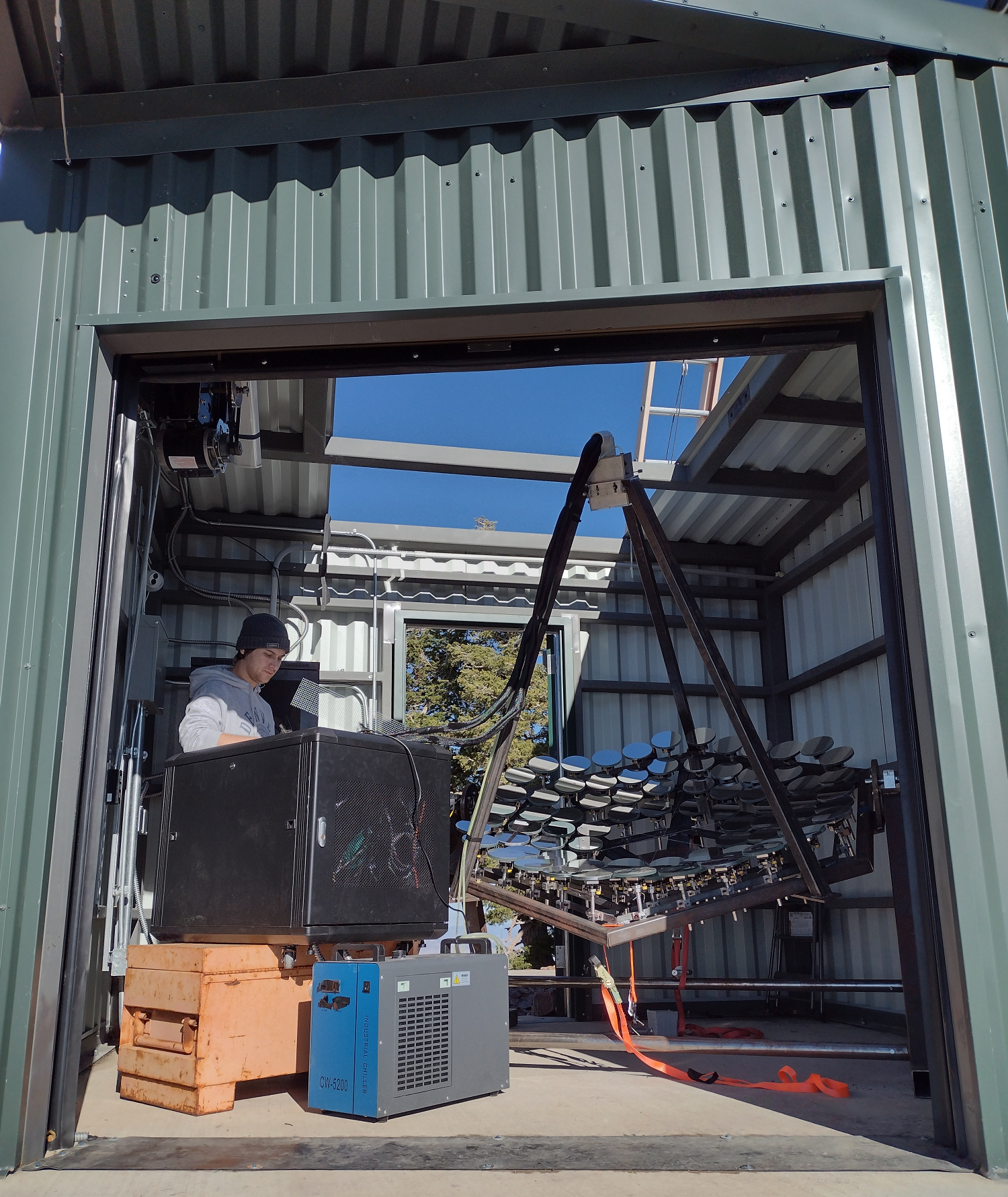}
\caption{The Demonstrator pointing at a $\sim30^\circ$ zenith angle to record air shower images from cosmic rays.}
\label{fig:pointingup}
\end{figure}

To show that the Demonstrator can record air-shower images, we removed the center sheathings of the roof and rotated the telescope up, pointing it at a zenith angle of about $30^\circ$ (see Figure \ref{fig:pointingup}).

\begin{figure}[!htb]
\centering
\includegraphics[angle=0,width=0.9\columnwidth]{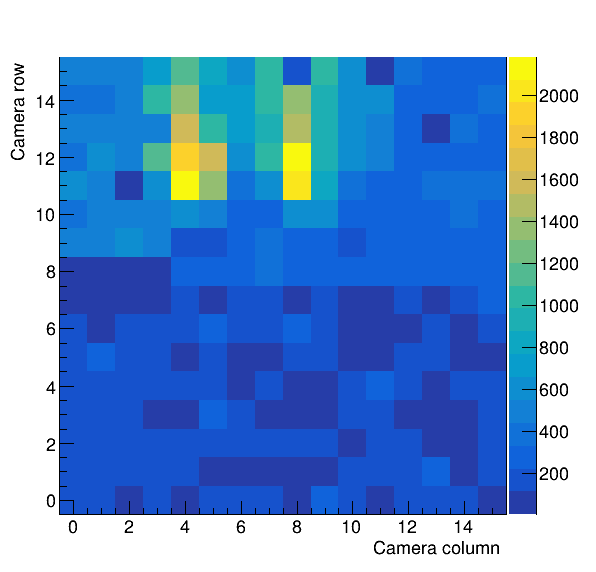}
\caption{Example of a recorded raw air-shower image. The colors indicate the recorded Cherenkov signal in each camera pixel in units of digital counts. \label{fig:airshower}}
\end{figure}

In that configuration, we recorded one hour of cosmic-ray data. One of the recorded images is shown in Figure \ref{fig:airshower}. The split image of the air shower is because the mirrors tilt in different directions when the telescope is rotated up due to sagging in the OSS and slack in the mirror mounts, causing a ``cross-eyed'' optics that projects an image at two positions in the camera. Due to the ``cross-eyed'' feature, we refrained from further analyzing the recorded air-shower images. However, the test shows that the Demonstrator triggers on air showers, and the readout properly records the Cherenkov signals.
 
\subsection{Noise Performance}

The camera signals are contaminated with electronic noise, SiPM dark counts, and photons from the night-sky background. The root-mean-square of the signal fluctuations is 1.9\,PE for pixels that view above the horizon, and 1.3\,PE for pixels below the horizon. Table \ref{tab:noise} details how the individual noise sources contribute to the signal fluctuations. For measuring the electronic noise, we biased the SiPMs below the breakdown voltage. The contribution from SiPM dark counts was measured at the nominal SiPM bias at night, with the building door closed, and the electronic noise was subtracted in quadrature. The NSB contribution was measured with an open door during a moonless night, and the measurement with the closed door was subtracted in quadrature. It is evident that the contribution from the NSB dominates. 

\begin{table}[h]
    \centering
    \begin{tabular}{c|cc}
        \textbf{Source} & \multicolumn{2}{c}{\textbf{Noise, PE}} \\
        \hline \hline
        Electronic noise & \multicolumn{2}{c}{0.32} \\
        SiPM dark counts & \multicolumn{2}{c}{0.47} \\
        
        & \textbf{Ground} & \textbf{Sky} \\
        
        NSB (no moon) & 1.11 & 1.80 \\
        \hline
        Total & 1.25 & 1.89 \\
    \end{tabular}
    \caption{Contributions to the fluctuations of the camera signals.}
    \label{tab:noise}
\end{table}

\section{Observing with the Demonstrator}
Observations with the Demonstrator are performed remotely. The observer initiates the observations by powering up the system, configuring the camera, and initiating the readout. Once the operator has verified that all systems operate within nominal parameters, they log out of the system, and the Demonstrator autonomously records data and shuts itself down at the end of the night. The safe operation limits are listed in Table \ref{tab:dutycycle} and monitored by the on-site software. The code calculates the sun and moon positions in real time and monitors the weather station. The onsite software also monitors the connectivity to the Georgia Tech network, data storage, and data-taking parameters like the SiPM current, temperatures, and trigger rates. If the software detects that a limit is reached, it closes the door and commands the camera and readout to shut down. Realtime monitoring of data-taking is also possible through a public-facing webpage \url{https://trinity.physics.gatech.edu/performance/}. At night's end, the operator connects to the Demonstrator system to check that the telescope has properly shut down. Parallel to data taking, the data is automatically processed at Georgia Tech to generate nightly data summaries to assess the quality of the data and swiftly identify technical issues.

\begin{table}[h]
    \centering
    \begin{tabular}{c|c}
        \textbf{Condition} & \textbf{Requirement} \\\hline
        \hline
        Sun Elevation &  $< -15^\circ$ \\\hline
        Moon Elevation &  $<90^\circ$ when rising\\\hline
        Moon Phase & $< 90\%$\\\hline
        Clouds  Base & $> 11,000$\,above msl\\\hline
        Humidity & $<90\%$ \\\hline
        Wind Speed & $<8$\,m/s \\
    \end{tabular}
    \caption{Environmental conditions for observations with the Demonstrator.}
    \label{tab:dutycycle}
\end{table}

\subsection{Safety Procedures} 
We defined and implemented operating procedures that allow for reliable operations while safeguarding the instrument. We implemented three safety layers to protect the telescope:
\begin{itemize}
    \item The script responsible for data taking stops telescope operations if the sun and moon elevation are outside safe limits, in the presence of adverse weather, or other adverse conditions such as problems storing the data, loss of network connection, or out-of-limits camera parameters. In each instance the observer receives an automated e-mail notification.    
    \item A script running independently on the camera computer (CT-PC) calculates the time of sunrise and commands a shutdown if the sun is about to rise independent of the camera's state. Similarly, a script running in the background on the Control-PC initiates closing the roll-up door at the end of the night.
    \item As a final safeguard, the power supplies for the SiPMs are current limited to protect the front-end electronics. 
\end{itemize}

To protect against user errors, we developed standard operation procedures codified in checklists and enforced in standardized training sessions for observers. All observers receive training in telescope operation and are supervised by on-call experts during observations. The high degree of standardization allowed us to integrate a team of undergraduate students into the \emph{Trinity} Demonstrator program to carry out the majority of observations.

\subsection{Summary of Observations}

The telescope saw first light on October 3, 2023 and commissioning was completed in June 2024. Between June 1, 2024 and February 28, 2025, the \emph{Trinity} Demonstrator has observed 515.3 hours (see table \ref{tab:observations}). Observations are currently limited to mostly moonless nights, as we are developing procedures for operations under bright moonlight.

\begin{table}[h]
    \centering
    \begin{tabular}{c|c}
        \textbf{Observation} & \textbf{Time, hours} \\
        \hline \hline
        Total &  515.3\\
        TXS 0506+056&   18.7 \\
         NGC 1068 & 7.9\\        
    \end{tabular}
    \caption{Summary of total observing time from June 1, 2024, to February 28, 2025, and the fraction of that time when the two sources were within the \emph{Trinity} Demonstrator's field of view.}
    \label{tab:observations}
\end{table}

\section{Summary and Outlook\label{sec:Summary}} 

The \emph{Trinity} Demonstrator is the first phase towards the \emph{Trinity} PeV Neutrino Observatory. Its main goals are proof of the \emph{Trinity} concept and to serve as a platform to develop the analysis and technologies for the \emph{Trinity} Observatory. The Demonstrator is permanently pointing at $280^\circ$ azimuth in the direction where NGC 1068 and TXS 0506+056 cross the horizon.

In this paper, we have provided details about the design and operation of the Demonstrator telescope and low-level performance parameters obtained during its commissioning. After one year of operation, we conclude that the Demonstrator meets its design specifications.

The point spread function of the optics has a $2.1~\text{mm}$ containment radius. The signal chain has been calibrated and has a dynamic range of 100\,PE, and an intrinsic noise floor, including SiPM dark counts of 0.57\,PE. The camera and trigger response is flat fielded within 10\% and 5\%, respectively.

Fully calibrated and commissioned, the Demonstrator has transitioned into science observations. In parallel to regularly taking data, we are developing the simulation and analysis chain. Once the analysis and simulation chain is complete, we will verify the Demonstrator's sensitivities for diffuse and point neutrino sources and optimize the design of the \emph{Trinity} telescopes and the data analysis. 

Furthermore, we will use the Demonstrator to devise and implement strategies that will push observations well into bright moonlight periods to maximize the duty cycle of observations. Studies will include, but not be limited to, determining SiPM bias and trigger threshold settings that do not overwhelm the data acquisition rate while maintaining good sensitivity as a function of moon brightness, elevation, and azimuth angle.

With the experience gained by operating the Demonstrator, we are already moving towards the next phase of the \emph{Trinity} Observatory, the development of its first telescope, \emph{Trinity} One. As we transition from the Demonstrator to \emph{Trinity} One, the Demonstrator will become a platform to field-test camera, readout, and calibration hardware under development for \emph{Trinity} One.

Planned to be deployed on Frisco Peak, \emph{Trinity} One will be uniquely positioned to observe neutrinos above 1\,PeV from point sources with declinations between $-70^\circ$ and $+30^\circ$ and thus provide critical contributions to advance multi-messenger astronomy.

\section*{Acknowledgements}
We acknowledge the excellent work done by the Georgia Tech Montgomery Machining Mall staff. The National Science Foundation funded this work with the awards PHY-2112769 and PHY-2411666. 




\bibliographystyle{elsarticle-harv} 

\begin{thebibliography}{31}
\expandafter\ifx\csname natexlab\endcsname\relax\def\natexlab#1{#1}\fi
\providecommand{\url}[1]{\texttt{#1}}
\providecommand{\href}[2]{#2}
\providecommand{\path}[1]{#1}
\providecommand{\DOIprefix}{doi:}
\providecommand{\ArXivprefix}{arXiv:}
\providecommand{\URLprefix}{URL: }
\providecommand{\Pubmedprefix}{pmid:}
\providecommand{\doi}[1]{\href{http://dx.doi.org/#1}{\path{#1}}}
\providecommand{\Pubmed}[1]{\href{pmid:#1}{\path{#1}}}
\providecommand{\bibinfo}[2]{#2}
\ifx\xfnm\relax \def\xfnm[#1]{\unskip,\space#1}\fi
\bibitem[{Aab et~al.(2019)Aab, Abreu, Aglietta, Albuquerque, Albury, Allekotte, Almela, Castillo, Alvarez-Muñiz, Anastasi, Anchordoqui, Andrada, Andringa, Aramo, Asorey, Assis, Avila, Badescu, Bakalova, Balaceanu, Barbato, Luz, Baur, Becker, Bellido, Berat, Bertaina, Bertou, Biermann, Biteau, Blanco, Blazek, Bleve, Boháčová, Boncioli, Bonifazi, Borodai, Botti, Brack, Bretz, Bridgeman, Briechle, Buchholz, Bueno, Buitink, Buscemi, Caballero-Mora, Caccianiga, Calcagni, Cancio, Canfora, Caracas, Carceller, Caruso, Castellina, Catalani, Cataldi, Cazon, Cerda, Chinellato, Choi, Chudoba, Chytka, Clay, Cerutti, Colalillo, Coleman, Coluccia, Conceição, Condorelli, Consolati, Contreras, Convenga, Cooper, Coutu, Covault, Daniel, Dasso, Daumiller, Dawson, Day, de~Almeida, de~Jong, Mauro, Neto, Mitri, de~Oliveira, de~Souza, Debatin, Río, Deligny, Dhital, Matteo, Castro, Dobrigkeit, D’Olivo, Dorosti, Anjos, Dova, Dundovic, Ebr, Engel, Erdmann, Escobar, Etchegoyen, Falcke, Farmer, Farrar, Fauth, Fazzini, Feldbusch,
  Fenu, Ferreyro, Figueira, Filipčič, Freire, Fujii, Fuster, García, Gemmeke, Gesualdi, Gherghel-Lascu, Ghia, Giaccari, Giammarchi, Giller, Głas, Glombitza, Gobbi, Golup, Berisso, Vitale, Gongora, González, Goos, Góra, Gorgi, Gottowik, Grubb, Guarino, Guedes, Guido, Hahn, Halliday, Hampel, Hansen, Harari, Harrison, Harvey, Haungs, Hebbeker, Heck, Heimann, Hill, Hojvat, Holt, Homola, Hörandel, Horvath, Hrabovský, Huege, Hulsman, Insolia, Isar, Johnsen, Jurysek, Kääpä, Kampert, Keilhauer, Kemmerich, Kemp, Klages, Kleifges, Kleinfeller, Kuempel, Mezek, Awad, Lago, LaHurd, Lang, Legumina, de~Oliveira, Lenok, Letessier-Selvon, Lhenry-Yvon, Lippmann, Presti, Lopes, López, Casado, Lorek, Luce, Lucero, Malacari, Mancarella, Mandat, Manning, Manshanden, Mantsch, Mariazzi, Mariş, Marsella, Martello, Martinez, Bravo, Mastrodicasa, Mathes, Mathys, Matthews, Matthiae, Mayotte, Mazur, Medina-Tanco, Melo, Menshikov, Merenda, Michal, Micheletti, Miramonti, Mockler, Mollerach, Montanet, Morello, Morlino,
  Mostafá, Müller, Muller, Müller, Mussa, Namasaka, Nellen, Niculescu-Oglinzanu, Niechciol, Nitz, Nosek, Novotny, Nožka, Nucita, Núñez, Olinto, Palatka, Pallotta, Panetta, Papenbreer, Parente, Parra, Pech, Pedreira, Pekala, Pelayo, Peña-Rodriguez, Pereira, Perlin, Perrone, Peters, Petrera, Phuntsok, Pierog, Pimenta, Pirronello, Platino, Poh, Pont, Porowski, Pothast, Prado, Privitera, Prouza, Puyleart, Querchfeld, Quinn, Ramos-Pollan, Rautenberg, Ravignani, Reininghaus, Ridky, Riehn, Risse, Ristori, Rizi, de~Carvalho, Rojo, Roncoroni, Roth, Roulet, Rovero, Ruehl, Saffi, Saftoiu, Salamida, Salazar, Salina, Gomez, Sánchez, Santos, Santos, Sarazin, Sarmento, Sarmiento-Cano, Sato, Savina, Schauer, Scherini, Schieler, Schimassek, Schimp, Schlüter, Schmidt, Scholten, Schovánek, Schröder, Schröder, Schumacher, Sciutto, Scornavacche, Shellard, Sigl, Silli, Sima, Šmída, Snow, Sommers, Soriano, Souchard, Squartini, Stadelmaier, Stanca, Stanič, Stasielak, Stassi, Stolpovskiy, Streich, Suárez-Durán,
  Sudholz, Suomijärvi, Supanitsky, Šupík, Szadkowski, Taboada, Taborda, Tapia, Timmermans, Tobiska, Peixoto, Tomé, Elipe, Travaini, Travnicek, Trini, Tueros, Ulrich, Unger, Urban, Galicia, Valiño, Valore, Bodegom, Berg, Vliet, Varela, Cárdenas, Vásquez-Ramírez, Veberič, Ventura, Quispe, Verzi, Vicha, Villaseñor, Vink, Vorobiov, Wahlberg, Watson, Weber, Weindl, Wiedeński, Wiencke, Wilczyński, Winchen, Wirtz, Wittkowski, Wundheiler, Yang, Yushkov, Zas, Zavrtanik, Zavrtanik, Zehrer, Zepeda, Zimmermann, Ziolkowski and Zuccarello}]{Auger_Aab_2019}
\bibinfo{author}{Aab, A.}, \bibinfo{author}{Abreu, P.}, \bibinfo{author}{Aglietta, M.}, \bibinfo{author}{Albuquerque, I.}, \bibinfo{author}{Albury, J.}, \bibinfo{author}{Allekotte, I.}, \bibinfo{author}{Almela, A.}, \bibinfo{author}{Castillo, J.A.}, \bibinfo{author}{Alvarez-Muñiz, J.}, \bibinfo{author}{Anastasi, G.}, \bibinfo{author}{Anchordoqui, L.}, \bibinfo{author}{Andrada, B.}, \bibinfo{author}{Andringa, S.}, \bibinfo{author}{Aramo, C.}, \bibinfo{author}{Asorey, H.}, \bibinfo{author}{Assis, P.}, \bibinfo{author}{Avila, G.}, \bibinfo{author}{Badescu, A.}, \bibinfo{author}{Bakalova, A.}, \bibinfo{author}{Balaceanu, A.}, \bibinfo{author}{Barbato, F.}, \bibinfo{author}{Luz, R.B.}, \bibinfo{author}{Baur, S.}, \bibinfo{author}{Becker, K.}, \bibinfo{author}{Bellido, J.}, \bibinfo{author}{Berat, C.}, \bibinfo{author}{Bertaina, M.}, \bibinfo{author}{Bertou, X.}, \bibinfo{author}{Biermann, P.}, \bibinfo{author}{Biteau, J.}, \bibinfo{author}{Blanco, A.}, \bibinfo{author}{Blazek, J.}, \bibinfo{author}{Bleve, C.},
  \bibinfo{author}{Boháčová, M.}, \bibinfo{author}{Boncioli, D.}, \bibinfo{author}{Bonifazi, C.}, \bibinfo{author}{Borodai, N.}, \bibinfo{author}{Botti, A.}, \bibinfo{author}{Brack, J.}, \bibinfo{author}{Bretz, T.}, \bibinfo{author}{Bridgeman, A.}, \bibinfo{author}{Briechle, F.}, \bibinfo{author}{Buchholz, P.}, \bibinfo{author}{Bueno, A.}, \bibinfo{author}{Buitink, S.}, \bibinfo{author}{Buscemi, M.}, \bibinfo{author}{Caballero-Mora, K.}, \bibinfo{author}{Caccianiga, L.}, \bibinfo{author}{Calcagni, L.}, \bibinfo{author}{Cancio, A.}, \bibinfo{author}{Canfora, F.}, \bibinfo{author}{Caracas, I.}, \bibinfo{author}{Carceller, J.}, \bibinfo{author}{Caruso, R.}, \bibinfo{author}{Castellina, A.}, \bibinfo{author}{Catalani, F.}, \bibinfo{author}{Cataldi, G.}, \bibinfo{author}{Cazon, L.}, \bibinfo{author}{Cerda, M.}, \bibinfo{author}{Chinellato, J.}, \bibinfo{author}{Choi, K.}, \bibinfo{author}{Chudoba, J.}, \bibinfo{author}{Chytka, L.}, \bibinfo{author}{Clay, R.}, \bibinfo{author}{Cerutti, A.C.},
  \bibinfo{author}{Colalillo, R.}, \bibinfo{author}{Coleman, A.}, \bibinfo{author}{Coluccia, M.}, \bibinfo{author}{Conceição, R.}, \bibinfo{author}{Condorelli, A.}, \bibinfo{author}{Consolati, G.}, \bibinfo{author}{Contreras, F.}, \bibinfo{author}{Convenga, F.}, \bibinfo{author}{Cooper, M.}, \bibinfo{author}{Coutu, S.}, \bibinfo{author}{Covault, C.}, \bibinfo{author}{Daniel, B.}, \bibinfo{author}{Dasso, S.}, \bibinfo{author}{Daumiller, K.}, \bibinfo{author}{Dawson, B.}, \bibinfo{author}{Day, J.}, \bibinfo{author}{de~Almeida, R.}, \bibinfo{author}{de~Jong, S.}, \bibinfo{author}{Mauro, G.D.}, \bibinfo{author}{Neto, J.d.M.}, \bibinfo{author}{Mitri, I.D.}, \bibinfo{author}{de~Oliveira, J.}, \bibinfo{author}{de~Souza, V.}, \bibinfo{author}{Debatin, J.}, \bibinfo{author}{Río, M.d.}, \bibinfo{author}{Deligny, O.}, \bibinfo{author}{Dhital, N.}, \bibinfo{author}{Matteo, A.D.}, \bibinfo{author}{Castro, M.D.}, \bibinfo{author}{Dobrigkeit, C.}, \bibinfo{author}{D’Olivo, J.}, \bibinfo{author}{Dorosti, Q.},
  \bibinfo{author}{Anjos, R.d.}, \bibinfo{author}{Dova, M.}, \bibinfo{author}{Dundovic, A.}, \bibinfo{author}{Ebr, J.}, \bibinfo{author}{Engel, R.}, \bibinfo{author}{Erdmann, M.}, \bibinfo{author}{Escobar, C.}, \bibinfo{author}{Etchegoyen, A.}, \bibinfo{author}{Falcke, H.}, \bibinfo{author}{Farmer, J.}, \bibinfo{author}{Farrar, G.}, \bibinfo{author}{Fauth, A.}, \bibinfo{author}{Fazzini, N.}, \bibinfo{author}{Feldbusch, F.}, \bibinfo{author}{Fenu, F.}, \bibinfo{author}{Ferreyro, L.}, \bibinfo{author}{Figueira, J.}, \bibinfo{author}{Filipčič, A.}, \bibinfo{author}{Freire, M.}, \bibinfo{author}{Fujii, T.}, \bibinfo{author}{Fuster, A.}, \bibinfo{author}{García, B.}, \bibinfo{author}{Gemmeke, H.}, \bibinfo{author}{Gesualdi, F.}, \bibinfo{author}{Gherghel-Lascu, A.}, \bibinfo{author}{Ghia, P.}, \bibinfo{author}{Giaccari, U.}, \bibinfo{author}{Giammarchi, M.}, \bibinfo{author}{Giller, M.}, \bibinfo{author}{Głas, D.}, \bibinfo{author}{Glombitza, J.}, \bibinfo{author}{Gobbi, F.}, \bibinfo{author}{Golup, G.},
  \bibinfo{author}{Berisso, M.G.}, \bibinfo{author}{Vitale, P.G.}, \bibinfo{author}{Gongora, J.}, \bibinfo{author}{González, N.}, \bibinfo{author}{Goos, I.}, \bibinfo{author}{Góra, D.}, \bibinfo{author}{Gorgi, A.}, \bibinfo{author}{Gottowik, M.}, \bibinfo{author}{Grubb, T.}, \bibinfo{author}{Guarino, F.}, \bibinfo{author}{Guedes, G.}, \bibinfo{author}{Guido, E.}, \bibinfo{author}{Hahn, S.}, \bibinfo{author}{Halliday, R.}, \bibinfo{author}{Hampel, M.}, \bibinfo{author}{Hansen, P.}, \bibinfo{author}{Harari, D.}, \bibinfo{author}{Harrison, T.}, \bibinfo{author}{Harvey, V.}, \bibinfo{author}{Haungs, A.}, \bibinfo{author}{Hebbeker, T.}, \bibinfo{author}{Heck, D.}, \bibinfo{author}{Heimann, P.}, \bibinfo{author}{Hill, G.}, \bibinfo{author}{Hojvat, C.}, \bibinfo{author}{Holt, E.}, \bibinfo{author}{Homola, P.}, \bibinfo{author}{Hörandel, J.}, \bibinfo{author}{Horvath, P.}, \bibinfo{author}{Hrabovský, M.}, \bibinfo{author}{Huege, T.}, \bibinfo{author}{Hulsman, J.}, \bibinfo{author}{Insolia, A.},
  \bibinfo{author}{Isar, P.}, \bibinfo{author}{Johnsen, J.}, \bibinfo{author}{Jurysek, J.}, \bibinfo{author}{Kääpä, A.}, \bibinfo{author}{Kampert, K.}, \bibinfo{author}{Keilhauer, B.}, \bibinfo{author}{Kemmerich, N.}, \bibinfo{author}{Kemp, J.}, \bibinfo{author}{Klages, H.}, \bibinfo{author}{Kleifges, M.}, \bibinfo{author}{Kleinfeller, J.}, \bibinfo{author}{Kuempel, D.}, \bibinfo{author}{Mezek, G.K.}, \bibinfo{author}{Awad, A.K.}, \bibinfo{author}{Lago, B.}, \bibinfo{author}{LaHurd, D.}, \bibinfo{author}{Lang, R.}, \bibinfo{author}{Legumina, R.}, \bibinfo{author}{de~Oliveira, M.L.}, \bibinfo{author}{Lenok, V.}, \bibinfo{author}{Letessier-Selvon, A.}, \bibinfo{author}{Lhenry-Yvon, I.}, \bibinfo{author}{Lippmann, O.}, \bibinfo{author}{Presti, D.L.}, \bibinfo{author}{Lopes, L.}, \bibinfo{author}{López, R.}, \bibinfo{author}{Casado, A.L.}, \bibinfo{author}{Lorek, R.}, \bibinfo{author}{Luce, Q.}, \bibinfo{author}{Lucero, A.}, \bibinfo{author}{Malacari, M.}, \bibinfo{author}{Mancarella, G.},
  \bibinfo{author}{Mandat, D.}, \bibinfo{author}{Manning, B.}, \bibinfo{author}{Manshanden, J.}, \bibinfo{author}{Mantsch, P.}, \bibinfo{author}{Mariazzi, A.}, \bibinfo{author}{Mariş, I.}, \bibinfo{author}{Marsella, G.}, \bibinfo{author}{Martello, D.}, \bibinfo{author}{Martinez, H.}, \bibinfo{author}{Bravo, O.M.}, \bibinfo{author}{Mastrodicasa, M.}, \bibinfo{author}{Mathes, H.}, \bibinfo{author}{Mathys, S.}, \bibinfo{author}{Matthews, J.}, \bibinfo{author}{Matthiae, G.}, \bibinfo{author}{Mayotte, E.}, \bibinfo{author}{Mazur, P.}, \bibinfo{author}{Medina-Tanco, G.}, \bibinfo{author}{Melo, D.}, \bibinfo{author}{Menshikov, A.}, \bibinfo{author}{Merenda, K.D.}, \bibinfo{author}{Michal, S.}, \bibinfo{author}{Micheletti, M.}, \bibinfo{author}{Miramonti, L.}, \bibinfo{author}{Mockler, D.}, \bibinfo{author}{Mollerach, S.}, \bibinfo{author}{Montanet, F.}, \bibinfo{author}{Morello, C.}, \bibinfo{author}{Morlino, G.}, \bibinfo{author}{Mostafá, M.}, \bibinfo{author}{Müller, A.}, \bibinfo{author}{Muller, M.},
  \bibinfo{author}{Müller, S.}, \bibinfo{author}{Mussa, R.}, \bibinfo{author}{Namasaka, W.}, \bibinfo{author}{Nellen, L.}, \bibinfo{author}{Niculescu-Oglinzanu, M.}, \bibinfo{author}{Niechciol, M.}, \bibinfo{author}{Nitz, D.}, \bibinfo{author}{Nosek, D.}, \bibinfo{author}{Novotny, V.}, \bibinfo{author}{Nožka, L.}, \bibinfo{author}{Nucita, A.}, \bibinfo{author}{Núñez, L.}, \bibinfo{author}{Olinto, A.}, \bibinfo{author}{Palatka, M.}, \bibinfo{author}{Pallotta, J.}, \bibinfo{author}{Panetta, M.}, \bibinfo{author}{Papenbreer, P.}, \bibinfo{author}{Parente, G.}, \bibinfo{author}{Parra, A.}, \bibinfo{author}{Pech, M.}, \bibinfo{author}{Pedreira, F.}, \bibinfo{author}{Pekala, J.}, \bibinfo{author}{Pelayo, R.}, \bibinfo{author}{Peña-Rodriguez, J.}, \bibinfo{author}{Pereira, L.}, \bibinfo{author}{Perlin, M.}, \bibinfo{author}{Perrone, L.}, \bibinfo{author}{Peters, C.}, \bibinfo{author}{Petrera, S.}, \bibinfo{author}{Phuntsok, J.}, \bibinfo{author}{Pierog, T.}, \bibinfo{author}{Pimenta, M.},
  \bibinfo{author}{Pirronello, V.}, \bibinfo{author}{Platino, M.}, \bibinfo{author}{Poh, J.}, \bibinfo{author}{Pont, B.}, \bibinfo{author}{Porowski, C.}, \bibinfo{author}{Pothast, M.}, \bibinfo{author}{Prado, R.}, \bibinfo{author}{Privitera, P.}, \bibinfo{author}{Prouza, M.}, \bibinfo{author}{Puyleart, A.}, \bibinfo{author}{Querchfeld, S.}, \bibinfo{author}{Quinn, S.}, \bibinfo{author}{Ramos-Pollan, R.}, \bibinfo{author}{Rautenberg, J.}, \bibinfo{author}{Ravignani, D.}, \bibinfo{author}{Reininghaus, M.}, \bibinfo{author}{Ridky, J.}, \bibinfo{author}{Riehn, F.}, \bibinfo{author}{Risse, M.}, \bibinfo{author}{Ristori, P.}, \bibinfo{author}{Rizi, V.}, \bibinfo{author}{de~Carvalho, W.R.}, \bibinfo{author}{Rojo, J.R.}, \bibinfo{author}{Roncoroni, M.}, \bibinfo{author}{Roth, M.}, \bibinfo{author}{Roulet, E.}, \bibinfo{author}{Rovero, A.}, \bibinfo{author}{Ruehl, P.}, \bibinfo{author}{Saffi, S.}, \bibinfo{author}{Saftoiu, A.}, \bibinfo{author}{Salamida, F.}, \bibinfo{author}{Salazar, H.}, \bibinfo{author}{Salina,
  G.}, \bibinfo{author}{Gomez, J.S.}, \bibinfo{author}{Sánchez, F.}, \bibinfo{author}{Santos, E.}, \bibinfo{author}{Santos, E.}, \bibinfo{author}{Sarazin, F.}, \bibinfo{author}{Sarmento, R.}, \bibinfo{author}{Sarmiento-Cano, C.}, \bibinfo{author}{Sato, R.}, \bibinfo{author}{Savina, P.}, \bibinfo{author}{Schauer, M.}, \bibinfo{author}{Scherini, V.}, \bibinfo{author}{Schieler, H.}, \bibinfo{author}{Schimassek, M.}, \bibinfo{author}{Schimp, M.}, \bibinfo{author}{Schlüter, F.}, \bibinfo{author}{Schmidt, D.}, \bibinfo{author}{Scholten, O.}, \bibinfo{author}{Schovánek, P.}, \bibinfo{author}{Schröder, F.}, \bibinfo{author}{Schröder, S.}, \bibinfo{author}{Schumacher, J.}, \bibinfo{author}{Sciutto, S.}, \bibinfo{author}{Scornavacche, M.}, \bibinfo{author}{Shellard, R.}, \bibinfo{author}{Sigl, G.}, \bibinfo{author}{Silli, G.}, \bibinfo{author}{Sima, O.}, \bibinfo{author}{Šmída, R.}, \bibinfo{author}{Snow, G.}, \bibinfo{author}{Sommers, P.}, \bibinfo{author}{Soriano, J.}, \bibinfo{author}{Souchard, J.},
  \bibinfo{author}{Squartini, R.}, \bibinfo{author}{Stadelmaier, M.}, \bibinfo{author}{Stanca, D.}, \bibinfo{author}{Stanič, S.}, \bibinfo{author}{Stasielak, J.}, \bibinfo{author}{Stassi, P.}, \bibinfo{author}{Stolpovskiy, M.}, \bibinfo{author}{Streich, A.}, \bibinfo{author}{Suárez-Durán, M.}, \bibinfo{author}{Sudholz, T.}, \bibinfo{author}{Suomijärvi, T.}, \bibinfo{author}{Supanitsky, A.}, \bibinfo{author}{Šupík, J.}, \bibinfo{author}{Szadkowski, Z.}, \bibinfo{author}{Taboada, A.}, \bibinfo{author}{Taborda, O.}, \bibinfo{author}{Tapia, A.}, \bibinfo{author}{Timmermans, C.}, \bibinfo{author}{Tobiska, P.}, \bibinfo{author}{Peixoto, C.T.}, \bibinfo{author}{Tomé, B.}, \bibinfo{author}{Elipe, G.T.}, \bibinfo{author}{Travaini, A.}, \bibinfo{author}{Travnicek, P.}, \bibinfo{author}{Trini, M.}, \bibinfo{author}{Tueros, M.}, \bibinfo{author}{Ulrich, R.}, \bibinfo{author}{Unger, M.}, \bibinfo{author}{Urban, M.}, \bibinfo{author}{Galicia, J.V.}, \bibinfo{author}{Valiño, I.}, \bibinfo{author}{Valore, L.},
  \bibinfo{author}{Bodegom, P.v.}, \bibinfo{author}{Berg, A.v.d.}, \bibinfo{author}{Vliet, A.v.}, \bibinfo{author}{Varela, E.}, \bibinfo{author}{Cárdenas, B.V.}, \bibinfo{author}{Vásquez-Ramírez, A.}, \bibinfo{author}{Veberič, D.}, \bibinfo{author}{Ventura, C.}, \bibinfo{author}{Quispe, I.V.}, \bibinfo{author}{Verzi, V.}, \bibinfo{author}{Vicha, J.}, \bibinfo{author}{Villaseñor, L.}, \bibinfo{author}{Vink, J.}, \bibinfo{author}{Vorobiov, S.}, \bibinfo{author}{Wahlberg, H.}, \bibinfo{author}{Watson, A.}, \bibinfo{author}{Weber, M.}, \bibinfo{author}{Weindl, A.}, \bibinfo{author}{Wiedeński, M.}, \bibinfo{author}{Wiencke, L.}, \bibinfo{author}{Wilczyński, H.}, \bibinfo{author}{Winchen, T.}, \bibinfo{author}{Wirtz, M.}, \bibinfo{author}{Wittkowski, D.}, \bibinfo{author}{Wundheiler, B.}, \bibinfo{author}{Yang, L.}, \bibinfo{author}{Yushkov, A.}, \bibinfo{author}{Zas, E.}, \bibinfo{author}{Zavrtanik, D.}, \bibinfo{author}{Zavrtanik, M.}, \bibinfo{author}{Zehrer, L.}, \bibinfo{author}{Zepeda, A.},
  \bibinfo{author}{Zimmermann, B.}, \bibinfo{author}{Ziolkowski, M.}, \bibinfo{author}{Zuccarello, F.}, \bibinfo{year}{2019}.
\newblock \bibinfo{title}{Limits on point-like sources of ultra-high-energy neutrinos with the pierre auger observatory}.
\newblock \bibinfo{journal}{Journal of Cosmology and Astroparticle Physics} \bibinfo{volume}{2019}, \bibinfo{pages}{004–004}.
\newblock \URLprefix \url{http://dx.doi.org/10.1088/1475-7516/2019/11/004}, \DOIprefix\doi{10.1088/1475-7516/2019/11/004}.
\bibitem[{Aartsen et~al.(2018)Aartsen, Ackermann, Adams, Aguilar, Ahlers, Ahrens, Al~Samarai, Altmann, Andeen, Anderson, Ansseau, Anton, Argüelles, Auffenberg, Axani, Backes, Bagherpour, Bai, Barbano, Barron, Barwick, Baum, Bay, Beatty, Becker~Tjus, Becker, BenZvi, Berley, Bernardini, Besson, Binder, Bindig, Blaufuss, Blot, Bohm, Börner, Bos, Böser, Botner, Bourbeau, Bourbeau, Bradascio, Braun, Brenzke, Bretz, Bron, Brostean-Kaiser, Burgman, Busse, Carver, Cheung, Chirkin, Christov, Clark, Classen, Collin, Conrad, Coppin, Correa, Cowen, Cross, Dave, Day, de~André, De~Clercq, DeLaunay, Dembinski, Deoskar, De~Ridder, Desiati, de~Vries, de~Wasseige, de~With, DeYoung, Díaz-Vélez, di~Lorenzo, Dujmovic, Dumm, Dunkman, Dvorak, Eberhardt, Ehrhardt, Eichmann, Eller, Evenson, Fahey, Fazely, Felde, Filimonov, Finley, Flis, Franckowiak, Friedman, Fritz, Gaisser, Gallagher, Ganster, Gerhardt, Ghorbani, Giang, Glauch, Glüsenkamp, Goldschmidt, Gonzalez, Grant, Griffith, Haack, Hallgren, Halve, Halzen, Hanson,
  Hebecker, Heereman, Helbing, Hellauer, Hickford, Hignight, Hill, Hoffman, Hoffmann, Hoinka, Hokanson-Fasig, Hoshina, Huang, Huber, Hultqvist, Hünnefeld, Hussain, In, Iovine, Ishihara, Jacobi, Japaridze, Jeong, Jero, Jones, Kalaczynski, Kang, Kappes, Kappesser, Karg, Karle, Katz, Kauer, Keivani, Kelley, Kheirandish, Kim, Kintscher, Kiryluk, Kittler, Klein, Koirala, Kolanoski, Köpke, Kopper, Kopper, Koschinsky, Koskinen, Kowalski, Krings, Kroll, Krückl, Kunwar, Kurahashi, Kyriacou, Labare, Lanfranchi, Larson, Lauber, Leonard, Leuermann, Liu, Lohfink, Lozano~Mariscal, Lu, Lünemann, Luszczak, Madsen, Maggi, Mahn, Makino, Mancina, Mariş, Maruyama, Mase, Maunu, Meagher, Medici, Meier, Menne, Merino, Meures, Miarecki, Micallef, Momenté, Montaruli, Moore, Moulai, Nagai, Nahnhauer, Nakarmi, Naumann, Neer, Niederhausen, Nowicki, Nygren, Obertacke~Pollmann, Olivas, O’Murchadha, O’Sullivan, Palczewski, Pandya, Pankova, Peiffer, Pepper, Pérez de~los Heros, Pieloth, Pinat, Pizzuto, Plum, Price, Przybylski,
  Raab, Rädel, Rameez, Rauch, Rawlins, Rea, Reimann, Relethford, Renzi, Resconi, Rhode, Richman, Robertson, Rongen, Rott, Ruhe, Ryckbosch, Rysewyk, Safa, Sanchez~Herrera, Sandrock, Sandroos, Santander, Sarkar, Sarkar, Satalecka, Schaufel, Schlunder, Schmidt, Schneider, Schoenen, Schöneberg, Schumacher, Sclafani, Seckel, Seunarine, Soedingrekso, Soldin, Song, Spiczak, Spiering, Stachurska, Stamatikos, Stanev, Stasik, Stein, Stettner, Steuer, Stezelberger, Stokstad, Stößl, Strotjohann, Stuttard, Sullivan, Sutherland, Taboada, Tenholt, Ter-Antonyan, Terliuk, Tilav, Toale, Tobin, Tönnis, Toscano, Tosi, Tselengidou, Tung, Turcati, Turley, Ty, Unger, Usner, Vandenbroucke, Van~Driessche, van Eijk, van Eijndhoven, Vanheule, van Santen, Vraeghe, Walck, Wallace, Wallraff, Wandler, Wandkowsky, Watson, Waza, Weaver, Weiss, Wendt, Werthebach, Westerhoff, Whelan, Whitehorn, Wiebe, Wiebusch, Wille, Williams, Wills, Wolf, Wood, Wood, Woolsey, Woschnagg, Wrede, Xu, Xu, Xu, Yanez, Yodh, Yoshida and
  Yuan}]{IceCube_Aartsen_2018}
\bibinfo{author}{Aartsen, M.}, \bibinfo{author}{Ackermann, M.}, \bibinfo{author}{Adams, J.}, \bibinfo{author}{Aguilar, J.}, \bibinfo{author}{Ahlers, M.}, \bibinfo{author}{Ahrens, M.}, \bibinfo{author}{Al~Samarai, I.}, \bibinfo{author}{Altmann, D.}, \bibinfo{author}{Andeen, K.}, \bibinfo{author}{Anderson, T.}, \bibinfo{author}{Ansseau, I.}, \bibinfo{author}{Anton, G.}, \bibinfo{author}{Argüelles, C.}, \bibinfo{author}{Auffenberg, J.}, \bibinfo{author}{Axani, S.}, \bibinfo{author}{Backes, P.}, \bibinfo{author}{Bagherpour, H.}, \bibinfo{author}{Bai, X.}, \bibinfo{author}{Barbano, A.}, \bibinfo{author}{Barron, J.}, \bibinfo{author}{Barwick, S.}, \bibinfo{author}{Baum, V.}, \bibinfo{author}{Bay, R.}, \bibinfo{author}{Beatty, J.}, \bibinfo{author}{Becker~Tjus, J.}, \bibinfo{author}{Becker, K.H.}, \bibinfo{author}{BenZvi, S.}, \bibinfo{author}{Berley, D.}, \bibinfo{author}{Bernardini, E.}, \bibinfo{author}{Besson, D.}, \bibinfo{author}{Binder, G.}, \bibinfo{author}{Bindig, D.}, \bibinfo{author}{Blaufuss, E.},
  \bibinfo{author}{Blot, S.}, \bibinfo{author}{Bohm, C.}, \bibinfo{author}{Börner, M.}, \bibinfo{author}{Bos, F.}, \bibinfo{author}{Böser, S.}, \bibinfo{author}{Botner, O.}, \bibinfo{author}{Bourbeau, E.}, \bibinfo{author}{Bourbeau, J.}, \bibinfo{author}{Bradascio, F.}, \bibinfo{author}{Braun, J.}, \bibinfo{author}{Brenzke, M.}, \bibinfo{author}{Bretz, H.P.}, \bibinfo{author}{Bron, S.}, \bibinfo{author}{Brostean-Kaiser, J.}, \bibinfo{author}{Burgman, A.}, \bibinfo{author}{Busse, R.}, \bibinfo{author}{Carver, T.}, \bibinfo{author}{Cheung, E.}, \bibinfo{author}{Chirkin, D.}, \bibinfo{author}{Christov, A.}, \bibinfo{author}{Clark, K.}, \bibinfo{author}{Classen, L.}, \bibinfo{author}{Collin, G.}, \bibinfo{author}{Conrad, J.}, \bibinfo{author}{Coppin, P.}, \bibinfo{author}{Correa, P.}, \bibinfo{author}{Cowen, D.}, \bibinfo{author}{Cross, R.}, \bibinfo{author}{Dave, P.}, \bibinfo{author}{Day, M.}, \bibinfo{author}{de~André, J.}, \bibinfo{author}{De~Clercq, C.}, \bibinfo{author}{DeLaunay, J.},
  \bibinfo{author}{Dembinski, H.}, \bibinfo{author}{Deoskar, K.}, \bibinfo{author}{De~Ridder, S.}, \bibinfo{author}{Desiati, P.}, \bibinfo{author}{de~Vries, K.}, \bibinfo{author}{de~Wasseige, G.}, \bibinfo{author}{de~With, M.}, \bibinfo{author}{DeYoung, T.}, \bibinfo{author}{Díaz-Vélez, J.}, \bibinfo{author}{di~Lorenzo, V.}, \bibinfo{author}{Dujmovic, H.}, \bibinfo{author}{Dumm, J.}, \bibinfo{author}{Dunkman, M.}, \bibinfo{author}{Dvorak, E.}, \bibinfo{author}{Eberhardt, B.}, \bibinfo{author}{Ehrhardt, T.}, \bibinfo{author}{Eichmann, B.}, \bibinfo{author}{Eller, P.}, \bibinfo{author}{Evenson, P.}, \bibinfo{author}{Fahey, S.}, \bibinfo{author}{Fazely, A.}, \bibinfo{author}{Felde, J.}, \bibinfo{author}{Filimonov, K.}, \bibinfo{author}{Finley, C.}, \bibinfo{author}{Flis, S.}, \bibinfo{author}{Franckowiak, A.}, \bibinfo{author}{Friedman, E.}, \bibinfo{author}{Fritz, A.}, \bibinfo{author}{Gaisser, T.}, \bibinfo{author}{Gallagher, J.}, \bibinfo{author}{Ganster, E.}, \bibinfo{author}{Gerhardt, L.},
  \bibinfo{author}{Ghorbani, K.}, \bibinfo{author}{Giang, W.}, \bibinfo{author}{Glauch, T.}, \bibinfo{author}{Glüsenkamp, T.}, \bibinfo{author}{Goldschmidt, A.}, \bibinfo{author}{Gonzalez, J.}, \bibinfo{author}{Grant, D.}, \bibinfo{author}{Griffith, Z.}, \bibinfo{author}{Haack, C.}, \bibinfo{author}{Hallgren, A.}, \bibinfo{author}{Halve, L.}, \bibinfo{author}{Halzen, F.}, \bibinfo{author}{Hanson, K.}, \bibinfo{author}{Hebecker, D.}, \bibinfo{author}{Heereman, D.}, \bibinfo{author}{Helbing, K.}, \bibinfo{author}{Hellauer, R.}, \bibinfo{author}{Hickford, S.}, \bibinfo{author}{Hignight, J.}, \bibinfo{author}{Hill, G.}, \bibinfo{author}{Hoffman, K.}, \bibinfo{author}{Hoffmann, R.}, \bibinfo{author}{Hoinka, T.}, \bibinfo{author}{Hokanson-Fasig, B.}, \bibinfo{author}{Hoshina, K.}, \bibinfo{author}{Huang, F.}, \bibinfo{author}{Huber, M.}, \bibinfo{author}{Hultqvist, K.}, \bibinfo{author}{Hünnefeld, M.}, \bibinfo{author}{Hussain, R.}, \bibinfo{author}{In, S.}, \bibinfo{author}{Iovine, N.},
  \bibinfo{author}{Ishihara, A.}, \bibinfo{author}{Jacobi, E.}, \bibinfo{author}{Japaridze, G.}, \bibinfo{author}{Jeong, M.}, \bibinfo{author}{Jero, K.}, \bibinfo{author}{Jones, B.}, \bibinfo{author}{Kalaczynski, P.}, \bibinfo{author}{Kang, W.}, \bibinfo{author}{Kappes, A.}, \bibinfo{author}{Kappesser, D.}, \bibinfo{author}{Karg, T.}, \bibinfo{author}{Karle, A.}, \bibinfo{author}{Katz, U.}, \bibinfo{author}{Kauer, M.}, \bibinfo{author}{Keivani, A.}, \bibinfo{author}{Kelley, J.}, \bibinfo{author}{Kheirandish, A.}, \bibinfo{author}{Kim, J.}, \bibinfo{author}{Kintscher, T.}, \bibinfo{author}{Kiryluk, J.}, \bibinfo{author}{Kittler, T.}, \bibinfo{author}{Klein, S.}, \bibinfo{author}{Koirala, R.}, \bibinfo{author}{Kolanoski, H.}, \bibinfo{author}{Köpke, L.}, \bibinfo{author}{Kopper, C.}, \bibinfo{author}{Kopper, S.}, \bibinfo{author}{Koschinsky, J.}, \bibinfo{author}{Koskinen, D.}, \bibinfo{author}{Kowalski, M.}, \bibinfo{author}{Krings, K.}, \bibinfo{author}{Kroll, M.}, \bibinfo{author}{Krückl, G.},
  \bibinfo{author}{Kunwar, S.}, \bibinfo{author}{Kurahashi, N.}, \bibinfo{author}{Kyriacou, A.}, \bibinfo{author}{Labare, M.}, \bibinfo{author}{Lanfranchi, J.}, \bibinfo{author}{Larson, M.}, \bibinfo{author}{Lauber, F.}, \bibinfo{author}{Leonard, K.}, \bibinfo{author}{Leuermann, M.}, \bibinfo{author}{Liu, Q.}, \bibinfo{author}{Lohfink, E.}, \bibinfo{author}{Lozano~Mariscal, C.}, \bibinfo{author}{Lu, L.}, \bibinfo{author}{Lünemann, J.}, \bibinfo{author}{Luszczak, W.}, \bibinfo{author}{Madsen, J.}, \bibinfo{author}{Maggi, G.}, \bibinfo{author}{Mahn, K.}, \bibinfo{author}{Makino, Y.}, \bibinfo{author}{Mancina, S.}, \bibinfo{author}{Mariş, I.}, \bibinfo{author}{Maruyama, R.}, \bibinfo{author}{Mase, K.}, \bibinfo{author}{Maunu, R.}, \bibinfo{author}{Meagher, K.}, \bibinfo{author}{Medici, M.}, \bibinfo{author}{Meier, M.}, \bibinfo{author}{Menne, T.}, \bibinfo{author}{Merino, G.}, \bibinfo{author}{Meures, T.}, \bibinfo{author}{Miarecki, S.}, \bibinfo{author}{Micallef, J.}, \bibinfo{author}{Momenté, G.},
  \bibinfo{author}{Montaruli, T.}, \bibinfo{author}{Moore, R.}, \bibinfo{author}{Moulai, M.}, \bibinfo{author}{Nagai, R.}, \bibinfo{author}{Nahnhauer, R.}, \bibinfo{author}{Nakarmi, P.}, \bibinfo{author}{Naumann, U.}, \bibinfo{author}{Neer, G.}, \bibinfo{author}{Niederhausen, H.}, \bibinfo{author}{Nowicki, S.}, \bibinfo{author}{Nygren, D.}, \bibinfo{author}{Obertacke~Pollmann, A.}, \bibinfo{author}{Olivas, A.}, \bibinfo{author}{O’Murchadha, A.}, \bibinfo{author}{O’Sullivan, E.}, \bibinfo{author}{Palczewski, T.}, \bibinfo{author}{Pandya, H.}, \bibinfo{author}{Pankova, D.}, \bibinfo{author}{Peiffer, P.}, \bibinfo{author}{Pepper, J.}, \bibinfo{author}{Pérez de~los Heros, C.}, \bibinfo{author}{Pieloth, D.}, \bibinfo{author}{Pinat, E.}, \bibinfo{author}{Pizzuto, A.}, \bibinfo{author}{Plum, M.}, \bibinfo{author}{Price, P.}, \bibinfo{author}{Przybylski, G.}, \bibinfo{author}{Raab, C.}, \bibinfo{author}{Rädel, L.}, \bibinfo{author}{Rameez, M.}, \bibinfo{author}{Rauch, L.}, \bibinfo{author}{Rawlins, K.},
  \bibinfo{author}{Rea, I.}, \bibinfo{author}{Reimann, R.}, \bibinfo{author}{Relethford, B.}, \bibinfo{author}{Renzi, G.}, \bibinfo{author}{Resconi, E.}, \bibinfo{author}{Rhode, W.}, \bibinfo{author}{Richman, M.}, \bibinfo{author}{Robertson, S.}, \bibinfo{author}{Rongen, M.}, \bibinfo{author}{Rott, C.}, \bibinfo{author}{Ruhe, T.}, \bibinfo{author}{Ryckbosch, D.}, \bibinfo{author}{Rysewyk, D.}, \bibinfo{author}{Safa, I.}, \bibinfo{author}{Sanchez~Herrera, S.}, \bibinfo{author}{Sandrock, A.}, \bibinfo{author}{Sandroos, J.}, \bibinfo{author}{Santander, M.}, \bibinfo{author}{Sarkar, S.}, \bibinfo{author}{Sarkar, S.}, \bibinfo{author}{Satalecka, K.}, \bibinfo{author}{Schaufel, M.}, \bibinfo{author}{Schlunder, P.}, \bibinfo{author}{Schmidt, T.}, \bibinfo{author}{Schneider, A.}, \bibinfo{author}{Schoenen, S.}, \bibinfo{author}{Schöneberg, S.}, \bibinfo{author}{Schumacher, L.}, \bibinfo{author}{Sclafani, S.}, \bibinfo{author}{Seckel, D.}, \bibinfo{author}{Seunarine, S.}, \bibinfo{author}{Soedingrekso, J.},
  \bibinfo{author}{Soldin, D.}, \bibinfo{author}{Song, M.}, \bibinfo{author}{Spiczak, G.}, \bibinfo{author}{Spiering, C.}, \bibinfo{author}{Stachurska, J.}, \bibinfo{author}{Stamatikos, M.}, \bibinfo{author}{Stanev, T.}, \bibinfo{author}{Stasik, A.}, \bibinfo{author}{Stein, R.}, \bibinfo{author}{Stettner, J.}, \bibinfo{author}{Steuer, A.}, \bibinfo{author}{Stezelberger, T.}, \bibinfo{author}{Stokstad, R.}, \bibinfo{author}{Stößl, A.}, \bibinfo{author}{Strotjohann, N.}, \bibinfo{author}{Stuttard, T.}, \bibinfo{author}{Sullivan, G.}, \bibinfo{author}{Sutherland, M.}, \bibinfo{author}{Taboada, I.}, \bibinfo{author}{Tenholt, F.}, \bibinfo{author}{Ter-Antonyan, S.}, \bibinfo{author}{Terliuk, A.}, \bibinfo{author}{Tilav, S.}, \bibinfo{author}{Toale, P.}, \bibinfo{author}{Tobin, M.}, \bibinfo{author}{Tönnis, C.}, \bibinfo{author}{Toscano, S.}, \bibinfo{author}{Tosi, D.}, \bibinfo{author}{Tselengidou, M.}, \bibinfo{author}{Tung, C.}, \bibinfo{author}{Turcati, A.}, \bibinfo{author}{Turley, C.}, \bibinfo{author}{Ty,
  B.}, \bibinfo{author}{Unger, E.}, \bibinfo{author}{Usner, M.}, \bibinfo{author}{Vandenbroucke, J.}, \bibinfo{author}{Van~Driessche, W.}, \bibinfo{author}{van Eijk, D.}, \bibinfo{author}{van Eijndhoven, N.}, \bibinfo{author}{Vanheule, S.}, \bibinfo{author}{van Santen, J.}, \bibinfo{author}{Vraeghe, M.}, \bibinfo{author}{Walck, C.}, \bibinfo{author}{Wallace, A.}, \bibinfo{author}{Wallraff, M.}, \bibinfo{author}{Wandler, F.}, \bibinfo{author}{Wandkowsky, N.}, \bibinfo{author}{Watson, T.}, \bibinfo{author}{Waza, A.}, \bibinfo{author}{Weaver, C.}, \bibinfo{author}{Weiss, M.}, \bibinfo{author}{Wendt, C.}, \bibinfo{author}{Werthebach, J.}, \bibinfo{author}{Westerhoff, S.}, \bibinfo{author}{Whelan, B.}, \bibinfo{author}{Whitehorn, N.}, \bibinfo{author}{Wiebe, K.}, \bibinfo{author}{Wiebusch, C.}, \bibinfo{author}{Wille, L.}, \bibinfo{author}{Williams, D.}, \bibinfo{author}{Wills, L.}, \bibinfo{author}{Wolf, M.}, \bibinfo{author}{Wood, J.}, \bibinfo{author}{Wood, T.}, \bibinfo{author}{Woolsey, E.},
  \bibinfo{author}{Woschnagg, K.}, \bibinfo{author}{Wrede, G.}, \bibinfo{author}{Xu, D.}, \bibinfo{author}{Xu, X.}, \bibinfo{author}{Xu, Y.}, \bibinfo{author}{Yanez, J.}, \bibinfo{author}{Yodh, G.}, \bibinfo{author}{Yoshida, S.}, \bibinfo{author}{Yuan, T.}, \bibinfo{year}{2018}.
\newblock \bibinfo{title}{Differential limit on the extremely-high-energy cosmic neutrino flux in the presence of astrophysical background from nine years of icecube data}.
\newblock \bibinfo{journal}{Physical Review D} \bibinfo{volume}{98}.
\newblock \URLprefix \url{http://dx.doi.org/10.1103/PhysRevD.98.062003}, \DOIprefix\doi{10.1103/physrevd.98.062003}.
\bibitem[{Aartsen et~al.(2013)Aartsen, Abbasi, Abdou, Ackermann, Adams, Aguilar, Ahlers, Altmann, Auffenberg, Bai, Baker, Barwick, Baum, Bay, Beatty, Bechet, Tjus, Becker, Benabderrahmane, BenZvi, Berghaus, Berley, Bernardini, Bernhard, Bertrand, Besson, Binder, Bindig, Bissok, Blaufuss, Blumenthal, Boersma, Bohaichuk, Bohm, Bose, Böser, Botner, Brayeur, Bretz, Brown, Bruijn, Brunner, Carson, Casey, Casier, Chirkin, Christov, Christy, Clark, Clevermann, Coenders, Cohen, Cowen, Silva, Danninger, Daughhetee, Davis, Day, Clercq, Ridder, Desiati, de~Vries, de~With, DeYoung, Díaz-Vélez, Dunkman, Eagan, Eberhardt, Eichmann, Eisch, Ellsworth, Euler, Evenson, Fadiran, Fazely, Fedynitch, Feintzeig, Feusels, Filimonov, Finley, Fischer-Wasels, Flis, Franckowiak, Frantzen, Fuchs, Gaisser, Gallagher, Gerhardt, Gladstone, Glüsenkamp, Goldschmidt, Golup, Gonzalez, Goodman, Góra, Grandmont, Grant, Groß, Ha, Ismail, Hallen, Hallgren, Halzen, Hanson, Heereman, Heinen, Helbing, Hellauer, Hickford, Hill, Hoffman, Hoffmann,
  Homeier, Hoshina, Huelsnitz, Hulth, Hultqvist, Hussain, Ishihara, Jacobi, Jacobsen, Jagielski, Japaridze, Jero, Jlelati, Kaminsky, Kappes, Karg, Karle, Kelley, Kiryluk, Kläs, Klein, Köhne, Kohnen, Kolanoski, Köpke, Kopper, Kopper, Koskinen, Kowalski, Krasberg, Krings, Kroll, Kunnen, Kurahashi, Kuwabara, Labare, Landsman, Larson, Lesiak-Bzdak, Leuermann, Leute, Lünemann, Madsen, Maggi, Maruyama, Mase, Matis, McNally, Meagher, Merck, Meures, Miarecki, Middell, Milke, Miller, Mohrmann, Montaruli, Morse, Nahnhauer, Naumann, Niederhausen, Nowicki, Nygren, Obertacke, Odrowski, Olivas, O'Murchadha, Paul, Pepper, de~los Heros, Pfendner, Pieloth, Pinat, Posselt, Price, Przybylski, Rädel, Rameez, Rawlins, Redl, Reimann, Resconi, Rhode, Ribordy, Richman, Riedel, Rodrigues, Rott, Ruhe, Ruzybayev, Ryckbosch, Saba, Salameh, Sander, Santander, Sarkar, Schatto, Scheriau, Schmidt, Schmitz, Schoenen, Schöneberg, Schönwald, Schukraft, Schulte, Schulz, Seckel, Sestayo, Seunarine, Shanidze, Sheremata, Smith, Soldin,
  Spiczak, Spiering, Stamatikos, Stanev, Stasik, Stezelberger, Stokstad, Stößl, Strahler, Ström, Sullivan, Taavola, Taboada, Tamburro, Tepe, Ter-Antonyan, Tešić, Tilav, Toale, Toscano, Unger, Usner, van Eijndhoven, Overloop, van Santen, Vehring, Voge, Vraeghe, Walck, Waldenmaier, Wallraff, Weaver, Wellons, Wendt, Westerhoff, Whitehorn, Wiebe, Wiebusch, Williams, Wissing, Wolf, Wood, Woschnagg, Xu, Xu, Yanez, Yodh, Yoshida, Zarzhitsky, Ziemann, Zierke and Zoll}]{astrophysical2013}
\bibinfo{author}{Aartsen, M.G.}, \bibinfo{author}{Abbasi, R.}, \bibinfo{author}{Abdou, Y.}, \bibinfo{author}{Ackermann, M.}, \bibinfo{author}{Adams, J.}, \bibinfo{author}{Aguilar, J.A.}, \bibinfo{author}{Ahlers, M.}, \bibinfo{author}{Altmann, D.}, \bibinfo{author}{Auffenberg, J.}, \bibinfo{author}{Bai, X.}, \bibinfo{author}{Baker, M.}, \bibinfo{author}{Barwick, S.W.}, \bibinfo{author}{Baum, V.}, \bibinfo{author}{Bay, R.}, \bibinfo{author}{Beatty, J.J.}, \bibinfo{author}{Bechet, S.}, \bibinfo{author}{Tjus, J.B.}, \bibinfo{author}{Becker, K.H.}, \bibinfo{author}{Benabderrahmane, M.L.}, \bibinfo{author}{BenZvi, S.}, \bibinfo{author}{Berghaus, P.}, \bibinfo{author}{Berley, D.}, \bibinfo{author}{Bernardini, E.}, \bibinfo{author}{Bernhard, A.}, \bibinfo{author}{Bertrand, D.}, \bibinfo{author}{Besson, D.Z.}, \bibinfo{author}{Binder, G.}, \bibinfo{author}{Bindig, D.}, \bibinfo{author}{Bissok, M.}, \bibinfo{author}{Blaufuss, E.}, \bibinfo{author}{Blumenthal, J.}, \bibinfo{author}{Boersma, D.J.},
  \bibinfo{author}{Bohaichuk, S.}, \bibinfo{author}{Bohm, C.}, \bibinfo{author}{Bose, D.}, \bibinfo{author}{Böser, S.}, \bibinfo{author}{Botner, O.}, \bibinfo{author}{Brayeur, L.}, \bibinfo{author}{Bretz, H.P.}, \bibinfo{author}{Brown, A.M.}, \bibinfo{author}{Bruijn, R.}, \bibinfo{author}{Brunner, J.}, \bibinfo{author}{Carson, M.}, \bibinfo{author}{Casey, J.}, \bibinfo{author}{Casier, M.}, \bibinfo{author}{Chirkin, D.}, \bibinfo{author}{Christov, A.}, \bibinfo{author}{Christy, B.}, \bibinfo{author}{Clark, K.}, \bibinfo{author}{Clevermann, F.}, \bibinfo{author}{Coenders, S.}, \bibinfo{author}{Cohen, S.}, \bibinfo{author}{Cowen, D.F.}, \bibinfo{author}{Silva, A.H.C.}, \bibinfo{author}{Danninger, M.}, \bibinfo{author}{Daughhetee, J.}, \bibinfo{author}{Davis, J.C.}, \bibinfo{author}{Day, M.}, \bibinfo{author}{Clercq, C.D.}, \bibinfo{author}{Ridder, S.D.}, \bibinfo{author}{Desiati, P.}, \bibinfo{author}{de~Vries, K.D.}, \bibinfo{author}{de~With, M.}, \bibinfo{author}{DeYoung, T.}, \bibinfo{author}{Díaz-Vélez,
  J.C.}, \bibinfo{author}{Dunkman, M.}, \bibinfo{author}{Eagan, R.}, \bibinfo{author}{Eberhardt, B.}, \bibinfo{author}{Eichmann, B.}, \bibinfo{author}{Eisch, J.}, \bibinfo{author}{Ellsworth, R.W.}, \bibinfo{author}{Euler, S.}, \bibinfo{author}{Evenson, P.A.}, \bibinfo{author}{Fadiran, O.}, \bibinfo{author}{Fazely, A.R.}, \bibinfo{author}{Fedynitch, A.}, \bibinfo{author}{Feintzeig, J.}, \bibinfo{author}{Feusels, T.}, \bibinfo{author}{Filimonov, K.}, \bibinfo{author}{Finley, C.}, \bibinfo{author}{Fischer-Wasels, T.}, \bibinfo{author}{Flis, S.}, \bibinfo{author}{Franckowiak, A.}, \bibinfo{author}{Frantzen, K.}, \bibinfo{author}{Fuchs, T.}, \bibinfo{author}{Gaisser, T.K.}, \bibinfo{author}{Gallagher, J.}, \bibinfo{author}{Gerhardt, L.}, \bibinfo{author}{Gladstone, L.}, \bibinfo{author}{Glüsenkamp, T.}, \bibinfo{author}{Goldschmidt, A.}, \bibinfo{author}{Golup, G.}, \bibinfo{author}{Gonzalez, J.G.}, \bibinfo{author}{Goodman, J.A.}, \bibinfo{author}{Góra, D.}, \bibinfo{author}{Grandmont, D.T.},
  \bibinfo{author}{Grant, D.}, \bibinfo{author}{Groß, A.}, \bibinfo{author}{Ha, C.}, \bibinfo{author}{Ismail, A.H.}, \bibinfo{author}{Hallen, P.}, \bibinfo{author}{Hallgren, A.}, \bibinfo{author}{Halzen, F.}, \bibinfo{author}{Hanson, K.}, \bibinfo{author}{Heereman, D.}, \bibinfo{author}{Heinen, D.}, \bibinfo{author}{Helbing, K.}, \bibinfo{author}{Hellauer, R.}, \bibinfo{author}{Hickford, S.}, \bibinfo{author}{Hill, G.C.}, \bibinfo{author}{Hoffman, K.D.}, \bibinfo{author}{Hoffmann, R.}, \bibinfo{author}{Homeier, A.}, \bibinfo{author}{Hoshina, K.}, \bibinfo{author}{Huelsnitz, W.}, \bibinfo{author}{Hulth, P.O.}, \bibinfo{author}{Hultqvist, K.}, \bibinfo{author}{Hussain, S.}, \bibinfo{author}{Ishihara, A.}, \bibinfo{author}{Jacobi, E.}, \bibinfo{author}{Jacobsen, J.}, \bibinfo{author}{Jagielski, K.}, \bibinfo{author}{Japaridze, G.S.}, \bibinfo{author}{Jero, K.}, \bibinfo{author}{Jlelati, O.}, \bibinfo{author}{Kaminsky, B.}, \bibinfo{author}{Kappes, A.}, \bibinfo{author}{Karg, T.}, \bibinfo{author}{Karle, A.},
  \bibinfo{author}{Kelley, J.L.}, \bibinfo{author}{Kiryluk, J.}, \bibinfo{author}{Kläs, J.}, \bibinfo{author}{Klein, S.R.}, \bibinfo{author}{Köhne, J.H.}, \bibinfo{author}{Kohnen, G.}, \bibinfo{author}{Kolanoski, H.}, \bibinfo{author}{Köpke, L.}, \bibinfo{author}{Kopper, C.}, \bibinfo{author}{Kopper, S.}, \bibinfo{author}{Koskinen, D.J.}, \bibinfo{author}{Kowalski, M.}, \bibinfo{author}{Krasberg, M.}, \bibinfo{author}{Krings, K.}, \bibinfo{author}{Kroll, G.}, \bibinfo{author}{Kunnen, J.}, \bibinfo{author}{Kurahashi, N.}, \bibinfo{author}{Kuwabara, T.}, \bibinfo{author}{Labare, M.}, \bibinfo{author}{Landsman, H.}, \bibinfo{author}{Larson, M.J.}, \bibinfo{author}{Lesiak-Bzdak, M.}, \bibinfo{author}{Leuermann, M.}, \bibinfo{author}{Leute, J.}, \bibinfo{author}{Lünemann, J.}, \bibinfo{author}{Madsen, J.}, \bibinfo{author}{Maggi, G.}, \bibinfo{author}{Maruyama, R.}, \bibinfo{author}{Mase, K.}, \bibinfo{author}{Matis, H.S.}, \bibinfo{author}{McNally, F.}, \bibinfo{author}{Meagher, K.}, \bibinfo{author}{Merck,
  M.}, \bibinfo{author}{Meures, T.}, \bibinfo{author}{Miarecki, S.}, \bibinfo{author}{Middell, E.}, \bibinfo{author}{Milke, N.}, \bibinfo{author}{Miller, J.}, \bibinfo{author}{Mohrmann, L.}, \bibinfo{author}{Montaruli, T.}, \bibinfo{author}{Morse, R.}, \bibinfo{author}{Nahnhauer, R.}, \bibinfo{author}{Naumann, U.}, \bibinfo{author}{Niederhausen, H.}, \bibinfo{author}{Nowicki, S.C.}, \bibinfo{author}{Nygren, D.R.}, \bibinfo{author}{Obertacke, A.}, \bibinfo{author}{Odrowski, S.}, \bibinfo{author}{Olivas, A.}, \bibinfo{author}{O'Murchadha, A.}, \bibinfo{author}{Paul, L.}, \bibinfo{author}{Pepper, J.A.}, \bibinfo{author}{de~los Heros, C.P.}, \bibinfo{author}{Pfendner, C.}, \bibinfo{author}{Pieloth, D.}, \bibinfo{author}{Pinat, E.}, \bibinfo{author}{Posselt, J.}, \bibinfo{author}{Price, P.B.}, \bibinfo{author}{Przybylski, G.T.}, \bibinfo{author}{Rädel, L.}, \bibinfo{author}{Rameez, M.}, \bibinfo{author}{Rawlins, K.}, \bibinfo{author}{Redl, P.}, \bibinfo{author}{Reimann, R.}, \bibinfo{author}{Resconi, E.},
  \bibinfo{author}{Rhode, W.}, \bibinfo{author}{Ribordy, M.}, \bibinfo{author}{Richman, M.}, \bibinfo{author}{Riedel, B.}, \bibinfo{author}{Rodrigues, J.P.}, \bibinfo{author}{Rott, C.}, \bibinfo{author}{Ruhe, T.}, \bibinfo{author}{Ruzybayev, B.}, \bibinfo{author}{Ryckbosch, D.}, \bibinfo{author}{Saba, S.M.}, \bibinfo{author}{Salameh, T.}, \bibinfo{author}{Sander, H.G.}, \bibinfo{author}{Santander, M.}, \bibinfo{author}{Sarkar, S.}, \bibinfo{author}{Schatto, K.}, \bibinfo{author}{Scheriau, F.}, \bibinfo{author}{Schmidt, T.}, \bibinfo{author}{Schmitz, M.}, \bibinfo{author}{Schoenen, S.}, \bibinfo{author}{Schöneberg, S.}, \bibinfo{author}{Schönwald, A.}, \bibinfo{author}{Schukraft, A.}, \bibinfo{author}{Schulte, L.}, \bibinfo{author}{Schulz, O.}, \bibinfo{author}{Seckel, D.}, \bibinfo{author}{Sestayo, Y.}, \bibinfo{author}{Seunarine, S.}, \bibinfo{author}{Shanidze, R.}, \bibinfo{author}{Sheremata, C.}, \bibinfo{author}{Smith, M.W.E.}, \bibinfo{author}{Soldin, D.}, \bibinfo{author}{Spiczak, G.M.},
  \bibinfo{author}{Spiering, C.}, \bibinfo{author}{Stamatikos, M.}, \bibinfo{author}{Stanev, T.}, \bibinfo{author}{Stasik, A.}, \bibinfo{author}{Stezelberger, T.}, \bibinfo{author}{Stokstad, R.G.}, \bibinfo{author}{Stößl, A.}, \bibinfo{author}{Strahler, E.A.}, \bibinfo{author}{Ström, R.}, \bibinfo{author}{Sullivan, G.W.}, \bibinfo{author}{Taavola, H.}, \bibinfo{author}{Taboada, I.}, \bibinfo{author}{Tamburro, A.}, \bibinfo{author}{Tepe, A.}, \bibinfo{author}{Ter-Antonyan, S.}, \bibinfo{author}{Tešić, G.}, \bibinfo{author}{Tilav, S.}, \bibinfo{author}{Toale, P.A.}, \bibinfo{author}{Toscano, S.}, \bibinfo{author}{Unger, E.}, \bibinfo{author}{Usner, M.}, \bibinfo{author}{van Eijndhoven, N.}, \bibinfo{author}{Overloop, A.V.}, \bibinfo{author}{van Santen, J.}, \bibinfo{author}{Vehring, M.}, \bibinfo{author}{Voge, M.}, \bibinfo{author}{Vraeghe, M.}, \bibinfo{author}{Walck, C.}, \bibinfo{author}{Waldenmaier, T.}, \bibinfo{author}{Wallraff, M.}, \bibinfo{author}{Weaver, C.}, \bibinfo{author}{Wellons, M.},
  \bibinfo{author}{Wendt, C.}, \bibinfo{author}{Westerhoff, S.}, \bibinfo{author}{Whitehorn, N.}, \bibinfo{author}{Wiebe, K.}, \bibinfo{author}{Wiebusch, C.H.}, \bibinfo{author}{Williams, D.R.}, \bibinfo{author}{Wissing, H.}, \bibinfo{author}{Wolf, M.}, \bibinfo{author}{Wood, T.R.}, \bibinfo{author}{Woschnagg, K.}, \bibinfo{author}{Xu, D.L.}, \bibinfo{author}{Xu, X.W.}, \bibinfo{author}{Yanez, J.P.}, \bibinfo{author}{Yodh, G.}, \bibinfo{author}{Yoshida, S.}, \bibinfo{author}{Zarzhitsky, P.}, \bibinfo{author}{Ziemann, J.}, \bibinfo{author}{Zierke, S.}, \bibinfo{author}{Zoll, M.}, \bibinfo{year}{2013}.
\newblock \bibinfo{title}{Evidence for high-energy extraterrestrial neutrinos at the icecube detector}.
\newblock \bibinfo{journal}{Science} \bibinfo{volume}{342}, \bibinfo{pages}{1242856--1242856}.
\newblock \URLprefix \url{http://www.sciencemag.org/cgi/doi/10.1126/science.1242856}, \DOIprefix\doi{10.1126/science.1242856}.
\bibitem[{Aartsen et~al.(2021)Aartsen, Abbasi, Ackermann, Adams, Aguilar, Ahlers, Ahrens, Alispach, Allison, Amin, Andeen, Anderson, Ansseau, Anton, Argüelles, Arlen, Auffenberg, Axani, Bagherpour, Bai, Balagopal~V, Barbano, Bartos, Bastian, Basu, Baum, Baur, Bay, Beatty, Becker, Tjus, BenZvi, Berley, Bernardini, Besson, Binder, Bindig, Blaufuss, Blot, Bohm, Bohmer, Böser, Botner, Böttcher, Bourbeau, Bourbeau, Bradascio, Braun, Bron, Brostean-Kaiser, Burgman, Burley, Buscher, Busse, Bustamante, Campana, Carnie-Bronca, Carver, Chen, Chen, Cheung, Chirkin, Choi, Clark, Clark, Classen, Coleman, Collin, Connolly, Conrad, Coppin, Correa, Cowen, Cross, Dave, Deaconu, De~Clercq, DeLaunay, De~Kockere, Dembinski, Deoskar, De~Ridder, Desai, Desiati, de~Vries, de~Wasseige, de~With, DeYoung, Dharani, Diaz, Díaz-Vélez, Dujmovic, Dunkman, DuVernois, Dvorak, Ehrhardt, Eller, Engel, Evans, Evenson, Fahey, Farrag, Fazely, Felde, Fienberg, Filimonov, Finley, Fischer, Fox, Franckowiak, Friedman, Fritz, Gaisser, Gallagher,
  Ganster, Garcia-Fernandez, Garrappa, Gartner, Gerhard, Gernhaeuser, Ghadimi, Glaser, Glauch, Glüsenkamp, Goldschmidt, Gonzalez, Goswami, Grant, Grégoire, Griffith, Griswold, Gündüz, Haack, Hallgren, Halliday, Halve, Halzen, Hanson, Hanson, Hardin, Haugen, Haungs, Hauser, Hebecker, Heinen, Heix, Helbing, Hellauer, Henningsen, Hickford, Hignight, Hill, Hill, Hoffman, Hoffmann, Hoffmann, Hoinka, Hokanson-Fasig, Holzapfel, Hoshina, Huang, Huber, Huber, Huege, Hughes, Hultqvist, Hünnefeld, Hussain, In, Iovine, Ishihara, Jansson, Japaridze, Jeong, Jones, Jonske, Joppe, Kalekin, Kang, Kang, Kang, Kappes, Kappesser, Karg, Karl, Karle, Katori, Katz, Kauer, Keivani, Kellermann, Kelley, Kheirandish, Kim, Kin, Kintscher, Kiryluk, Kittler, Kleifges, Klein, Koirala, Kolanoski, Köpke, Kopper, Kopper, Koskinen, Koundal, Kovacevich, Kowalski, Krauss, Krings, Krückl, Kulacz, Kurahashi, Gualda, Lahmann, Lanfranchi, Larson, Latif, Lauber, Lazar, Leonard, Leszczyńska, Li, Liu, Lohfink, LoSecco, Mariscal, Lu, Lucarelli,
  Ludwig, Lünemann, Luszczak, Lyu, Ma, Madsen, Maggi, Mahn, Makino, Mallik, Mancina, Mandalia, Mariş, Marka, Marka, Maruyama, Mase, Maunu, McNally, Meagher, Medina, Meier, Meighen-Berger, Merz, Meyers, Micallef, Mockler, Momenté, Montaruli, Moore, Morse, Moulai, Muth, Naab, Nagai, Nam, Nauman, Necker, Neer, Nelles, Nguyen, Niederhausen, Nisa, Nowicki, Nygren, Oberla, Pollmann, Oehler, Olivas, O’Sullivan, Pan, Pandya, Pankova, Papp, Park, Parker, Paudel, Peiffer, Pérez de~los Heros, Petersen, Philippen, Pieloth, Pieper, Pinfold, Pizzuto, Plaisier, Plum, Popovych, Porcelli, Rodriguez, Price, Przybylski, Raab, Raissi, Rameez, Rauch, Rawlins, Rea, Rehman, Reimann, Renschler, Renzi, Resconi, Reusch, Rhode, Richman, Riedel, Riegel, Roberts, Robertson, Roellinghoff, Rongen, Rott, Ruhe, Ryckbosch, Cantu, Safa, Herrera, Sandrock, Sandroos, Sandstrom, Santander, Sarkar, Sarkar, Satalecka, Scharf, Schaufel, Schieler, Schlunder, Schmidt, Schneider, Schneider, Schröder, Schumacher, Sclafani, Seckel, Seunarine,
  Shaevitz, Sharma, Shefali, Silva, Smith, Smithers, Snihur, Soedingrekso, Soldin, Söldner-Rembold, Song, Southall, Spiczak, Spiering, Stachurska, Stamatikos, Stanev, Stein, Stettner, Steuer, Stezelberger, Stokstad, Strotjohann, Stürwald, Stuttard, Sullivan, Taboada, Taketa, Tanaka, Tenholt, Ter-Antonyan, Terliuk, Tilav, Tollefson, Tomankova, Tönnis, Torres, Toscano, Tosi, Trettin, Tselengidou, Tung, Turcati, Turcotte, Turley, Twagirayezu, Ty, Unger, Elorrieta, Vandenbroucke, van Eijk, van Eijndhoven, Vannerom, van Santen, Veberic, Verpoest, Vieregg, Vraeghe, Walck, Watson, Weaver, Weindl, Weinstock, Weiss, Weldert, Welling, Wendt, Werthebach, Whitehorn, Wiebe, Wiebusch, Williams, Wissel, Wolf, Wood, Woschnagg, Wrede, Wren, Wulff, Xu, Xu, Yanez, Yoshida, Yuan, Zhang, Zierke and Zöcklein}]{gen2_Aartsen_2021}
\bibinfo{author}{Aartsen, M.G.}, \bibinfo{author}{Abbasi, R.}, \bibinfo{author}{Ackermann, M.}, \bibinfo{author}{Adams, J.}, \bibinfo{author}{Aguilar, J.A.}, \bibinfo{author}{Ahlers, M.}, \bibinfo{author}{Ahrens, M.}, \bibinfo{author}{Alispach, C.}, \bibinfo{author}{Allison, P.}, \bibinfo{author}{Amin, N.M.}, \bibinfo{author}{Andeen, K.}, \bibinfo{author}{Anderson, T.}, \bibinfo{author}{Ansseau, I.}, \bibinfo{author}{Anton, G.}, \bibinfo{author}{Argüelles, C.}, \bibinfo{author}{Arlen, T.C.}, \bibinfo{author}{Auffenberg, J.}, \bibinfo{author}{Axani, S.}, \bibinfo{author}{Bagherpour, H.}, \bibinfo{author}{Bai, X.}, \bibinfo{author}{Balagopal~V, A.}, \bibinfo{author}{Barbano, A.}, \bibinfo{author}{Bartos, I.}, \bibinfo{author}{Bastian, B.}, \bibinfo{author}{Basu, V.}, \bibinfo{author}{Baum, V.}, \bibinfo{author}{Baur, S.}, \bibinfo{author}{Bay, R.}, \bibinfo{author}{Beatty, J.J.}, \bibinfo{author}{Becker, K.H.}, \bibinfo{author}{Tjus, J.B.}, \bibinfo{author}{BenZvi, S.}, \bibinfo{author}{Berley, D.},
  \bibinfo{author}{Bernardini, E.}, \bibinfo{author}{Besson, D.Z.}, \bibinfo{author}{Binder, G.}, \bibinfo{author}{Bindig, D.}, \bibinfo{author}{Blaufuss, E.}, \bibinfo{author}{Blot, S.}, \bibinfo{author}{Bohm, C.}, \bibinfo{author}{Bohmer, M.}, \bibinfo{author}{Böser, S.}, \bibinfo{author}{Botner, O.}, \bibinfo{author}{Böttcher, J.}, \bibinfo{author}{Bourbeau, E.}, \bibinfo{author}{Bourbeau, J.}, \bibinfo{author}{Bradascio, F.}, \bibinfo{author}{Braun, J.}, \bibinfo{author}{Bron, S.}, \bibinfo{author}{Brostean-Kaiser, J.}, \bibinfo{author}{Burgman, A.}, \bibinfo{author}{Burley, R.T.}, \bibinfo{author}{Buscher, J.}, \bibinfo{author}{Busse, R.S.}, \bibinfo{author}{Bustamante, M.}, \bibinfo{author}{Campana, M.A.}, \bibinfo{author}{Carnie-Bronca, E.G.}, \bibinfo{author}{Carver, T.}, \bibinfo{author}{Chen, C.}, \bibinfo{author}{Chen, P.}, \bibinfo{author}{Cheung, E.}, \bibinfo{author}{Chirkin, D.}, \bibinfo{author}{Choi, S.}, \bibinfo{author}{Clark, B.A.}, \bibinfo{author}{Clark, K.}, \bibinfo{author}{Classen,
  L.}, \bibinfo{author}{Coleman, A.}, \bibinfo{author}{Collin, G.H.}, \bibinfo{author}{Connolly, A.}, \bibinfo{author}{Conrad, J.M.}, \bibinfo{author}{Coppin, P.}, \bibinfo{author}{Correa, P.}, \bibinfo{author}{Cowen, D.F.}, \bibinfo{author}{Cross, R.}, \bibinfo{author}{Dave, P.}, \bibinfo{author}{Deaconu, C.}, \bibinfo{author}{De~Clercq, C.}, \bibinfo{author}{DeLaunay, J.J.}, \bibinfo{author}{De~Kockere, S.}, \bibinfo{author}{Dembinski, H.}, \bibinfo{author}{Deoskar, K.}, \bibinfo{author}{De~Ridder, S.}, \bibinfo{author}{Desai, A.}, \bibinfo{author}{Desiati, P.}, \bibinfo{author}{de~Vries, K.D.}, \bibinfo{author}{de~Wasseige, G.}, \bibinfo{author}{de~With, M.}, \bibinfo{author}{DeYoung, T.}, \bibinfo{author}{Dharani, S.}, \bibinfo{author}{Diaz, A.}, \bibinfo{author}{Díaz-Vélez, J.C.}, \bibinfo{author}{Dujmovic, H.}, \bibinfo{author}{Dunkman, M.}, \bibinfo{author}{DuVernois, M.A.}, \bibinfo{author}{Dvorak, E.}, \bibinfo{author}{Ehrhardt, T.}, \bibinfo{author}{Eller, P.}, \bibinfo{author}{Engel, R.},
  \bibinfo{author}{Evans, J.J.}, \bibinfo{author}{Evenson, P.A.}, \bibinfo{author}{Fahey, S.}, \bibinfo{author}{Farrag, K.}, \bibinfo{author}{Fazely, A.R.}, \bibinfo{author}{Felde, J.}, \bibinfo{author}{Fienberg, A.T.}, \bibinfo{author}{Filimonov, K.}, \bibinfo{author}{Finley, C.}, \bibinfo{author}{Fischer, L.}, \bibinfo{author}{Fox, D.}, \bibinfo{author}{Franckowiak, A.}, \bibinfo{author}{Friedman, E.}, \bibinfo{author}{Fritz, A.}, \bibinfo{author}{Gaisser, T.K.}, \bibinfo{author}{Gallagher, J.}, \bibinfo{author}{Ganster, E.}, \bibinfo{author}{Garcia-Fernandez, D.}, \bibinfo{author}{Garrappa, S.}, \bibinfo{author}{Gartner, A.}, \bibinfo{author}{Gerhard, L.}, \bibinfo{author}{Gernhaeuser, R.}, \bibinfo{author}{Ghadimi, A.}, \bibinfo{author}{Glaser, C.}, \bibinfo{author}{Glauch, T.}, \bibinfo{author}{Glüsenkamp, T.}, \bibinfo{author}{Goldschmidt, A.}, \bibinfo{author}{Gonzalez, J.G.}, \bibinfo{author}{Goswami, S.}, \bibinfo{author}{Grant, D.}, \bibinfo{author}{Grégoire, T.}, \bibinfo{author}{Griffith, Z.},
  \bibinfo{author}{Griswold, S.}, \bibinfo{author}{Gündüz, M.}, \bibinfo{author}{Haack, C.}, \bibinfo{author}{Hallgren, A.}, \bibinfo{author}{Halliday, R.}, \bibinfo{author}{Halve, L.}, \bibinfo{author}{Halzen, F.}, \bibinfo{author}{Hanson, J.C.}, \bibinfo{author}{Hanson, K.}, \bibinfo{author}{Hardin, J.}, \bibinfo{author}{Haugen, J.}, \bibinfo{author}{Haungs, A.}, \bibinfo{author}{Hauser, S.}, \bibinfo{author}{Hebecker, D.}, \bibinfo{author}{Heinen, D.}, \bibinfo{author}{Heix, P.}, \bibinfo{author}{Helbing, K.}, \bibinfo{author}{Hellauer, R.}, \bibinfo{author}{Henningsen, F.}, \bibinfo{author}{Hickford, S.}, \bibinfo{author}{Hignight, J.}, \bibinfo{author}{Hill, C.}, \bibinfo{author}{Hill, G.C.}, \bibinfo{author}{Hoffman, K.D.}, \bibinfo{author}{Hoffmann, B.}, \bibinfo{author}{Hoffmann, R.}, \bibinfo{author}{Hoinka, T.}, \bibinfo{author}{Hokanson-Fasig, B.}, \bibinfo{author}{Holzapfel, K.}, \bibinfo{author}{Hoshina, K.}, \bibinfo{author}{Huang, F.}, \bibinfo{author}{Huber, M.}, \bibinfo{author}{Huber, T.},
  \bibinfo{author}{Huege, T.}, \bibinfo{author}{Hughes, K.}, \bibinfo{author}{Hultqvist, K.}, \bibinfo{author}{Hünnefeld, M.}, \bibinfo{author}{Hussain, R.}, \bibinfo{author}{In, S.}, \bibinfo{author}{Iovine, N.}, \bibinfo{author}{Ishihara, A.}, \bibinfo{author}{Jansson, M.}, \bibinfo{author}{Japaridze, G.S.}, \bibinfo{author}{Jeong, M.}, \bibinfo{author}{Jones, B.J.P.}, \bibinfo{author}{Jonske, F.}, \bibinfo{author}{Joppe, R.}, \bibinfo{author}{Kalekin, O.}, \bibinfo{author}{Kang, D.}, \bibinfo{author}{Kang, W.}, \bibinfo{author}{Kang, X.}, \bibinfo{author}{Kappes, A.}, \bibinfo{author}{Kappesser, D.}, \bibinfo{author}{Karg, T.}, \bibinfo{author}{Karl, M.}, \bibinfo{author}{Karle, A.}, \bibinfo{author}{Katori, T.}, \bibinfo{author}{Katz, U.}, \bibinfo{author}{Kauer, M.}, \bibinfo{author}{Keivani, A.}, \bibinfo{author}{Kellermann, M.}, \bibinfo{author}{Kelley, J.L.}, \bibinfo{author}{Kheirandish, A.}, \bibinfo{author}{Kim, J.}, \bibinfo{author}{Kin, K.}, \bibinfo{author}{Kintscher, T.},
  \bibinfo{author}{Kiryluk, J.}, \bibinfo{author}{Kittler, T.}, \bibinfo{author}{Kleifges, M.}, \bibinfo{author}{Klein, S.R.}, \bibinfo{author}{Koirala, R.}, \bibinfo{author}{Kolanoski, H.}, \bibinfo{author}{Köpke, L.}, \bibinfo{author}{Kopper, C.}, \bibinfo{author}{Kopper, S.}, \bibinfo{author}{Koskinen, D.J.}, \bibinfo{author}{Koundal, P.}, \bibinfo{author}{Kovacevich, M.}, \bibinfo{author}{Kowalski, M.}, \bibinfo{author}{Krauss, C.B.}, \bibinfo{author}{Krings, K.}, \bibinfo{author}{Krückl, G.}, \bibinfo{author}{Kulacz, N.}, \bibinfo{author}{Kurahashi, N.}, \bibinfo{author}{Gualda, C.L.}, \bibinfo{author}{Lahmann, R.}, \bibinfo{author}{Lanfranchi, J.L.}, \bibinfo{author}{Larson, M.J.}, \bibinfo{author}{Latif, U.}, \bibinfo{author}{Lauber, F.}, \bibinfo{author}{Lazar, J.P.}, \bibinfo{author}{Leonard, K.}, \bibinfo{author}{Leszczyńska, A.}, \bibinfo{author}{Li, Y.}, \bibinfo{author}{Liu, Q.R.}, \bibinfo{author}{Lohfink, E.}, \bibinfo{author}{LoSecco, J.}, \bibinfo{author}{Mariscal, C.J.L.},
  \bibinfo{author}{Lu, L.}, \bibinfo{author}{Lucarelli, F.}, \bibinfo{author}{Ludwig, A.}, \bibinfo{author}{Lünemann, J.}, \bibinfo{author}{Luszczak, W.}, \bibinfo{author}{Lyu, Y.}, \bibinfo{author}{Ma, W.Y.}, \bibinfo{author}{Madsen, J.}, \bibinfo{author}{Maggi, G.}, \bibinfo{author}{Mahn, K.B.M.}, \bibinfo{author}{Makino, Y.}, \bibinfo{author}{Mallik, P.}, \bibinfo{author}{Mancina, S.}, \bibinfo{author}{Mandalia, S.}, \bibinfo{author}{Mariş, I.C.}, \bibinfo{author}{Marka, S.}, \bibinfo{author}{Marka, Z.}, \bibinfo{author}{Maruyama, R.}, \bibinfo{author}{Mase, K.}, \bibinfo{author}{Maunu, R.}, \bibinfo{author}{McNally, F.}, \bibinfo{author}{Meagher, K.}, \bibinfo{author}{Medina, A.}, \bibinfo{author}{Meier, M.}, \bibinfo{author}{Meighen-Berger, S.}, \bibinfo{author}{Merz, J.}, \bibinfo{author}{Meyers, Z.S.}, \bibinfo{author}{Micallef, J.}, \bibinfo{author}{Mockler, D.}, \bibinfo{author}{Momenté, G.}, \bibinfo{author}{Montaruli, T.}, \bibinfo{author}{Moore, R.W.}, \bibinfo{author}{Morse, R.},
  \bibinfo{author}{Moulai, M.}, \bibinfo{author}{Muth, P.}, \bibinfo{author}{Naab, R.}, \bibinfo{author}{Nagai, R.}, \bibinfo{author}{Nam, J.}, \bibinfo{author}{Nauman, U.}, \bibinfo{author}{Necker, J.}, \bibinfo{author}{Neer, G.}, \bibinfo{author}{Nelles, A.}, \bibinfo{author}{Nguyen, L.V.}, \bibinfo{author}{Niederhausen, H.}, \bibinfo{author}{Nisa, M.U.}, \bibinfo{author}{Nowicki, S.C.}, \bibinfo{author}{Nygren, D.R.}, \bibinfo{author}{Oberla, E.}, \bibinfo{author}{Pollmann, A.O.}, \bibinfo{author}{Oehler, M.}, \bibinfo{author}{Olivas, A.}, \bibinfo{author}{O’Sullivan, E.}, \bibinfo{author}{Pan, Y.}, \bibinfo{author}{Pandya, H.}, \bibinfo{author}{Pankova, D.V.}, \bibinfo{author}{Papp, L.}, \bibinfo{author}{Park, N.}, \bibinfo{author}{Parker, G.K.}, \bibinfo{author}{Paudel, E.N.}, \bibinfo{author}{Peiffer, P.}, \bibinfo{author}{Pérez de~los Heros, C.}, \bibinfo{author}{Petersen, T.C.}, \bibinfo{author}{Philippen, S.}, \bibinfo{author}{Pieloth, D.}, \bibinfo{author}{Pieper, S.}, \bibinfo{author}{Pinfold,
  J.L.}, \bibinfo{author}{Pizzuto, A.}, \bibinfo{author}{Plaisier, I.}, \bibinfo{author}{Plum, M.}, \bibinfo{author}{Popovych, Y.}, \bibinfo{author}{Porcelli, A.}, \bibinfo{author}{Rodriguez, M.P.}, \bibinfo{author}{Price, P.B.}, \bibinfo{author}{Przybylski, G.T.}, \bibinfo{author}{Raab, C.}, \bibinfo{author}{Raissi, A.}, \bibinfo{author}{Rameez, M.}, \bibinfo{author}{Rauch, L.}, \bibinfo{author}{Rawlins, K.}, \bibinfo{author}{Rea, I.C.}, \bibinfo{author}{Rehman, A.}, \bibinfo{author}{Reimann, R.}, \bibinfo{author}{Renschler, M.}, \bibinfo{author}{Renzi, G.}, \bibinfo{author}{Resconi, E.}, \bibinfo{author}{Reusch, S.}, \bibinfo{author}{Rhode, W.}, \bibinfo{author}{Richman, M.}, \bibinfo{author}{Riedel, B.}, \bibinfo{author}{Riegel, M.}, \bibinfo{author}{Roberts, E.J.}, \bibinfo{author}{Robertson, S.}, \bibinfo{author}{Roellinghoff, G.}, \bibinfo{author}{Rongen, M.}, \bibinfo{author}{Rott, C.}, \bibinfo{author}{Ruhe, T.}, \bibinfo{author}{Ryckbosch, D.}, \bibinfo{author}{Cantu, D.R.}, \bibinfo{author}{Safa,
  I.}, \bibinfo{author}{Herrera, S.E.S.}, \bibinfo{author}{Sandrock, A.}, \bibinfo{author}{Sandroos, J.}, \bibinfo{author}{Sandstrom, P.}, \bibinfo{author}{Santander, M.}, \bibinfo{author}{Sarkar, S.}, \bibinfo{author}{Sarkar, S.}, \bibinfo{author}{Satalecka, K.}, \bibinfo{author}{Scharf, M.}, \bibinfo{author}{Schaufel, M.}, \bibinfo{author}{Schieler, H.}, \bibinfo{author}{Schlunder, P.}, \bibinfo{author}{Schmidt, T.}, \bibinfo{author}{Schneider, A.}, \bibinfo{author}{Schneider, J.}, \bibinfo{author}{Schröder, F.G.}, \bibinfo{author}{Schumacher, L.}, \bibinfo{author}{Sclafani, S.}, \bibinfo{author}{Seckel, D.}, \bibinfo{author}{Seunarine, S.}, \bibinfo{author}{Shaevitz, M.H.}, \bibinfo{author}{Sharma, A.}, \bibinfo{author}{Shefali, S.}, \bibinfo{author}{Silva, M.}, \bibinfo{author}{Smith, D.}, \bibinfo{author}{Smithers, B.}, \bibinfo{author}{Snihur, R.}, \bibinfo{author}{Soedingrekso, J.}, \bibinfo{author}{Soldin, D.}, \bibinfo{author}{Söldner-Rembold, S.}, \bibinfo{author}{Song, M.},
  \bibinfo{author}{Southall, D.}, \bibinfo{author}{Spiczak, G.M.}, \bibinfo{author}{Spiering, C.}, \bibinfo{author}{Stachurska, J.}, \bibinfo{author}{Stamatikos, M.}, \bibinfo{author}{Stanev, T.}, \bibinfo{author}{Stein, R.}, \bibinfo{author}{Stettner, J.}, \bibinfo{author}{Steuer, A.}, \bibinfo{author}{Stezelberger, T.}, \bibinfo{author}{Stokstad, R.G.}, \bibinfo{author}{Strotjohann, N.L.}, \bibinfo{author}{Stürwald, T.}, \bibinfo{author}{Stuttard, T.}, \bibinfo{author}{Sullivan, G.W.}, \bibinfo{author}{Taboada, I.}, \bibinfo{author}{Taketa, A.}, \bibinfo{author}{Tanaka, H.K.M.}, \bibinfo{author}{Tenholt, F.}, \bibinfo{author}{Ter-Antonyan, S.}, \bibinfo{author}{Terliuk, A.}, \bibinfo{author}{Tilav, S.}, \bibinfo{author}{Tollefson, K.}, \bibinfo{author}{Tomankova, L.}, \bibinfo{author}{Tönnis, C.}, \bibinfo{author}{Torres, J.}, \bibinfo{author}{Toscano, S.}, \bibinfo{author}{Tosi, D.}, \bibinfo{author}{Trettin, A.}, \bibinfo{author}{Tselengidou, M.}, \bibinfo{author}{Tung, C.F.}, \bibinfo{author}{Turcati,
  A.}, \bibinfo{author}{Turcotte, R.}, \bibinfo{author}{Turley, C.F.}, \bibinfo{author}{Twagirayezu, J.P.}, \bibinfo{author}{Ty, B.}, \bibinfo{author}{Unger, E.}, \bibinfo{author}{Elorrieta, M.A.U.}, \bibinfo{author}{Vandenbroucke, J.}, \bibinfo{author}{van Eijk, D.}, \bibinfo{author}{van Eijndhoven, N.}, \bibinfo{author}{Vannerom, D.}, \bibinfo{author}{van Santen, J.}, \bibinfo{author}{Veberic, D.}, \bibinfo{author}{Verpoest, S.}, \bibinfo{author}{Vieregg, A.}, \bibinfo{author}{Vraeghe, M.}, \bibinfo{author}{Walck, C.}, \bibinfo{author}{Watson, T.B.}, \bibinfo{author}{Weaver, C.}, \bibinfo{author}{Weindl, A.}, \bibinfo{author}{Weinstock, L.}, \bibinfo{author}{Weiss, M.J.}, \bibinfo{author}{Weldert, J.}, \bibinfo{author}{Welling, C.}, \bibinfo{author}{Wendt, C.}, \bibinfo{author}{Werthebach, J.}, \bibinfo{author}{Whitehorn, N.}, \bibinfo{author}{Wiebe, K.}, \bibinfo{author}{Wiebusch, C.H.}, \bibinfo{author}{Williams, D.R.}, \bibinfo{author}{Wissel, S.A.}, \bibinfo{author}{Wolf, M.}, \bibinfo{author}{Wood,
  T.R.}, \bibinfo{author}{Woschnagg, K.}, \bibinfo{author}{Wrede, G.}, \bibinfo{author}{Wren, S.}, \bibinfo{author}{Wulff, J.}, \bibinfo{author}{Xu, X.W.}, \bibinfo{author}{Xu, Y.}, \bibinfo{author}{Yanez, J.P.}, \bibinfo{author}{Yoshida, S.}, \bibinfo{author}{Yuan, T.}, \bibinfo{author}{Zhang, Z.}, \bibinfo{author}{Zierke, S.}, \bibinfo{author}{Zöcklein, M.}, \bibinfo{year}{2021}.
\newblock \bibinfo{title}{Icecube-gen2: the window to the extreme universe}.
\newblock \bibinfo{journal}{Journal of Physics G: Nuclear and Particle Physics} \bibinfo{volume}{48}, \bibinfo{pages}{060501}.
\newblock \URLprefix \url{http://dx.doi.org/10.1088/1361-6471/abbd48}, \DOIprefix\doi{10.1088/1361-6471/abbd48}.
\bibitem[{Abarr et~al.(2021)Abarr, Allison, Ammerman~Yebra, Alvarez-Muñiz, Beatty, Besson, Chen, Chen, Xie, Clem, Connolly, Cremonesi, Deaconu, Flaherty, Frikken, Gorham, Hast, Hornhuber, Huang, Hughes, Hynous, Ku, Kuo, Liu, Martin, Miki, Nam, Nichol, Nishimura, Novikov, Nozdrina, Oberla, Prohira, Prechelt, Rauch, Roberts, Romero-Wolf, Russell, Seckel, Shiao, Smith, Southall, Varner, Vieregg, Wang, Wang, Wissel, Young, Zas and Zeolla}]{PUEO_Abarr_2021}
\bibinfo{author}{Abarr, Q.}, \bibinfo{author}{Allison, P.}, \bibinfo{author}{Ammerman~Yebra, J.}, \bibinfo{author}{Alvarez-Muñiz, J.}, \bibinfo{author}{Beatty, J.}, \bibinfo{author}{Besson, D.}, \bibinfo{author}{Chen, P.}, \bibinfo{author}{Chen, Y.}, \bibinfo{author}{Xie, C.}, \bibinfo{author}{Clem, J.}, \bibinfo{author}{Connolly, A.}, \bibinfo{author}{Cremonesi, L.}, \bibinfo{author}{Deaconu, C.}, \bibinfo{author}{Flaherty, J.}, \bibinfo{author}{Frikken, D.}, \bibinfo{author}{Gorham, P.}, \bibinfo{author}{Hast, C.}, \bibinfo{author}{Hornhuber, C.}, \bibinfo{author}{Huang, J.}, \bibinfo{author}{Hughes, K.}, \bibinfo{author}{Hynous, A.}, \bibinfo{author}{Ku, Y.}, \bibinfo{author}{Kuo, C.Y.}, \bibinfo{author}{Liu, T.}, \bibinfo{author}{Martin, Z.}, \bibinfo{author}{Miki, C.}, \bibinfo{author}{Nam, J.}, \bibinfo{author}{Nichol, R.}, \bibinfo{author}{Nishimura, K.}, \bibinfo{author}{Novikov, A.}, \bibinfo{author}{Nozdrina, A.}, \bibinfo{author}{Oberla, E.}, \bibinfo{author}{Prohira, S.},
  \bibinfo{author}{Prechelt, R.}, \bibinfo{author}{Rauch, B.}, \bibinfo{author}{Roberts, J.}, \bibinfo{author}{Romero-Wolf, A.}, \bibinfo{author}{Russell, J.}, \bibinfo{author}{Seckel, D.}, \bibinfo{author}{Shiao, J.}, \bibinfo{author}{Smith, D.}, \bibinfo{author}{Southall, D.}, \bibinfo{author}{Varner, G.}, \bibinfo{author}{Vieregg, A.}, \bibinfo{author}{Wang, S.H.}, \bibinfo{author}{Wang, Y.H.}, \bibinfo{author}{Wissel, S.}, \bibinfo{author}{Young, R.}, \bibinfo{author}{Zas, E.}, \bibinfo{author}{Zeolla, A.}, \bibinfo{year}{2021}.
\newblock \bibinfo{title}{The payload for ultrahigh energy observations (pueo): a white paper}.
\newblock \bibinfo{journal}{Journal of Instrumentation} \bibinfo{volume}{16}, \bibinfo{pages}{P08035}.
\newblock \URLprefix \url{http://dx.doi.org/10.1088/1748-0221/16/08/P08035}, \DOIprefix\doi{10.1088/1748-0221/16/08/p08035}.
\bibitem[{Abbasi et~al.(2022)Abbasi, Ackermann, Adams, Aguilar, Ahlers, Ahrens, Alameddine, Alispach, Alves, Amin, Andeen, Anderson, Anton, Argüelles, Ashida, Axani, Bai, Balagopal~V., Barbano, Barwick, Bastian, Basu, Baur, Bay, Beatty, Becker, Becker~Tjus, Bellenghi, BenZvi, Berley, Bernardini, Besson, Binder, Bindig, Blaufuss, Blot, Boddenberg, Bontempo, Borowka, Böser, Botner, Böttcher, Bourbeau, Bradascio, Braun, Brinson, Bron, Brostean-Kaiser, Browne, Burgman, Burley, Busse, Campana, Carnie-Bronca, Chen, Chen, Chirkin, Choi, Clark, Clark, Classen, Coleman, Collin, Conrad, Coppin, Correa, Cowen, Cross, Dappen, Dave, De~Clercq, DeLaunay, Delgado~López, Dembinski, Deoskar, Desai, Desiati, de~Vries, de~Wasseige, de~With, DeYoung, Diaz, Díaz-Vélez, Dittmer, Dujmovic, Dunkman, DuVernois, Dvorak, Ehrhardt, Eller, Engel, Erpenbeck, Evans, Evenson, Fan, Fazely, Fedynitch, Feigl, Fiedlschuster, Fienberg, Filimonov, Finley, Fischer, Fox, Franckowiak, Friedman, Fritz, Fürst, Gaisser, Gallagher, Ganster,
  Garcia, Garrappa, Gerhardt, Ghadimi, Glaser, Glauch, Glüsenkamp, Goldschmidt, Gonzalez, Goswami, Grant, Grégoire, Griswold, Günther, Gutjahr, Haack, Hallgren, Halliday, Halve, Halzen, Ha~Minh, Hanson, Hardin, Harnisch, Haungs, Hebecker, Helbing, Henningsen, Hettinger, Hickford, Hignight, Hill, Hill, Hoffman, Hoffmann, Hokanson-Fasig, Hoshina, Huang, Huber, Huber, Hultqvist, Hünnefeld, Hussain, Hymon, In, Iovine, Ishihara, Jansson, Japaridze, Jeong, Jin, Jones, Kang, Kang, Kang, Kappes, Kappesser, Kardum, Karg, Karl, Karle, Katz, Kauer, Kellermann, Kelley, Kheirandish, Kin, Kintscher, Kiryluk, Klein, Koirala, Kolanoski, Kontrimas, Köpke, Kopper, Kopper, Koskinen, Koundal, Kovacevich, Kowalski, Kozynets, Kun, Kurahashi, Lad, Lagunas~Gualda, Lanfranchi, Larson, Lauber, Lazar, Lee, Leonard, Leszczyńska, Li, Lincetto, Liu, Liubarska, Lohfink, Lozano~Mariscal, Lu, Lucarelli, Ludwig, Luszczak, Lyu, Ma, Madsen, Mahn, Makino, Mancina, Mariş, Martinez-Soler, Maruyama, Mase, McElroy, McNally, Mead, Meagher,
  Mechbal, Medina, Meier, Meighen-Berger, Micallef, Mockler, Montaruli, Moore, Morse, Moulai, Naab, Nagai, Nahnhauer, Naumann, Necker, Nguyen, Niederhausen, Nisa, Nowicki, Nygren, Obertacke~Pollmann, Oehler, Oeyen, Olivas, O’Sullivan, Pandya, Pankova, Park, Parker, Paudel, Paul, Pérez de~los Heros, Peters, Peterson, Philippen, Pieper, Pittermann, Pizzuto, Plum, Popovych, Porcelli, Prado~Rodriguez, Price, Pries, Przybylski, Raab, Rack-Helleis, Raissi, Rameez, Rawlins, Rea, Rehman, Reichherzer, Reimann, Renzi, Resconi, Reusch, Rhode, Richman, Riedel, Roberts, Robertson, Roellinghoff, Rongen, Rott, Ruhe, Ryckbosch, Rysewyk~Cantu, Safa, Saffer, Sanchez~Herrera, Sandrock, Sandroos, Santander, Sarkar, Sarkar, Satalecka, Schaufel, Schieler, Schindler, Schmidt, Schneider, Schneider, Schröder, Schumacher, Schwefer, Sclafani, Seckel, Seunarine, Sharma, Shefali, Silva, Skrzypek, Smithers, Snihur, Soedingrekso, Soldin, Spannfellner, Spiczak, Spiering, Stachurska, Stamatikos, Stanev, Stein, Stettner, Steuer,
  Stezelberger, Stokstad, Stürwald, Stuttard, Sullivan, Taboada, Ter-Antonyan, Tilav, Tischbein, Tollefson, Tönnis, Toscano, Tosi, Trettin, Tselengidou, Tung, Turcati, Turcotte, Turley, Twagirayezu, Ty, Unland~Elorrieta, Valtonen-Mattila, Vandenbroucke, van Eijndhoven, Vannerom, van Santen, Verpoest, Walck, Watson, Weaver, Weigel, Weindl, Weiss, Weldert, Wendt, Werthebach, Weyrauch, Whitehorn, Wiebusch, Williams, Wolf, Woschnagg, Wrede, Wulff, Xu, Yanez, Yoshida, Yu, Yuan, Zhang and Zhelnin}]{NGC2022}
\bibinfo{author}{Abbasi, R.}, \bibinfo{author}{Ackermann, M.}, \bibinfo{author}{Adams, J.}, \bibinfo{author}{Aguilar, J.A.}, \bibinfo{author}{Ahlers, M.}, \bibinfo{author}{Ahrens, M.}, \bibinfo{author}{Alameddine, J.M.}, \bibinfo{author}{Alispach, C.}, \bibinfo{author}{Alves, A.A.}, \bibinfo{author}{Amin, N.M.}, \bibinfo{author}{Andeen, K.}, \bibinfo{author}{Anderson, T.}, \bibinfo{author}{Anton, G.}, \bibinfo{author}{Argüelles, C.}, \bibinfo{author}{Ashida, Y.}, \bibinfo{author}{Axani, S.}, \bibinfo{author}{Bai, X.}, \bibinfo{author}{Balagopal~V., A.}, \bibinfo{author}{Barbano, A.}, \bibinfo{author}{Barwick, S.W.}, \bibinfo{author}{Bastian, B.}, \bibinfo{author}{Basu, V.}, \bibinfo{author}{Baur, S.}, \bibinfo{author}{Bay, R.}, \bibinfo{author}{Beatty, J.J.}, \bibinfo{author}{Becker, K.H.}, \bibinfo{author}{Becker~Tjus, J.}, \bibinfo{author}{Bellenghi, C.}, \bibinfo{author}{BenZvi, S.}, \bibinfo{author}{Berley, D.}, \bibinfo{author}{Bernardini, E.}, \bibinfo{author}{Besson, D.Z.}, \bibinfo{author}{Binder,
  G.}, \bibinfo{author}{Bindig, D.}, \bibinfo{author}{Blaufuss, E.}, \bibinfo{author}{Blot, S.}, \bibinfo{author}{Boddenberg, M.}, \bibinfo{author}{Bontempo, F.}, \bibinfo{author}{Borowka, J.}, \bibinfo{author}{Böser, S.}, \bibinfo{author}{Botner, O.}, \bibinfo{author}{Böttcher, J.}, \bibinfo{author}{Bourbeau, E.}, \bibinfo{author}{Bradascio, F.}, \bibinfo{author}{Braun, J.}, \bibinfo{author}{Brinson, B.}, \bibinfo{author}{Bron, S.}, \bibinfo{author}{Brostean-Kaiser, J.}, \bibinfo{author}{Browne, S.}, \bibinfo{author}{Burgman, A.}, \bibinfo{author}{Burley, R.T.}, \bibinfo{author}{Busse, R.S.}, \bibinfo{author}{Campana, M.A.}, \bibinfo{author}{Carnie-Bronca, E.G.}, \bibinfo{author}{Chen, C.}, \bibinfo{author}{Chen, Z.}, \bibinfo{author}{Chirkin, D.}, \bibinfo{author}{Choi, K.}, \bibinfo{author}{Clark, B.A.}, \bibinfo{author}{Clark, K.}, \bibinfo{author}{Classen, L.}, \bibinfo{author}{Coleman, A.}, \bibinfo{author}{Collin, G.H.}, \bibinfo{author}{Conrad, J.M.}, \bibinfo{author}{Coppin, P.},
  \bibinfo{author}{Correa, P.}, \bibinfo{author}{Cowen, D.F.}, \bibinfo{author}{Cross, R.}, \bibinfo{author}{Dappen, C.}, \bibinfo{author}{Dave, P.}, \bibinfo{author}{De~Clercq, C.}, \bibinfo{author}{DeLaunay, J.J.}, \bibinfo{author}{Delgado~López, D.}, \bibinfo{author}{Dembinski, H.}, \bibinfo{author}{Deoskar, K.}, \bibinfo{author}{Desai, A.}, \bibinfo{author}{Desiati, P.}, \bibinfo{author}{de~Vries, K.D.}, \bibinfo{author}{de~Wasseige, G.}, \bibinfo{author}{de~With, M.}, \bibinfo{author}{DeYoung, T.}, \bibinfo{author}{Diaz, A.}, \bibinfo{author}{Díaz-Vélez, J.C.}, \bibinfo{author}{Dittmer, M.}, \bibinfo{author}{Dujmovic, H.}, \bibinfo{author}{Dunkman, M.}, \bibinfo{author}{DuVernois, M.A.}, \bibinfo{author}{Dvorak, E.}, \bibinfo{author}{Ehrhardt, T.}, \bibinfo{author}{Eller, P.}, \bibinfo{author}{Engel, R.}, \bibinfo{author}{Erpenbeck, H.}, \bibinfo{author}{Evans, J.}, \bibinfo{author}{Evenson, P.A.}, \bibinfo{author}{Fan, K.L.}, \bibinfo{author}{Fazely, A.R.}, \bibinfo{author}{Fedynitch, A.},
  \bibinfo{author}{Feigl, N.}, \bibinfo{author}{Fiedlschuster, S.}, \bibinfo{author}{Fienberg, A.T.}, \bibinfo{author}{Filimonov, K.}, \bibinfo{author}{Finley, C.}, \bibinfo{author}{Fischer, L.}, \bibinfo{author}{Fox, D.}, \bibinfo{author}{Franckowiak, A.}, \bibinfo{author}{Friedman, E.}, \bibinfo{author}{Fritz, A.}, \bibinfo{author}{Fürst, P.}, \bibinfo{author}{Gaisser, T.K.}, \bibinfo{author}{Gallagher, J.}, \bibinfo{author}{Ganster, E.}, \bibinfo{author}{Garcia, A.}, \bibinfo{author}{Garrappa, S.}, \bibinfo{author}{Gerhardt, L.}, \bibinfo{author}{Ghadimi, A.}, \bibinfo{author}{Glaser, C.}, \bibinfo{author}{Glauch, T.}, \bibinfo{author}{Glüsenkamp, T.}, \bibinfo{author}{Goldschmidt, A.}, \bibinfo{author}{Gonzalez, J.G.}, \bibinfo{author}{Goswami, S.}, \bibinfo{author}{Grant, D.}, \bibinfo{author}{Grégoire, T.}, \bibinfo{author}{Griswold, S.}, \bibinfo{author}{Günther, C.}, \bibinfo{author}{Gutjahr, P.}, \bibinfo{author}{Haack, C.}, \bibinfo{author}{Hallgren, A.}, \bibinfo{author}{Halliday, R.},
  \bibinfo{author}{Halve, L.}, \bibinfo{author}{Halzen, F.}, \bibinfo{author}{Ha~Minh, M.}, \bibinfo{author}{Hanson, K.}, \bibinfo{author}{Hardin, J.}, \bibinfo{author}{Harnisch, A.A.}, \bibinfo{author}{Haungs, A.}, \bibinfo{author}{Hebecker, D.}, \bibinfo{author}{Helbing, K.}, \bibinfo{author}{Henningsen, F.}, \bibinfo{author}{Hettinger, E.C.}, \bibinfo{author}{Hickford, S.}, \bibinfo{author}{Hignight, J.}, \bibinfo{author}{Hill, C.}, \bibinfo{author}{Hill, G.C.}, \bibinfo{author}{Hoffman, K.D.}, \bibinfo{author}{Hoffmann, R.}, \bibinfo{author}{Hokanson-Fasig, B.}, \bibinfo{author}{Hoshina, K.}, \bibinfo{author}{Huang, F.}, \bibinfo{author}{Huber, M.}, \bibinfo{author}{Huber, T.}, \bibinfo{author}{Hultqvist, K.}, \bibinfo{author}{Hünnefeld, M.}, \bibinfo{author}{Hussain, R.}, \bibinfo{author}{Hymon, K.}, \bibinfo{author}{In, S.}, \bibinfo{author}{Iovine, N.}, \bibinfo{author}{Ishihara, A.}, \bibinfo{author}{Jansson, M.}, \bibinfo{author}{Japaridze, G.S.}, \bibinfo{author}{Jeong, M.}, \bibinfo{author}{Jin,
  M.}, \bibinfo{author}{Jones, B.J.P.}, \bibinfo{author}{Kang, D.}, \bibinfo{author}{Kang, W.}, \bibinfo{author}{Kang, X.}, \bibinfo{author}{Kappes, A.}, \bibinfo{author}{Kappesser, D.}, \bibinfo{author}{Kardum, L.}, \bibinfo{author}{Karg, T.}, \bibinfo{author}{Karl, M.}, \bibinfo{author}{Karle, A.}, \bibinfo{author}{Katz, U.}, \bibinfo{author}{Kauer, M.}, \bibinfo{author}{Kellermann, M.}, \bibinfo{author}{Kelley, J.L.}, \bibinfo{author}{Kheirandish, A.}, \bibinfo{author}{Kin, K.}, \bibinfo{author}{Kintscher, T.}, \bibinfo{author}{Kiryluk, J.}, \bibinfo{author}{Klein, S.R.}, \bibinfo{author}{Koirala, R.}, \bibinfo{author}{Kolanoski, H.}, \bibinfo{author}{Kontrimas, T.}, \bibinfo{author}{Köpke, L.}, \bibinfo{author}{Kopper, C.}, \bibinfo{author}{Kopper, S.}, \bibinfo{author}{Koskinen, D.J.}, \bibinfo{author}{Koundal, P.}, \bibinfo{author}{Kovacevich, M.}, \bibinfo{author}{Kowalski, M.}, \bibinfo{author}{Kozynets, T.}, \bibinfo{author}{Kun, E.}, \bibinfo{author}{Kurahashi, N.}, \bibinfo{author}{Lad, N.},
  \bibinfo{author}{Lagunas~Gualda, C.}, \bibinfo{author}{Lanfranchi, J.L.}, \bibinfo{author}{Larson, M.J.}, \bibinfo{author}{Lauber, F.}, \bibinfo{author}{Lazar, J.P.}, \bibinfo{author}{Lee, J.W.}, \bibinfo{author}{Leonard, K.}, \bibinfo{author}{Leszczyńska, A.}, \bibinfo{author}{Li, Y.}, \bibinfo{author}{Lincetto, M.}, \bibinfo{author}{Liu, Q.R.}, \bibinfo{author}{Liubarska, M.}, \bibinfo{author}{Lohfink, E.}, \bibinfo{author}{Lozano~Mariscal, C.J.}, \bibinfo{author}{Lu, L.}, \bibinfo{author}{Lucarelli, F.}, \bibinfo{author}{Ludwig, A.}, \bibinfo{author}{Luszczak, W.}, \bibinfo{author}{Lyu, Y.}, \bibinfo{author}{Ma, W.Y.}, \bibinfo{author}{Madsen, J.}, \bibinfo{author}{Mahn, K.B.M.}, \bibinfo{author}{Makino, Y.}, \bibinfo{author}{Mancina, S.}, \bibinfo{author}{Mariş, I.C.}, \bibinfo{author}{Martinez-Soler, I.}, \bibinfo{author}{Maruyama, R.}, \bibinfo{author}{Mase, K.}, \bibinfo{author}{McElroy, T.}, \bibinfo{author}{McNally, F.}, \bibinfo{author}{Mead, J.V.}, \bibinfo{author}{Meagher, K.},
  \bibinfo{author}{Mechbal, S.}, \bibinfo{author}{Medina, A.}, \bibinfo{author}{Meier, M.}, \bibinfo{author}{Meighen-Berger, S.}, \bibinfo{author}{Micallef, J.}, \bibinfo{author}{Mockler, D.}, \bibinfo{author}{Montaruli, T.}, \bibinfo{author}{Moore, R.W.}, \bibinfo{author}{Morse, R.}, \bibinfo{author}{Moulai, M.}, \bibinfo{author}{Naab, R.}, \bibinfo{author}{Nagai, R.}, \bibinfo{author}{Nahnhauer, R.}, \bibinfo{author}{Naumann, U.}, \bibinfo{author}{Necker, J.}, \bibinfo{author}{Nguyen, L.V.}, \bibinfo{author}{Niederhausen, H.}, \bibinfo{author}{Nisa, M.U.}, \bibinfo{author}{Nowicki, S.C.}, \bibinfo{author}{Nygren, D.}, \bibinfo{author}{Obertacke~Pollmann, A.}, \bibinfo{author}{Oehler, M.}, \bibinfo{author}{Oeyen, B.}, \bibinfo{author}{Olivas, A.}, \bibinfo{author}{O’Sullivan, E.}, \bibinfo{author}{Pandya, H.}, \bibinfo{author}{Pankova, D.V.}, \bibinfo{author}{Park, N.}, \bibinfo{author}{Parker, G.K.}, \bibinfo{author}{Paudel, E.N.}, \bibinfo{author}{Paul, L.}, \bibinfo{author}{Pérez de~los Heros, C.},
  \bibinfo{author}{Peters, L.}, \bibinfo{author}{Peterson, J.}, \bibinfo{author}{Philippen, S.}, \bibinfo{author}{Pieper, S.}, \bibinfo{author}{Pittermann, M.}, \bibinfo{author}{Pizzuto, A.}, \bibinfo{author}{Plum, M.}, \bibinfo{author}{Popovych, Y.}, \bibinfo{author}{Porcelli, A.}, \bibinfo{author}{Prado~Rodriguez, M.}, \bibinfo{author}{Price, P.B.}, \bibinfo{author}{Pries, B.}, \bibinfo{author}{Przybylski, G.T.}, \bibinfo{author}{Raab, C.}, \bibinfo{author}{Rack-Helleis, J.}, \bibinfo{author}{Raissi, A.}, \bibinfo{author}{Rameez, M.}, \bibinfo{author}{Rawlins, K.}, \bibinfo{author}{Rea, I.C.}, \bibinfo{author}{Rehman, A.}, \bibinfo{author}{Reichherzer, P.}, \bibinfo{author}{Reimann, R.}, \bibinfo{author}{Renzi, G.}, \bibinfo{author}{Resconi, E.}, \bibinfo{author}{Reusch, S.}, \bibinfo{author}{Rhode, W.}, \bibinfo{author}{Richman, M.}, \bibinfo{author}{Riedel, B.}, \bibinfo{author}{Roberts, E.J.}, \bibinfo{author}{Robertson, S.}, \bibinfo{author}{Roellinghoff, G.}, \bibinfo{author}{Rongen, M.},
  \bibinfo{author}{Rott, C.}, \bibinfo{author}{Ruhe, T.}, \bibinfo{author}{Ryckbosch, D.}, \bibinfo{author}{Rysewyk~Cantu, D.}, \bibinfo{author}{Safa, I.}, \bibinfo{author}{Saffer, J.}, \bibinfo{author}{Sanchez~Herrera, S.E.}, \bibinfo{author}{Sandrock, A.}, \bibinfo{author}{Sandroos, J.}, \bibinfo{author}{Santander, M.}, \bibinfo{author}{Sarkar, S.}, \bibinfo{author}{Sarkar, S.}, \bibinfo{author}{Satalecka, K.}, \bibinfo{author}{Schaufel, M.}, \bibinfo{author}{Schieler, H.}, \bibinfo{author}{Schindler, S.}, \bibinfo{author}{Schmidt, T.}, \bibinfo{author}{Schneider, A.}, \bibinfo{author}{Schneider, J.}, \bibinfo{author}{Schröder, F.G.}, \bibinfo{author}{Schumacher, L.}, \bibinfo{author}{Schwefer, G.}, \bibinfo{author}{Sclafani, S.}, \bibinfo{author}{Seckel, D.}, \bibinfo{author}{Seunarine, S.}, \bibinfo{author}{Sharma, A.}, \bibinfo{author}{Shefali, S.}, \bibinfo{author}{Silva, M.}, \bibinfo{author}{Skrzypek, B.}, \bibinfo{author}{Smithers, B.}, \bibinfo{author}{Snihur, R.}, \bibinfo{author}{Soedingrekso,
  J.}, \bibinfo{author}{Soldin, D.}, \bibinfo{author}{Spannfellner, C.}, \bibinfo{author}{Spiczak, G.M.}, \bibinfo{author}{Spiering, C.}, \bibinfo{author}{Stachurska, J.}, \bibinfo{author}{Stamatikos, M.}, \bibinfo{author}{Stanev, T.}, \bibinfo{author}{Stein, R.}, \bibinfo{author}{Stettner, J.}, \bibinfo{author}{Steuer, A.}, \bibinfo{author}{Stezelberger, T.}, \bibinfo{author}{Stokstad, R.}, \bibinfo{author}{Stürwald, T.}, \bibinfo{author}{Stuttard, T.}, \bibinfo{author}{Sullivan, G.W.}, \bibinfo{author}{Taboada, I.}, \bibinfo{author}{Ter-Antonyan, S.}, \bibinfo{author}{Tilav, S.}, \bibinfo{author}{Tischbein, F.}, \bibinfo{author}{Tollefson, K.}, \bibinfo{author}{Tönnis, C.}, \bibinfo{author}{Toscano, S.}, \bibinfo{author}{Tosi, D.}, \bibinfo{author}{Trettin, A.}, \bibinfo{author}{Tselengidou, M.}, \bibinfo{author}{Tung, C.F.}, \bibinfo{author}{Turcati, A.}, \bibinfo{author}{Turcotte, R.}, \bibinfo{author}{Turley, C.F.}, \bibinfo{author}{Twagirayezu, J.P.}, \bibinfo{author}{Ty, B.},
  \bibinfo{author}{Unland~Elorrieta, M.A.}, \bibinfo{author}{Valtonen-Mattila, N.}, \bibinfo{author}{Vandenbroucke, J.}, \bibinfo{author}{van Eijndhoven, N.}, \bibinfo{author}{Vannerom, D.}, \bibinfo{author}{van Santen, J.}, \bibinfo{author}{Verpoest, S.}, \bibinfo{author}{Walck, C.}, \bibinfo{author}{Watson, T.B.}, \bibinfo{author}{Weaver, C.}, \bibinfo{author}{Weigel, P.}, \bibinfo{author}{Weindl, A.}, \bibinfo{author}{Weiss, M.J.}, \bibinfo{author}{Weldert, J.}, \bibinfo{author}{Wendt, C.}, \bibinfo{author}{Werthebach, J.}, \bibinfo{author}{Weyrauch, M.}, \bibinfo{author}{Whitehorn, N.}, \bibinfo{author}{Wiebusch, C.H.}, \bibinfo{author}{Williams, D.R.}, \bibinfo{author}{Wolf, M.}, \bibinfo{author}{Woschnagg, K.}, \bibinfo{author}{Wrede, G.}, \bibinfo{author}{Wulff, J.}, \bibinfo{author}{Xu, X.W.}, \bibinfo{author}{Yanez, J.P.}, \bibinfo{author}{Yoshida, S.}, \bibinfo{author}{Yu, S.}, \bibinfo{author}{Yuan, T.}, \bibinfo{author}{Zhang, Z.}, \bibinfo{author}{Zhelnin, P.}, \bibinfo{year}{2022}.
\newblock \bibinfo{title}{Evidence for neutrino emission from the nearby active galaxy ngc 1068}.
\newblock \bibinfo{journal}{Science} \bibinfo{volume}{378}, \bibinfo{pages}{538–543}.
\newblock \URLprefix \url{http://dx.doi.org/10.1126/science.abg3395}, \DOIprefix\doi{10.1126/science.abg3395}.
\bibitem[{Abbasi et~al.(2023)Abbasi, Ackermann, Adams, Aguilar, Ahlers, Ahrens, Alameddine, Alves, Amin, Andeen, Anderson, Anton, Argüelles, Ashida, Athanasiadou, Axani, Bai, Balagopal~V., Barwick, Basu, Baur, Bay, Beatty, Becker, Tjus, Beise, Bellenghi, Benda, BenZvi, Berley, Bernardini, Besson, Binder, Bindig, Blaufuss, Blot, Boddenberg, Bontempo, Book, Borowka, Böser, Botner, Böttcher, Bourbeau, Bradascio, Braun, Brinson, Bron, Brostean-Kaiser, Burley, Busse, Campana, Carnie-Bronca, Chen, Chen, Chirkin, Choi, Clark, Clark, Classen, Coleman, Collin, Connolly, Conrad, Coppin, Correa, Cowen, Cross, Dappen, Dave, De~Clercq, DeLaunay, López, Dembinski, Deoskar, Desai, Desiati, de~Vries, de~Wasseige, DeYoung, Diaz, Díaz-Vélez, Dittmer, Dujmovic, Dunkman, DuVernois, Ehrhardt, Eller, Engel, Erpenbeck, Evans, Evenson, Fan, Fazely, Fedynitch, Feigl, Fiedlschuster, Fienberg, Finley, Fischer, Fox, Franckowiak, Friedman, Fritz, Fürst, Gaisser, Gallagher, Ganster, Garcia, Garrappa, Gerhardt, Ghadimi, Glaser,
  Glauch, Glüsenkamp, Goehlke, Goldschmidt, Gonzalez, Goswami, Grant, Grégoire, Griswold, Günther, Gutjahr, Haack, Hallgren, Halliday, Halve, Halzen, Minh, Hanson, Hardin, Harnisch, Haungs, Helbing, Henningsen, Hettinger, Hickford, Hignight, Hill, Hill, Hoffman, Hoshina, Hou, Huang, Huber, Huber, Hultqvist, Hünnefeld, Hussain, Hymon, In, Iovine, Ishihara, Jansson, Japaridze, Jeong, Jin, Jones, Kang, Kang, Kang, Kappes, Kappesser, Kardum, Karg, Karl, Karle, Katz, Kauer, Kellermann, Kelley, Kheirandish, Kin, Kiryluk, Klein, Kochocki, Koirala, Kolanoski, Kontrimas, Köpke, Kopper, Kopper, Koskinen, Koundal, Kovacevich, Kowalski, Kozynets, Krupczak, Kun, Kurahashi, Lad, Gualda, Lanfranchi, Larson, Lauber, Lazar, Lee, Leonard, Leszczyńska, Li, Lincetto, Liu, Liubarska, Lohfink, Mariscal, Lu, Lucarelli, Ludwig, Luszczak, Lyu, Ma, Madsen, Mahn, Makino, Mancina, Mariş, Martinez-Soler, Maruyama, McCarthy, McElroy, McNally, Mead, Meagher, Mechbal, Medina, Meier, Meighen-Berger, Merckx, Micallef, Mockler,
  Montaruli, Moore, Morik, Morse, Moulai, Mukherjee, Naab, Nagai, Nahnhauer, Naumann, Necker, Nguyen, Niederhausen, Nisa, Nowicki, Nygren, Pollmann, Oehler, Oeyen, Olivas, O’Sullivan, Pandya, Pankova, Park, Parker, Paudel, Paul, de~los Heros, Peters, Peterson, Philippen, Pieper, Pizzuto, Plum, Popovych, Porcelli, Rodriguez, Pries, Przybylski, Raab, Rack-Helleis, Raissi, Rameez, Rawlins, Rea, Rechav, Rehman, Reichherzer, Reimann, Renzi, Resconi, Reusch, Rhode, Richman, Riedel, Roberts, Robertson, Roellinghoff, Rongen, Rott, Ruhe, Ryckbosch, Cantu, Safa, Saffer, Salazar-Gallegos, Sampathkumar, Herrera, Sandrock, Santander, Sarkar, Sarkar, Satalecka, Schaufel, Schieler, Schindler, Schmidt, Schneider, Schneider, Schröder, Schumacher, Schwefer, Sclafani, Seckel, Seunarine, Sharma, Shefali, Shimizu, Silva, Skrzypek, Smithers, Snihur, Soedingrekso, Sogaard, Soldin, Spannfellner, Spiczak, Spiering, Stamatikos, Stanev, Stein, Stettner, Stezelberger, Stokstad, Stürwald, Stuttard, Sullivan, Taboada, Ter-Antonyan,
  Thwaites, Tilav, Tischbein, Tollefson, Tönnis, Toscano, Tosi, Trettin, Tselengidou, Tung, Turcati, Turcotte, Turley, Twagirayezu, Ty, Elorrieta, Valtonen-Mattila, Vandenbroucke, van Eijndhoven, Vannerom, van Santen, Veitch-Michaelis, Verpoest, Walck, Wang, Watson, Weaver, Weigel, Weindl, Weiss, Weldert, Wendt, Werthebach, Weyrauch, Whitehorn, Wiebusch, Willey, Williams, Wolf, Wrede, Wulff, Xu, Yanez, Yildizci, Yoshida, Yu, Yuan, Zhang and Zhelnin}]{galacticplaneIceCube2023}
\bibinfo{author}{Abbasi, R.}, \bibinfo{author}{Ackermann, M.}, \bibinfo{author}{Adams, J.}, \bibinfo{author}{Aguilar, J.A.}, \bibinfo{author}{Ahlers, M.}, \bibinfo{author}{Ahrens, M.}, \bibinfo{author}{Alameddine, J.M.}, \bibinfo{author}{Alves, A.A.}, \bibinfo{author}{Amin, N.M.}, \bibinfo{author}{Andeen, K.}, \bibinfo{author}{Anderson, T.}, \bibinfo{author}{Anton, G.}, \bibinfo{author}{Argüelles, C.}, \bibinfo{author}{Ashida, Y.}, \bibinfo{author}{Athanasiadou, S.}, \bibinfo{author}{Axani, S.}, \bibinfo{author}{Bai, X.}, \bibinfo{author}{Balagopal~V., A.}, \bibinfo{author}{Barwick, S.W.}, \bibinfo{author}{Basu, V.}, \bibinfo{author}{Baur, S.}, \bibinfo{author}{Bay, R.}, \bibinfo{author}{Beatty, J.J.}, \bibinfo{author}{Becker, K.H.}, \bibinfo{author}{Tjus, J.B.}, \bibinfo{author}{Beise, J.}, \bibinfo{author}{Bellenghi, C.}, \bibinfo{author}{Benda, S.}, \bibinfo{author}{BenZvi, S.}, \bibinfo{author}{Berley, D.}, \bibinfo{author}{Bernardini, E.}, \bibinfo{author}{Besson, D.Z.}, \bibinfo{author}{Binder, G.},
  \bibinfo{author}{Bindig, D.}, \bibinfo{author}{Blaufuss, E.}, \bibinfo{author}{Blot, S.}, \bibinfo{author}{Boddenberg, M.}, \bibinfo{author}{Bontempo, F.}, \bibinfo{author}{Book, J.Y.}, \bibinfo{author}{Borowka, J.}, \bibinfo{author}{Böser, S.}, \bibinfo{author}{Botner, O.}, \bibinfo{author}{Böttcher, J.}, \bibinfo{author}{Bourbeau, E.}, \bibinfo{author}{Bradascio, F.}, \bibinfo{author}{Braun, J.}, \bibinfo{author}{Brinson, B.}, \bibinfo{author}{Bron, S.}, \bibinfo{author}{Brostean-Kaiser, J.}, \bibinfo{author}{Burley, R.T.}, \bibinfo{author}{Busse, R.S.}, \bibinfo{author}{Campana, M.A.}, \bibinfo{author}{Carnie-Bronca, E.G.}, \bibinfo{author}{Chen, C.}, \bibinfo{author}{Chen, Z.}, \bibinfo{author}{Chirkin, D.}, \bibinfo{author}{Choi, K.}, \bibinfo{author}{Clark, B.A.}, \bibinfo{author}{Clark, K.}, \bibinfo{author}{Classen, L.}, \bibinfo{author}{Coleman, A.}, \bibinfo{author}{Collin, G.H.}, \bibinfo{author}{Connolly, A.}, \bibinfo{author}{Conrad, J.M.}, \bibinfo{author}{Coppin, P.},
  \bibinfo{author}{Correa, P.}, \bibinfo{author}{Cowen, D.F.}, \bibinfo{author}{Cross, R.}, \bibinfo{author}{Dappen, C.}, \bibinfo{author}{Dave, P.}, \bibinfo{author}{De~Clercq, C.}, \bibinfo{author}{DeLaunay, J.J.}, \bibinfo{author}{López, D.D.}, \bibinfo{author}{Dembinski, H.}, \bibinfo{author}{Deoskar, K.}, \bibinfo{author}{Desai, A.}, \bibinfo{author}{Desiati, P.}, \bibinfo{author}{de~Vries, K.D.}, \bibinfo{author}{de~Wasseige, G.}, \bibinfo{author}{DeYoung, T.}, \bibinfo{author}{Diaz, A.}, \bibinfo{author}{Díaz-Vélez, J.C.}, \bibinfo{author}{Dittmer, M.}, \bibinfo{author}{Dujmovic, H.}, \bibinfo{author}{Dunkman, M.}, \bibinfo{author}{DuVernois, M.A.}, \bibinfo{author}{Ehrhardt, T.}, \bibinfo{author}{Eller, P.}, \bibinfo{author}{Engel, R.}, \bibinfo{author}{Erpenbeck, H.}, \bibinfo{author}{Evans, J.}, \bibinfo{author}{Evenson, P.A.}, \bibinfo{author}{Fan, K.L.}, \bibinfo{author}{Fazely, A.R.}, \bibinfo{author}{Fedynitch, A.}, \bibinfo{author}{Feigl, N.}, \bibinfo{author}{Fiedlschuster, S.},
  \bibinfo{author}{Fienberg, A.T.}, \bibinfo{author}{Finley, C.}, \bibinfo{author}{Fischer, L.}, \bibinfo{author}{Fox, D.}, \bibinfo{author}{Franckowiak, A.}, \bibinfo{author}{Friedman, E.}, \bibinfo{author}{Fritz, A.}, \bibinfo{author}{Fürst, P.}, \bibinfo{author}{Gaisser, T.K.}, \bibinfo{author}{Gallagher, J.}, \bibinfo{author}{Ganster, E.}, \bibinfo{author}{Garcia, A.}, \bibinfo{author}{Garrappa, S.}, \bibinfo{author}{Gerhardt, L.}, \bibinfo{author}{Ghadimi, A.}, \bibinfo{author}{Glaser, C.}, \bibinfo{author}{Glauch, T.}, \bibinfo{author}{Glüsenkamp, T.}, \bibinfo{author}{Goehlke, N.}, \bibinfo{author}{Goldschmidt, A.}, \bibinfo{author}{Gonzalez, J.G.}, \bibinfo{author}{Goswami, S.}, \bibinfo{author}{Grant, D.}, \bibinfo{author}{Grégoire, T.}, \bibinfo{author}{Griswold, S.}, \bibinfo{author}{Günther, C.}, \bibinfo{author}{Gutjahr, P.}, \bibinfo{author}{Haack, C.}, \bibinfo{author}{Hallgren, A.}, \bibinfo{author}{Halliday, R.}, \bibinfo{author}{Halve, L.}, \bibinfo{author}{Halzen, F.},
  \bibinfo{author}{Minh, M.H.}, \bibinfo{author}{Hanson, K.}, \bibinfo{author}{Hardin, J.}, \bibinfo{author}{Harnisch, A.A.}, \bibinfo{author}{Haungs, A.}, \bibinfo{author}{Helbing, K.}, \bibinfo{author}{Henningsen, F.}, \bibinfo{author}{Hettinger, E.C.}, \bibinfo{author}{Hickford, S.}, \bibinfo{author}{Hignight, J.}, \bibinfo{author}{Hill, C.}, \bibinfo{author}{Hill, G.C.}, \bibinfo{author}{Hoffman, K.D.}, \bibinfo{author}{Hoshina, K.}, \bibinfo{author}{Hou, W.}, \bibinfo{author}{Huang, F.}, \bibinfo{author}{Huber, M.}, \bibinfo{author}{Huber, T.}, \bibinfo{author}{Hultqvist, K.}, \bibinfo{author}{Hünnefeld, M.}, \bibinfo{author}{Hussain, R.}, \bibinfo{author}{Hymon, K.}, \bibinfo{author}{In, S.}, \bibinfo{author}{Iovine, N.}, \bibinfo{author}{Ishihara, A.}, \bibinfo{author}{Jansson, M.}, \bibinfo{author}{Japaridze, G.S.}, \bibinfo{author}{Jeong, M.}, \bibinfo{author}{Jin, M.}, \bibinfo{author}{Jones, B.J.P.}, \bibinfo{author}{Kang, D.}, \bibinfo{author}{Kang, W.}, \bibinfo{author}{Kang, X.},
  \bibinfo{author}{Kappes, A.}, \bibinfo{author}{Kappesser, D.}, \bibinfo{author}{Kardum, L.}, \bibinfo{author}{Karg, T.}, \bibinfo{author}{Karl, M.}, \bibinfo{author}{Karle, A.}, \bibinfo{author}{Katz, U.}, \bibinfo{author}{Kauer, M.}, \bibinfo{author}{Kellermann, M.}, \bibinfo{author}{Kelley, J.L.}, \bibinfo{author}{Kheirandish, A.}, \bibinfo{author}{Kin, K.}, \bibinfo{author}{Kiryluk, J.}, \bibinfo{author}{Klein, S.R.}, \bibinfo{author}{Kochocki, A.}, \bibinfo{author}{Koirala, R.}, \bibinfo{author}{Kolanoski, H.}, \bibinfo{author}{Kontrimas, T.}, \bibinfo{author}{Köpke, L.}, \bibinfo{author}{Kopper, C.}, \bibinfo{author}{Kopper, S.}, \bibinfo{author}{Koskinen, D.J.}, \bibinfo{author}{Koundal, P.}, \bibinfo{author}{Kovacevich, M.}, \bibinfo{author}{Kowalski, M.}, \bibinfo{author}{Kozynets, T.}, \bibinfo{author}{Krupczak, E.}, \bibinfo{author}{Kun, E.}, \bibinfo{author}{Kurahashi, N.}, \bibinfo{author}{Lad, N.}, \bibinfo{author}{Gualda, C.L.}, \bibinfo{author}{Lanfranchi, J.L.}, \bibinfo{author}{Larson,
  M.J.}, \bibinfo{author}{Lauber, F.}, \bibinfo{author}{Lazar, J.P.}, \bibinfo{author}{Lee, J.W.}, \bibinfo{author}{Leonard, K.}, \bibinfo{author}{Leszczyńska, A.}, \bibinfo{author}{Li, Y.}, \bibinfo{author}{Lincetto, M.}, \bibinfo{author}{Liu, Q.R.}, \bibinfo{author}{Liubarska, M.}, \bibinfo{author}{Lohfink, E.}, \bibinfo{author}{Mariscal, C.J.L.}, \bibinfo{author}{Lu, L.}, \bibinfo{author}{Lucarelli, F.}, \bibinfo{author}{Ludwig, A.}, \bibinfo{author}{Luszczak, W.}, \bibinfo{author}{Lyu, Y.}, \bibinfo{author}{Ma, W.Y.}, \bibinfo{author}{Madsen, J.}, \bibinfo{author}{Mahn, K.B.M.}, \bibinfo{author}{Makino, Y.}, \bibinfo{author}{Mancina, S.}, \bibinfo{author}{Mariş, I.C.}, \bibinfo{author}{Martinez-Soler, I.}, \bibinfo{author}{Maruyama, R.}, \bibinfo{author}{McCarthy, S.}, \bibinfo{author}{McElroy, T.}, \bibinfo{author}{McNally, F.}, \bibinfo{author}{Mead, J.V.}, \bibinfo{author}{Meagher, K.}, \bibinfo{author}{Mechbal, S.}, \bibinfo{author}{Medina, A.}, \bibinfo{author}{Meier, M.},
  \bibinfo{author}{Meighen-Berger, S.}, \bibinfo{author}{Merckx, Y.}, \bibinfo{author}{Micallef, J.}, \bibinfo{author}{Mockler, D.}, \bibinfo{author}{Montaruli, T.}, \bibinfo{author}{Moore, R.W.}, \bibinfo{author}{Morik, K.}, \bibinfo{author}{Morse, R.}, \bibinfo{author}{Moulai, M.}, \bibinfo{author}{Mukherjee, T.}, \bibinfo{author}{Naab, R.}, \bibinfo{author}{Nagai, R.}, \bibinfo{author}{Nahnhauer, R.}, \bibinfo{author}{Naumann, U.}, \bibinfo{author}{Necker, J.}, \bibinfo{author}{Nguyen, L.V.}, \bibinfo{author}{Niederhausen, H.}, \bibinfo{author}{Nisa, M.U.}, \bibinfo{author}{Nowicki, S.C.}, \bibinfo{author}{Nygren, D.}, \bibinfo{author}{Pollmann, A.O.}, \bibinfo{author}{Oehler, M.}, \bibinfo{author}{Oeyen, B.}, \bibinfo{author}{Olivas, A.}, \bibinfo{author}{O’Sullivan, E.}, \bibinfo{author}{Pandya, H.}, \bibinfo{author}{Pankova, D.V.}, \bibinfo{author}{Park, N.}, \bibinfo{author}{Parker, G.K.}, \bibinfo{author}{Paudel, E.N.}, \bibinfo{author}{Paul, L.}, \bibinfo{author}{de~los Heros, C.P.},
  \bibinfo{author}{Peters, L.}, \bibinfo{author}{Peterson, J.}, \bibinfo{author}{Philippen, S.}, \bibinfo{author}{Pieper, S.}, \bibinfo{author}{Pizzuto, A.}, \bibinfo{author}{Plum, M.}, \bibinfo{author}{Popovych, Y.}, \bibinfo{author}{Porcelli, A.}, \bibinfo{author}{Rodriguez, M.P.}, \bibinfo{author}{Pries, B.}, \bibinfo{author}{Przybylski, G.T.}, \bibinfo{author}{Raab, C.}, \bibinfo{author}{Rack-Helleis, J.}, \bibinfo{author}{Raissi, A.}, \bibinfo{author}{Rameez, M.}, \bibinfo{author}{Rawlins, K.}, \bibinfo{author}{Rea, I.C.}, \bibinfo{author}{Rechav, Z.}, \bibinfo{author}{Rehman, A.}, \bibinfo{author}{Reichherzer, P.}, \bibinfo{author}{Reimann, R.}, \bibinfo{author}{Renzi, G.}, \bibinfo{author}{Resconi, E.}, \bibinfo{author}{Reusch, S.}, \bibinfo{author}{Rhode, W.}, \bibinfo{author}{Richman, M.}, \bibinfo{author}{Riedel, B.}, \bibinfo{author}{Roberts, E.J.}, \bibinfo{author}{Robertson, S.}, \bibinfo{author}{Roellinghoff, G.}, \bibinfo{author}{Rongen, M.}, \bibinfo{author}{Rott, C.}, \bibinfo{author}{Ruhe,
  T.}, \bibinfo{author}{Ryckbosch, D.}, \bibinfo{author}{Cantu, D.R.}, \bibinfo{author}{Safa, I.}, \bibinfo{author}{Saffer, J.}, \bibinfo{author}{Salazar-Gallegos, D.}, \bibinfo{author}{Sampathkumar, P.}, \bibinfo{author}{Herrera, S.E.S.}, \bibinfo{author}{Sandrock, A.}, \bibinfo{author}{Santander, M.}, \bibinfo{author}{Sarkar, S.}, \bibinfo{author}{Sarkar, S.}, \bibinfo{author}{Satalecka, K.}, \bibinfo{author}{Schaufel, M.}, \bibinfo{author}{Schieler, H.}, \bibinfo{author}{Schindler, S.}, \bibinfo{author}{Schmidt, T.}, \bibinfo{author}{Schneider, A.}, \bibinfo{author}{Schneider, J.}, \bibinfo{author}{Schröder, F.G.}, \bibinfo{author}{Schumacher, L.}, \bibinfo{author}{Schwefer, G.}, \bibinfo{author}{Sclafani, S.}, \bibinfo{author}{Seckel, D.}, \bibinfo{author}{Seunarine, S.}, \bibinfo{author}{Sharma, A.}, \bibinfo{author}{Shefali, S.}, \bibinfo{author}{Shimizu, N.}, \bibinfo{author}{Silva, M.}, \bibinfo{author}{Skrzypek, B.}, \bibinfo{author}{Smithers, B.}, \bibinfo{author}{Snihur, R.},
  \bibinfo{author}{Soedingrekso, J.}, \bibinfo{author}{Sogaard, A.}, \bibinfo{author}{Soldin, D.}, \bibinfo{author}{Spannfellner, C.}, \bibinfo{author}{Spiczak, G.M.}, \bibinfo{author}{Spiering, C.}, \bibinfo{author}{Stamatikos, M.}, \bibinfo{author}{Stanev, T.}, \bibinfo{author}{Stein, R.}, \bibinfo{author}{Stettner, J.}, \bibinfo{author}{Stezelberger, T.}, \bibinfo{author}{Stokstad, B.}, \bibinfo{author}{Stürwald, T.}, \bibinfo{author}{Stuttard, T.}, \bibinfo{author}{Sullivan, G.W.}, \bibinfo{author}{Taboada, I.}, \bibinfo{author}{Ter-Antonyan, S.}, \bibinfo{author}{Thwaites, J.}, \bibinfo{author}{Tilav, S.}, \bibinfo{author}{Tischbein, F.}, \bibinfo{author}{Tollefson, K.}, \bibinfo{author}{Tönnis, C.}, \bibinfo{author}{Toscano, S.}, \bibinfo{author}{Tosi, D.}, \bibinfo{author}{Trettin, A.}, \bibinfo{author}{Tselengidou, M.}, \bibinfo{author}{Tung, C.F.}, \bibinfo{author}{Turcati, A.}, \bibinfo{author}{Turcotte, R.}, \bibinfo{author}{Turley, C.F.}, \bibinfo{author}{Twagirayezu, J.P.}, \bibinfo{author}{Ty,
  B.}, \bibinfo{author}{Elorrieta, M.A.U.}, \bibinfo{author}{Valtonen-Mattila, N.}, \bibinfo{author}{Vandenbroucke, J.}, \bibinfo{author}{van Eijndhoven, N.}, \bibinfo{author}{Vannerom, D.}, \bibinfo{author}{van Santen, J.}, \bibinfo{author}{Veitch-Michaelis, J.}, \bibinfo{author}{Verpoest, S.}, \bibinfo{author}{Walck, C.}, \bibinfo{author}{Wang, W.}, \bibinfo{author}{Watson, T.B.}, \bibinfo{author}{Weaver, C.}, \bibinfo{author}{Weigel, P.}, \bibinfo{author}{Weindl, A.}, \bibinfo{author}{Weiss, M.J.}, \bibinfo{author}{Weldert, J.}, \bibinfo{author}{Wendt, C.}, \bibinfo{author}{Werthebach, J.}, \bibinfo{author}{Weyrauch, M.}, \bibinfo{author}{Whitehorn, N.}, \bibinfo{author}{Wiebusch, C.H.}, \bibinfo{author}{Willey, N.}, \bibinfo{author}{Williams, D.R.}, \bibinfo{author}{Wolf, M.}, \bibinfo{author}{Wrede, G.}, \bibinfo{author}{Wulff, J.}, \bibinfo{author}{Xu, X.W.}, \bibinfo{author}{Yanez, J.P.}, \bibinfo{author}{Yildizci, E.}, \bibinfo{author}{Yoshida, S.}, \bibinfo{author}{Yu, S.}, \bibinfo{author}{Yuan,
  T.}, \bibinfo{author}{Zhang, Z.}, \bibinfo{author}{Zhelnin, P.}, \bibinfo{year}{2023}.
\newblock \bibinfo{title}{Observation of high-energy neutrinos from the galactic plane}.
\newblock \bibinfo{journal}{Science} \bibinfo{volume}{380}, \bibinfo{pages}{1338–1343}.
\newblock \URLprefix \url{http://dx.doi.org/10.1126/science.adc9818}, \DOIprefix\doi{10.1126/science.adc9818}.
\bibitem[{Ackermann et~al.(2022)Ackermann, Bustamante, Lu, Otte, Reno, Wissel, Ackermann, Agarwalla, Alvarez-Muñiz, {Alves Batista}, Argüelles, Bustamante, Clark, Cummings, Das, Decoene, Denton, Dornic, Dzhilkibaev, Farzan, Garcia, Garzelli, Glaser, Heijboer, Hörandel, Illuminati, {Seon Jeong}, Kelley, Kelly, Kheirandish, Klein, Krizmanic, Larson, Lu, Murase, Narang, Otte, Prechelt, Prohira, Reno, Resconi, Santander, Valera, Vandenbroucke, {Vasil'evna Suvorova}, Wiencke, Wissel, Yoshida, Yuan, Zas, Zhelnin, Zhou, Anchordoqui, Ashida, Bagheri, Balagopal, Basu, Beatty, Bechtol, Bell, Bishop, Book, Brown, Burgman, Campana, Chau, Chen, Coleman, Connolly, Conrad, Correa, Creque-Sarbinowski, Cummings, Curtis-Ginsberg, Dasgupta, {De Kockere}, {de Vries}, Deaconu, Desai, DeYoung, {di Matteo}, Elsaesser, Fürst, Fan, Fedynitch, Fox, Ganster, Minh, Haack, Hallman, Halzen, Haungs, Ishihara, Judd, Karg, Karle, Katori, Kochocki, Kopper, Kowalski, Kravchenko, Kurahashi, Lamoureux, {León Vargas}, Lincetto, Liu, Madsen,
  Makino, Mammo, Marka, Mayotte, Meagher, Meier, Minh, Miramonti, Moulai, Mulrey, Muzio, Naab, Nelles, Nichols, Nozdrina, O'Sullivan, Osborne, Pandey, Paudel, Pizzuto, Plum, {Pobes Aranda}, Pyras, Raab, Rechav, Rojo, {Romero Matamala}, Santander, Savina, Schroeder, Schumacher, Sciutto, Sclafani, {Ful Hossain Seikh}, Silva, Singh, Smith, Spencer, Springer, Stachurska, Suvorova, Taboada, Toscano, Tueros, Twagirayezu, {van Eijndhoven}, Veres, Vieregg, Wang, Whitehorn, Winter, Yildizci and Yu}]{snowmass}
\bibinfo{author}{Ackermann, M.}, \bibinfo{author}{Bustamante, M.}, \bibinfo{author}{Lu, L.}, \bibinfo{author}{Otte, N.}, \bibinfo{author}{Reno, M.H.}, \bibinfo{author}{Wissel, S.}, \bibinfo{author}{Ackermann, M.}, \bibinfo{author}{Agarwalla, S.K.}, \bibinfo{author}{Alvarez-Muñiz, J.}, \bibinfo{author}{{Alves Batista}, R.}, \bibinfo{author}{Argüelles, C.A.}, \bibinfo{author}{Bustamante, M.}, \bibinfo{author}{Clark, B.A.}, \bibinfo{author}{Cummings, A.}, \bibinfo{author}{Das, S.}, \bibinfo{author}{Decoene, V.}, \bibinfo{author}{Denton, P.B.}, \bibinfo{author}{Dornic, D.}, \bibinfo{author}{Dzhilkibaev, Z.A.}, \bibinfo{author}{Farzan, Y.}, \bibinfo{author}{Garcia, A.}, \bibinfo{author}{Garzelli, M.V.}, \bibinfo{author}{Glaser, C.}, \bibinfo{author}{Heijboer, A.}, \bibinfo{author}{Hörandel, J.R.}, \bibinfo{author}{Illuminati, G.}, \bibinfo{author}{{Seon Jeong}, Y.}, \bibinfo{author}{Kelley, J.L.}, \bibinfo{author}{Kelly, K.J.}, \bibinfo{author}{Kheirandish, A.}, \bibinfo{author}{Klein, S.R.},
  \bibinfo{author}{Krizmanic, J.F.}, \bibinfo{author}{Larson, M.J.}, \bibinfo{author}{Lu, L.}, \bibinfo{author}{Murase, K.}, \bibinfo{author}{Narang, A.}, \bibinfo{author}{Otte, N.}, \bibinfo{author}{Prechelt, R.L.}, \bibinfo{author}{Prohira, S.}, \bibinfo{author}{Reno, M.H.}, \bibinfo{author}{Resconi, E.}, \bibinfo{author}{Santander, M.}, \bibinfo{author}{Valera, V.B.}, \bibinfo{author}{Vandenbroucke, J.}, \bibinfo{author}{{Vasil'evna Suvorova}, O.}, \bibinfo{author}{Wiencke, L.}, \bibinfo{author}{Wissel, S.}, \bibinfo{author}{Yoshida, S.}, \bibinfo{author}{Yuan, T.}, \bibinfo{author}{Zas, E.}, \bibinfo{author}{Zhelnin, P.}, \bibinfo{author}{Zhou, B.}, \bibinfo{author}{Anchordoqui, L.A.}, \bibinfo{author}{Ashida, Y.}, \bibinfo{author}{Bagheri, M.}, \bibinfo{author}{Balagopal, A.}, \bibinfo{author}{Basu, V.}, \bibinfo{author}{Beatty, J.}, \bibinfo{author}{Bechtol, K.}, \bibinfo{author}{Bell, N.}, \bibinfo{author}{Bishop, A.}, \bibinfo{author}{Book, J.}, \bibinfo{author}{Brown, A.}, \bibinfo{author}{Burgman,
  A.}, \bibinfo{author}{Campana, M.}, \bibinfo{author}{Chau, N.}, \bibinfo{author}{Chen, T.Y.}, \bibinfo{author}{Coleman, A.}, \bibinfo{author}{Connolly, A.}, \bibinfo{author}{Conrad, J.M.}, \bibinfo{author}{Correa, P.}, \bibinfo{author}{Creque-Sarbinowski, C.}, \bibinfo{author}{Cummings, A.}, \bibinfo{author}{Curtis-Ginsberg, Z.}, \bibinfo{author}{Dasgupta, P.}, \bibinfo{author}{{De Kockere}, S.}, \bibinfo{author}{{de Vries}, K.}, \bibinfo{author}{Deaconu, C.}, \bibinfo{author}{Desai, A.}, \bibinfo{author}{DeYoung, T.}, \bibinfo{author}{{di Matteo}, A.}, \bibinfo{author}{Elsaesser, D.}, \bibinfo{author}{Fürst, P.}, \bibinfo{author}{Fan, K.L.}, \bibinfo{author}{Fedynitch, A.}, \bibinfo{author}{Fox, D.}, \bibinfo{author}{Ganster, E.}, \bibinfo{author}{Minh, M.H.}, \bibinfo{author}{Haack, C.}, \bibinfo{author}{Hallman, S.}, \bibinfo{author}{Halzen, F.}, \bibinfo{author}{Haungs, A.}, \bibinfo{author}{Ishihara, A.}, \bibinfo{author}{Judd, E.}, \bibinfo{author}{Karg, T.}, \bibinfo{author}{Karle, A.},
  \bibinfo{author}{Katori, T.}, \bibinfo{author}{Kochocki, A.}, \bibinfo{author}{Kopper, C.}, \bibinfo{author}{Kowalski, M.}, \bibinfo{author}{Kravchenko, I.}, \bibinfo{author}{Kurahashi, N.}, \bibinfo{author}{Lamoureux, M.}, \bibinfo{author}{{León Vargas}, H.}, \bibinfo{author}{Lincetto, M.}, \bibinfo{author}{Liu, Q.}, \bibinfo{author}{Madsen, J.}, \bibinfo{author}{Makino, Y.}, \bibinfo{author}{Mammo, J.}, \bibinfo{author}{Marka, Z.}, \bibinfo{author}{Mayotte, E.}, \bibinfo{author}{Meagher, K.}, \bibinfo{author}{Meier, M.}, \bibinfo{author}{Minh, M.H.}, \bibinfo{author}{Miramonti, L.}, \bibinfo{author}{Moulai, M.}, \bibinfo{author}{Mulrey, K.}, \bibinfo{author}{Muzio, M.}, \bibinfo{author}{Naab, R.}, \bibinfo{author}{Nelles, A.}, \bibinfo{author}{Nichols, W.}, \bibinfo{author}{Nozdrina, A.}, \bibinfo{author}{O'Sullivan, E.}, \bibinfo{author}{Osborne, V.O.J.}, \bibinfo{author}{Pandey, V.}, \bibinfo{author}{Paudel, E.N.}, \bibinfo{author}{Pizzuto, A.}, \bibinfo{author}{Plum, M.}, \bibinfo{author}{{Pobes
  Aranda}, C.}, \bibinfo{author}{Pyras, L.}, \bibinfo{author}{Raab, C.}, \bibinfo{author}{Rechav, Z.}, \bibinfo{author}{Rojo, J.}, \bibinfo{author}{{Romero Matamala}, O.}, \bibinfo{author}{Santander, M.}, \bibinfo{author}{Savina, P.}, \bibinfo{author}{Schroeder, F.}, \bibinfo{author}{Schumacher, L.}, \bibinfo{author}{Sciutto, S.}, \bibinfo{author}{Sclafani, S.}, \bibinfo{author}{{Ful Hossain Seikh}, M.}, \bibinfo{author}{Silva, M.}, \bibinfo{author}{Singh, R.}, \bibinfo{author}{Smith, D.}, \bibinfo{author}{Spencer, S.T.}, \bibinfo{author}{Springer, R.W.}, \bibinfo{author}{Stachurska, J.}, \bibinfo{author}{Suvorova, O.}, \bibinfo{author}{Taboada, I.}, \bibinfo{author}{Toscano, S.}, \bibinfo{author}{Tueros, M.}, \bibinfo{author}{Twagirayezu, J.P.}, \bibinfo{author}{{van Eijndhoven}, N.}, \bibinfo{author}{Veres, P.}, \bibinfo{author}{Vieregg, A.}, \bibinfo{author}{Wang, W.}, \bibinfo{author}{Whitehorn, N.}, \bibinfo{author}{Winter, W.}, \bibinfo{author}{Yildizci, E.}, \bibinfo{author}{Yu, S.},
  \bibinfo{year}{2022}.
\newblock \bibinfo{title}{High-energy and ultra-high-energy neutrinos: A snowmass white paper}.
\newblock \bibinfo{journal}{Journal of High Energy Astrophysics} \bibinfo{volume}{36}, \bibinfo{pages}{55--110}.
\newblock \URLprefix \url{https://www.sciencedirect.com/science/article/pii/S2214404822000416}, \DOIprefix\doi{https://doi.org/10.1016/j.jheap.2022.08.001}.
\bibitem[{Aguilar et~al.(2021)Aguilar, Allison, Beatty, Bernhoff, Besson, Bingefors, Botner, Buitink, Carter, Clark, Connolly, Dasgupta, de~Kockere, de~Vries, Deaconu, DuVernois, Feigl, García-Fernández, Glaser, Hallgren, Hallmann, Hanson, Hendricks, Hokanson-Fasig, Hornhuber, Hughes, Karle, Kelley, Klein, Krebs, Lahmann, Magnuson, Meures, Meyers, Nelles, Novikov, Oberla, Oeyen, Pandya, Plaisier, Pyras, Ryckbosch, Scholten, Seckel, Smith, Southall, Torres, Toscano, Van Den~Broeck, van Eijndhoven, Vieregg, Welling, Wissel, Young and Zink}]{RNO_Aguilar_2021}
\bibinfo{author}{Aguilar, J.}, \bibinfo{author}{Allison, P.}, \bibinfo{author}{Beatty, J.}, \bibinfo{author}{Bernhoff, H.}, \bibinfo{author}{Besson, D.}, \bibinfo{author}{Bingefors, N.}, \bibinfo{author}{Botner, O.}, \bibinfo{author}{Buitink, S.}, \bibinfo{author}{Carter, K.}, \bibinfo{author}{Clark, B.}, \bibinfo{author}{Connolly, A.}, \bibinfo{author}{Dasgupta, P.}, \bibinfo{author}{de~Kockere, S.}, \bibinfo{author}{de~Vries, K.}, \bibinfo{author}{Deaconu, C.}, \bibinfo{author}{DuVernois, M.}, \bibinfo{author}{Feigl, N.}, \bibinfo{author}{García-Fernández, D.}, \bibinfo{author}{Glaser, C.}, \bibinfo{author}{Hallgren, A.}, \bibinfo{author}{Hallmann, S.}, \bibinfo{author}{Hanson, J.}, \bibinfo{author}{Hendricks, B.}, \bibinfo{author}{Hokanson-Fasig, B.}, \bibinfo{author}{Hornhuber, C.}, \bibinfo{author}{Hughes, K.}, \bibinfo{author}{Karle, A.}, \bibinfo{author}{Kelley, J.}, \bibinfo{author}{Klein, S.}, \bibinfo{author}{Krebs, R.}, \bibinfo{author}{Lahmann, R.}, \bibinfo{author}{Magnuson, M.},
  \bibinfo{author}{Meures, T.}, \bibinfo{author}{Meyers, Z.}, \bibinfo{author}{Nelles, A.}, \bibinfo{author}{Novikov, A.}, \bibinfo{author}{Oberla, E.}, \bibinfo{author}{Oeyen, B.}, \bibinfo{author}{Pandya, H.}, \bibinfo{author}{Plaisier, I.}, \bibinfo{author}{Pyras, L.}, \bibinfo{author}{Ryckbosch, D.}, \bibinfo{author}{Scholten, O.}, \bibinfo{author}{Seckel, D.}, \bibinfo{author}{Smith, D.}, \bibinfo{author}{Southall, D.}, \bibinfo{author}{Torres, J.}, \bibinfo{author}{Toscano, S.}, \bibinfo{author}{Van Den~Broeck, D.}, \bibinfo{author}{van Eijndhoven, N.}, \bibinfo{author}{Vieregg, A.}, \bibinfo{author}{Welling, C.}, \bibinfo{author}{Wissel, S.}, \bibinfo{author}{Young, R.}, \bibinfo{author}{Zink, A.}, \bibinfo{year}{2021}.
\newblock \bibinfo{title}{Design and sensitivity of the radio neutrino observatory in greenland (rno-g)}.
\newblock \bibinfo{journal}{Journal of Instrumentation} \bibinfo{volume}{16}, \bibinfo{pages}{P03025}.
\newblock \URLprefix \url{http://dx.doi.org/10.1088/1748-0221/16/03/P03025}, \DOIprefix\doi{10.1088/1748-0221/16/03/p03025}.
\bibitem[{Aiello et~al.(2025)Aiello, Albert, Alhebsi, Alshamsi, Garre, Ambrosone, Ameli, Andre, Anghinolfi, Aphecetche, Ardid, Ardid, Argüelles, Atmani, Aublin, Badaracco, Bailly-Salins, Bardačová, Baret, Bariego-Quintana, Becherini, Bendahman, Gualandi, Benhassi, Bennani, Benoit, Berbee, Bertin, Biagi, Boettcher, Bonanno, Bouasla, Boumaaza, Bouta, Bouwhuis, Bozza, Bozza, Brânzaş, Bretaudeau, Breuhaus, Bruijn, Brunner, Bruno, Buis, Buompane, Buson, Busto, Caiffi, Calvo, Capone, Carenini, Carretero, Cartraud, Castaldi, Cecchini, Celli, Cerisy, Chabab, Chen, Cherubini, Chiarusi, Circella, Cocimano, Coelho, Coleiro, Colonges, Condorelli, Coniglione, Coyle, Creusot, Cuttone, D’Amico, Dallier, Benedittis, Martino, Wasseige, Decoene, Rosso, Mauro, Palma, Diaz, Diego-Tortosa, Distefano, Domi, Donzaud, Dornic, Drakopoulou, Drouhin, Ducoin, Dvornický, Eberl, Eckerová, Eddymaoui, van Eeden, Eff, van Eijk, Bojaddaini, Hedri, Ellajosyula, Enzenhöfer, Ferrara, Filipovićv, Filippini, Franciotti, Fusco,
  Gagliardini, Gal, Méndez, Soto, Oliver, Geißelbrecht, Genton, Ghaddari, Gialanella, Gibson, Giorgio, Goos, Goswami, Gozzini, Gracia, Graf, Guidi, Guillon, Gutiérrez, Haack, van Haren, Heijboer, Hennig, Henry, Hernández-Rey, Ibnsalih, Ilioni, Illuminati, Joly, de~Jong, de~Jong, Jung, Kalaczyński, Kalekin, Kamp, Katz, Kistauri, Kopper, Kouchner, Kovalev, Kueviakoe, Kulikovskiy, Kvatadze, Labalme, Lahmann, Lamoureux, Lancelin, Larosa, Lastoria, Lazar, Lazo, Stum, Lehaut, Lemaitre, Leonora, Lessing, Levi, Lincetto, Clark, Longhitano, Lumb, Magnani, Majumdar, Malerba, Mamedov, Manfreda, Marconi, Margiotta, Marinelli, Markou, Martin, Marzaioli, Mastrodicasa, Mastroianni, Mauro, Miele, Migliozzi, Migneco, Mitsou, Mollo, Mongelli, Morales-Gallegos, Moussa, Mateo, Muller, Musone, Musumeci, Navas, Nayerhoda, Nicolau, Nkosi, Fearraigh, Oliviero, Orlando, Oukacha, Paesani, González, Papalashvili, Paries, Parisi, Gomez, Pastore, Păun, Păvălaş, Martínez, Perrin-Terrin, Pestel, Pestes, Pfeiffer, Piattelli,
  Plavin, Poirè, Popa, Pradier, Prado, Pulvirenti, Quiroz-Rangel, Randazzo, Razzaque, Rea, Real, Riccobene, Robinson, Romanov, Ros, Šaina, Greus, Samtleben, Losa, Sanfilippo, Sanguineti, Santonocito, Sapienza, Schmelling, Schnabel, Schumann, Schutte, Seneca, Sennan, Sevle, Sgura, Shanidze, Sharma, Shitov, Šimkovic, Simonelli, Sinopoulou, Spisso, Spurio, Stavropoulos, Štekl, Taiuti, Tayalati, Thiersen, Thoudam, e~Melo, Trocmé, Tsourapis, Tudorache, Tzamariudaki, Ukleja, Vacheret, Valsecchi, Elewyck, Vannoye, Vasileiadis, de~Sola, Verilhac, Veutro, Viola, Vivolo, van Vliet, Wen, de~Wolf, Lhenry-Yvon, Zavatarelli, Zegarelli, Zito, Zornoza, Zúñiga and Zywucka}]{Aiello2025}
\bibinfo{author}{Aiello, S.}, \bibinfo{author}{Albert, A.}, \bibinfo{author}{Alhebsi, A.R.}, \bibinfo{author}{Alshamsi, M.}, \bibinfo{author}{Garre, S.A.}, \bibinfo{author}{Ambrosone, A.}, \bibinfo{author}{Ameli, F.}, \bibinfo{author}{Andre, M.}, \bibinfo{author}{Anghinolfi, M.}, \bibinfo{author}{Aphecetche, L.}, \bibinfo{author}{Ardid, M.}, \bibinfo{author}{Ardid, S.}, \bibinfo{author}{Argüelles, C.}, \bibinfo{author}{Atmani, H.}, \bibinfo{author}{Aublin, J.}, \bibinfo{author}{Badaracco, F.}, \bibinfo{author}{Bailly-Salins, L.}, \bibinfo{author}{Bardačová, Z.}, \bibinfo{author}{Baret, B.}, \bibinfo{author}{Bariego-Quintana, A.}, \bibinfo{author}{Becherini, Y.}, \bibinfo{author}{Bendahman, M.}, \bibinfo{author}{Gualandi, F.B.}, \bibinfo{author}{Benhassi, M.}, \bibinfo{author}{Bennani, M.}, \bibinfo{author}{Benoit, D.M.}, \bibinfo{author}{Berbee, E.}, \bibinfo{author}{Bertin, V.}, \bibinfo{author}{Biagi, S.}, \bibinfo{author}{Boettcher, M.}, \bibinfo{author}{Bonanno, D.}, \bibinfo{author}{Bouasla, A.B.},
  \bibinfo{author}{Boumaaza, J.}, \bibinfo{author}{Bouta, M.}, \bibinfo{author}{Bouwhuis, M.}, \bibinfo{author}{Bozza, C.}, \bibinfo{author}{Bozza, R.M.}, \bibinfo{author}{Brânzaş, H.}, \bibinfo{author}{Bretaudeau, F.}, \bibinfo{author}{Breuhaus, M.}, \bibinfo{author}{Bruijn, R.}, \bibinfo{author}{Brunner, J.}, \bibinfo{author}{Bruno, R.}, \bibinfo{author}{Buis, E.}, \bibinfo{author}{Buompane, R.}, \bibinfo{author}{Buson, S.}, \bibinfo{author}{Busto, J.}, \bibinfo{author}{Caiffi, B.}, \bibinfo{author}{Calvo, D.}, \bibinfo{author}{Capone, A.}, \bibinfo{author}{Carenini, F.}, \bibinfo{author}{Carretero, V.}, \bibinfo{author}{Cartraud, T.}, \bibinfo{author}{Castaldi, P.}, \bibinfo{author}{Cecchini, V.}, \bibinfo{author}{Celli, S.}, \bibinfo{author}{Cerisy, L.}, \bibinfo{author}{Chabab, M.}, \bibinfo{author}{Chen, A.}, \bibinfo{author}{Cherubini, S.}, \bibinfo{author}{Chiarusi, T.}, \bibinfo{author}{Circella, M.}, \bibinfo{author}{Cocimano, R.}, \bibinfo{author}{Coelho, J.A.B.}, \bibinfo{author}{Coleiro, A.},
  \bibinfo{author}{Colonges, S.}, \bibinfo{author}{Condorelli, A.}, \bibinfo{author}{Coniglione, R.}, \bibinfo{author}{Coyle, P.}, \bibinfo{author}{Creusot, A.}, \bibinfo{author}{Cuttone, G.}, \bibinfo{author}{D’Amico, A.}, \bibinfo{author}{Dallier, R.}, \bibinfo{author}{Benedittis, A.D.}, \bibinfo{author}{Martino, B.D.}, \bibinfo{author}{Wasseige, G.D.}, \bibinfo{author}{Decoene, V.}, \bibinfo{author}{Rosso, I.D.}, \bibinfo{author}{Mauro, L.S.D.}, \bibinfo{author}{Palma, I.D.}, \bibinfo{author}{Diaz, A.F.}, \bibinfo{author}{Diego-Tortosa, D.}, \bibinfo{author}{Distefano, C.}, \bibinfo{author}{Domi, A.}, \bibinfo{author}{Donzaud, C.}, \bibinfo{author}{Dornic, D.}, \bibinfo{author}{Drakopoulou, E.}, \bibinfo{author}{Drouhin, D.}, \bibinfo{author}{Ducoin, J.G.}, \bibinfo{author}{Dvornický, R.}, \bibinfo{author}{Eberl, T.}, \bibinfo{author}{Eckerová, E.}, \bibinfo{author}{Eddymaoui, A.}, \bibinfo{author}{van Eeden, T.}, \bibinfo{author}{Eff, M.}, \bibinfo{author}{van Eijk, D.}, \bibinfo{author}{Bojaddaini,
  I.E.}, \bibinfo{author}{Hedri, S.E.}, \bibinfo{author}{Ellajosyula, V.}, \bibinfo{author}{Enzenhöfer, A.}, \bibinfo{author}{Ferrara, G.}, \bibinfo{author}{Filipovićv, M.D.}, \bibinfo{author}{Filippini, F.}, \bibinfo{author}{Franciotti, D.}, \bibinfo{author}{Fusco, L.A.}, \bibinfo{author}{Gagliardini, S.}, \bibinfo{author}{Gal, T.}, \bibinfo{author}{Méndez, J.G.}, \bibinfo{author}{Soto, A.G.}, \bibinfo{author}{Oliver, C.G.}, \bibinfo{author}{Geißelbrecht, N.}, \bibinfo{author}{Genton, E.}, \bibinfo{author}{Ghaddari, H.}, \bibinfo{author}{Gialanella, L.}, \bibinfo{author}{Gibson, B.K.}, \bibinfo{author}{Giorgio, E.}, \bibinfo{author}{Goos, I.}, \bibinfo{author}{Goswami, P.}, \bibinfo{author}{Gozzini, S.R.}, \bibinfo{author}{Gracia, R.}, \bibinfo{author}{Graf, K.}, \bibinfo{author}{Guidi, C.}, \bibinfo{author}{Guillon, B.}, \bibinfo{author}{Gutiérrez, M.}, \bibinfo{author}{Haack, C.}, \bibinfo{author}{van Haren, H.}, \bibinfo{author}{Heijboer, A.}, \bibinfo{author}{Hennig, L.}, \bibinfo{author}{Henry,
  S.}, \bibinfo{author}{Hernández-Rey, J.J.}, \bibinfo{author}{Ibnsalih, W.I.}, \bibinfo{author}{Ilioni, A.}, \bibinfo{author}{Illuminati, G.}, \bibinfo{author}{Joly, D.}, \bibinfo{author}{de~Jong, M.}, \bibinfo{author}{de~Jong, P.}, \bibinfo{author}{Jung, B.J.}, \bibinfo{author}{Kalaczyński, P.}, \bibinfo{author}{Kalekin, O.}, \bibinfo{author}{Kamp, N.}, \bibinfo{author}{Katz, U.F.}, \bibinfo{author}{Kistauri, G.}, \bibinfo{author}{Kopper, C.}, \bibinfo{author}{Kouchner, A.}, \bibinfo{author}{Kovalev, Y.Y.}, \bibinfo{author}{Kueviakoe, V.}, \bibinfo{author}{Kulikovskiy, V.}, \bibinfo{author}{Kvatadze, R.}, \bibinfo{author}{Labalme, M.}, \bibinfo{author}{Lahmann, R.}, \bibinfo{author}{Lamoureux, M.}, \bibinfo{author}{Lancelin, S.}, \bibinfo{author}{Larosa, G.}, \bibinfo{author}{Lastoria, C.}, \bibinfo{author}{Lazar, J.}, \bibinfo{author}{Lazo, A.}, \bibinfo{author}{Stum, S.L.}, \bibinfo{author}{Lehaut, G.}, \bibinfo{author}{Lemaitre, V.}, \bibinfo{author}{Leonora, E.}, \bibinfo{author}{Lessing, N.},
  \bibinfo{author}{Levi, G.}, \bibinfo{author}{Lincetto, M.}, \bibinfo{author}{Clark, M.L.}, \bibinfo{author}{Longhitano, F.}, \bibinfo{author}{Lumb, N.}, \bibinfo{author}{Magnani, F.}, \bibinfo{author}{Majumdar, J.}, \bibinfo{author}{Malerba, L.}, \bibinfo{author}{Mamedov, F.}, \bibinfo{author}{Manfreda, A.}, \bibinfo{author}{Marconi, M.}, \bibinfo{author}{Margiotta, A.}, \bibinfo{author}{Marinelli, A.}, \bibinfo{author}{Markou, C.}, \bibinfo{author}{Martin, L.}, \bibinfo{author}{Marzaioli, F.}, \bibinfo{author}{Mastrodicasa, M.}, \bibinfo{author}{Mastroianni, S.}, \bibinfo{author}{Mauro, J.}, \bibinfo{author}{Miele, G.}, \bibinfo{author}{Migliozzi, P.}, \bibinfo{author}{Migneco, E.}, \bibinfo{author}{Mitsou, M.L.}, \bibinfo{author}{Mollo, C.M.}, \bibinfo{author}{Mongelli, M.}, \bibinfo{author}{Morales-Gallegos, L.}, \bibinfo{author}{Moussa, A.}, \bibinfo{author}{Mateo, I.M.}, \bibinfo{author}{Muller, R.}, \bibinfo{author}{Musone, M.R.}, \bibinfo{author}{Musumeci, M.}, \bibinfo{author}{Navas, S.},
  \bibinfo{author}{Nayerhoda, A.}, \bibinfo{author}{Nicolau, C.A.}, \bibinfo{author}{Nkosi, B.}, \bibinfo{author}{Fearraigh, B.O.}, \bibinfo{author}{Oliviero, V.}, \bibinfo{author}{Orlando, A.}, \bibinfo{author}{Oukacha, E.}, \bibinfo{author}{Paesani, D.}, \bibinfo{author}{González, J.P.}, \bibinfo{author}{Papalashvili, G.}, \bibinfo{author}{Paries, C.}, \bibinfo{author}{Parisi, V.}, \bibinfo{author}{Gomez, E.J.P.}, \bibinfo{author}{Pastore, C.}, \bibinfo{author}{Păun, A.M.}, \bibinfo{author}{Păvălaş, G.E.}, \bibinfo{author}{Martínez, S.P.}, \bibinfo{author}{Perrin-Terrin, M.}, \bibinfo{author}{Pestel, V.}, \bibinfo{author}{Pestes, R.}, \bibinfo{author}{Pfeiffer, L.}, \bibinfo{author}{Piattelli, P.}, \bibinfo{author}{Plavin, A.}, \bibinfo{author}{Poirè, C.}, \bibinfo{author}{Popa, V.}, \bibinfo{author}{Pradier, T.}, \bibinfo{author}{Prado, J.}, \bibinfo{author}{Pulvirenti, S.}, \bibinfo{author}{Quiroz-Rangel, C.A.}, \bibinfo{author}{Randazzo, N.}, \bibinfo{author}{Razzaque, S.}, \bibinfo{author}{Rea,
  I.C.}, \bibinfo{author}{Real, D.}, \bibinfo{author}{Riccobene, G.}, \bibinfo{author}{Robinson, J.}, \bibinfo{author}{Romanov, A.}, \bibinfo{author}{Ros, E.}, \bibinfo{author}{Šaina, A.}, \bibinfo{author}{Greus, F.S.}, \bibinfo{author}{Samtleben, D.F.E.}, \bibinfo{author}{Losa, A.S.}, \bibinfo{author}{Sanfilippo, S.}, \bibinfo{author}{Sanguineti, M.}, \bibinfo{author}{Santonocito, D.}, \bibinfo{author}{Sapienza, P.}, \bibinfo{author}{Schmelling, J.}, \bibinfo{author}{Schnabel, J.}, \bibinfo{author}{Schumann, J.}, \bibinfo{author}{Schutte, H.M.}, \bibinfo{author}{Seneca, J.}, \bibinfo{author}{Sennan, N.}, \bibinfo{author}{Sevle, P.}, \bibinfo{author}{Sgura, I.}, \bibinfo{author}{Shanidze, R.}, \bibinfo{author}{Sharma, A.}, \bibinfo{author}{Shitov, Y.}, \bibinfo{author}{Šimkovic, F.}, \bibinfo{author}{Simonelli, A.}, \bibinfo{author}{Sinopoulou, A.}, \bibinfo{author}{Spisso, B.}, \bibinfo{author}{Spurio, M.}, \bibinfo{author}{Stavropoulos, D.}, \bibinfo{author}{Štekl, I.}, \bibinfo{author}{Taiuti, M.},
  \bibinfo{author}{Tayalati, Y.}, \bibinfo{author}{Thiersen, H.}, \bibinfo{author}{Thoudam, S.}, \bibinfo{author}{e~Melo, I.T.}, \bibinfo{author}{Trocmé, B.}, \bibinfo{author}{Tsourapis, V.}, \bibinfo{author}{Tudorache, A.}, \bibinfo{author}{Tzamariudaki, E.}, \bibinfo{author}{Ukleja, A.}, \bibinfo{author}{Vacheret, A.}, \bibinfo{author}{Valsecchi, V.}, \bibinfo{author}{Elewyck, V.V.}, \bibinfo{author}{Vannoye, G.}, \bibinfo{author}{Vasileiadis, G.}, \bibinfo{author}{de~Sola, F.V.}, \bibinfo{author}{Verilhac, C.}, \bibinfo{author}{Veutro, A.}, \bibinfo{author}{Viola, S.}, \bibinfo{author}{Vivolo, D.}, \bibinfo{author}{van Vliet, A.}, \bibinfo{author}{Wen, A.Y.}, \bibinfo{author}{de~Wolf, E.}, \bibinfo{author}{Lhenry-Yvon, I.}, \bibinfo{author}{Zavatarelli, S.}, \bibinfo{author}{Zegarelli, A.}, \bibinfo{author}{Zito, D.}, \bibinfo{author}{Zornoza, J.D.}, \bibinfo{author}{Zúñiga, J.}, \bibinfo{author}{Zywucka, N.}, \bibinfo{year}{2025}.
\newblock \bibinfo{title}{Observation of an ultra-high-energy cosmic neutrino with km3net}.
\newblock \bibinfo{journal}{Nature 2025 638:8050} \bibinfo{volume}{638}, \bibinfo{pages}{376--382}.
\newblock \URLprefix \url{https://www.nature.com/articles/s41586-024-08543-1}, \DOIprefix\doi{10.1038/s41586-024-08543-1}.
\bibitem[{Bagheri et~al.(2025)Bagheri, Gadamsetty, Gazda, Judd, Kuznetsov, Otte, Potts, {Romero Matamala}, Shapera, Sorell, Tandon and Wang}]{BAGHERI2025169999}
\bibinfo{author}{Bagheri, M.}, \bibinfo{author}{Gadamsetty, S.}, \bibinfo{author}{Gazda, E.}, \bibinfo{author}{Judd, E.}, \bibinfo{author}{Kuznetsov, E.}, \bibinfo{author}{Otte, A.N.}, \bibinfo{author}{Potts, M.}, \bibinfo{author}{{Romero Matamala}, O.}, \bibinfo{author}{Shapera, N.}, \bibinfo{author}{Sorell, J.}, \bibinfo{author}{Tandon, S.}, \bibinfo{author}{Wang, A.}, \bibinfo{year}{2025}.
\newblock \bibinfo{title}{The camera and readout for the trinity demonstrator and the euso-spb2 cherenkov telescope}.
\newblock \bibinfo{journal}{Nuclear Instruments and Methods in Physics Research Section A: Accelerators, Spectrometers, Detectors and Associated Equipment} \bibinfo{volume}{1070}, \bibinfo{pages}{169999}.
\newblock \URLprefix \url{https://www.sciencedirect.com/science/article/pii/S0168900224009252}, \DOIprefix\doi{https://doi.org/10.1016/j.nima.2024.169999}.
\bibitem[{{Brown} et~al.(2015){Brown}, {Armstrong}, {Chadwick}, {Daniel} and {White}}]{Brown2015ICRC}
\bibinfo{author}{{Brown}, A.}, \bibinfo{author}{{Armstrong}, T.}, \bibinfo{author}{{Chadwick}, P.}, \bibinfo{author}{{Daniel}, M.}, \bibinfo{author}{{White}, R.}, \bibinfo{year}{2015}.
\newblock \bibinfo{title}{{Flasher and muon-based calibration of the GCT telescopes proposed for the Cherenkov Telescope Array}}, in: \bibinfo{booktitle}{34th International Cosmic Ray Conference (ICRC2015)}, p. \bibinfo{pages}{934}.
\newblock \DOIprefix\doi{10.22323/1.236.0934}, \href{http://arxiv.org/abs/1509.00185}{{\tt arXiv:1509.00185}}.
\bibitem[{{Brown}(2018)}]{Brown2018}
\bibinfo{author}{{Brown}, A.M.}, \bibinfo{year}{2018}.
\newblock \bibinfo{title}{{On the prospects of cross-calibrating the Cherenkov Telescope Array with an airborne calibration platform}}.
\newblock \bibinfo{journal}{Astroparticle Physics} \bibinfo{volume}{97}, \bibinfo{pages}{69--79}.
\newblock \DOIprefix\doi{10.1016/j.astropartphys.2017.10.013}, \href{http://arxiv.org/abs/1711.01413}{{\tt arXiv:1711.01413}}.
\bibitem[{Brown et~al.(2021)Brown, Bagheri, Doro, Gazda, Kieda, Lin, Onel, Otte, Taboada and Wang}]{brown2021trinityimagingaircherenkov}
\bibinfo{author}{Brown, A.M.}, \bibinfo{author}{Bagheri, M.}, \bibinfo{author}{Doro, M.}, \bibinfo{author}{Gazda, E.}, \bibinfo{author}{Kieda, D.}, \bibinfo{author}{Lin, C.}, \bibinfo{author}{Onel, Y.}, \bibinfo{author}{Otte, N.}, \bibinfo{author}{Taboada, I.}, \bibinfo{author}{Wang, A.}, \bibinfo{year}{2021}.
\newblock \bibinfo{title}{Trinity: An imaging air cherenkov telescope to search for ultra-high-energy neutrinos}.
\newblock \URLprefix \url{https://arxiv.org/abs/2109.03125}, \href{http://arxiv.org/abs/2109.03125}{{\tt arXiv:2109.03125}}.
\bibitem[{{Brown} et~al.(2022){Brown}, {Muller}, {de Naurois} and {Clark}}]{Brown2022}
\bibinfo{author}{{Brown}, A.M.}, \bibinfo{author}{{Muller}, J.}, \bibinfo{author}{{de Naurois}, M.}, \bibinfo{author}{{Clark}, P.}, \bibinfo{year}{2022}.
\newblock \bibinfo{title}{{Inter-calibration of atmospheric Cherenkov telescopes with UAV-based airborne calibration system}}.
\newblock \bibinfo{journal}{Astroparticle Physics} \bibinfo{volume}{140}, \bibinfo{pages}{102695}.
\newblock \DOIprefix\doi{10.1016/j.astropartphys.2022.102695}, \href{http://arxiv.org/abs/2203.05839}{{\tt arXiv:2203.05839}}.
\bibitem[{{C. Pollacco et al.}(2018)}]{Pollacco2018}
\bibinfo{author}{{C. Pollacco et al.}, E.}, \bibinfo{year}{2018}.
\newblock \bibinfo{title}{Get: A generic electronics system for tpcs and nuclear physics instrumentation}.
\newblock \bibinfo{journal}{Nuclear Instruments and Methods in Physics Research Section A: Accelerators, Spectrometers, Detectors and Associated Equipment} \bibinfo{volume}{887}, \bibinfo{pages}{81--93}.
\newblock \DOIprefix\doi{10.1016/j.nima.2018.01.020}.
\bibitem[{Davies and Cotton(1957)}]{Davies1957}
\bibinfo{author}{Davies, J.M.}, \bibinfo{author}{Cotton, E.S.}, \bibinfo{year}{1957}.
\newblock \bibinfo{title}{Design of the quartermaster solar furnace}.
\newblock \bibinfo{journal}{Solar Energy} \bibinfo{volume}{1}, \bibinfo{pages}{16--22}.
\newblock \DOIprefix\doi{10.1016/0038-092X(57)90116-0}.
\bibitem[{Fargion(2001)}]{Fargion2001}
\bibinfo{author}{Fargion, D.}, \bibinfo{year}{2001}.
\newblock \bibinfo{title}{Upward and horizontal $\tau$ airshowers by uhe neutrinos}, \bibinfo{publisher}{JHEP}. pp. \bibinfo{pages}{hep--ph/0111289}.
\bibitem[{Gómez et~al.(2016)Gómez, Gascón, Fernández, Sanuy, Mauricio, Graciani and Sanchez}]{Gomez2016}
\bibinfo{author}{Gómez, S.}, \bibinfo{author}{Gascón, D.}, \bibinfo{author}{Fernández, G.}, \bibinfo{author}{Sanuy, A.}, \bibinfo{author}{Mauricio, J.}, \bibinfo{author}{Graciani, R.}, \bibinfo{author}{Sanchez, D.}, \bibinfo{year}{2016}.
\newblock \bibinfo{title}{Music: An 8 channel readout asic for sipm arrays}, \bibinfo{publisher}{International Society for Optics and Photonics}. p. \bibinfo{pages}{98990G}.
\newblock \DOIprefix\doi{10.1117/12.2231095}.
\bibitem[{Hillas(1985)}]{Hillas1985}
\bibinfo{author}{Hillas, A.M.}, \bibinfo{year}{1985}.
\newblock \bibinfo{title}{Cerenkov light images of eas produced by primary gamma rays}.
\newblock \bibinfo{journal}{Proceedings of the 19th International Cosmic Ray Conference} \bibinfo{volume}{3}, \bibinfo{pages}{445}.
\bibitem[{{IceCube Collaboration} et~al.(2018){IceCube Collaboration}, Aartsen, Ackermann, Adams, Aguilar, Ahlers, Ahrens, Samarai, Altmann, Andeen, Anderson, Ansseau, Anton, Argüelles, Arsioli, Auffenberg, Axani, Bagherpour, Bai, Barron, Barwick, Baum, Bay, Beatty, Becker, Tjus, BenZvi, Berley, Bernardini, Besson, Binder, Bindig, Blaufuss, Blot, Bohm, Boerner, Bos, Boeser, Botner, Bourbeau, Bourbeau, Bradascio, Braun, Brenzke, Bretz, Bron, Brostean-Kaiser, Burgman, Busse, Carver, Cheung, Chirkin, Christov, Clark, Classen, Coenders, Collin, Conrad, Coppin, Correa, Cowen, Cross, Dave, Day, de~André, Clercq, Delaunay, Dembinski, DeRidder, Desiati, de~Vries, DeWasseige, DeWith, DeYoung, Díaz-Vélez, Lorenzo, Dujmovic, Dumm, Dunkman, Dvorak, Eberhardt, Ehrhardt, Eichmann, Eller, Evenson, Fahey, Fazely, Felde, Filimonov, Finley, Flis, Franckowiak, Friedman, Fritz, Gaisser, Gallagher, Gerhardt, Ghorbani, Giommi, Glauch, Gluesenkamp, Goldschmidt, Gonzalez, Grant, Griffith, Haack, Hallgren, Halzen, Hanson,
  Hebecker, Heereman, Helbing, Hellauer, Hickford, Hignight, Hill, Hoffman, Hoffmann, Hoinka, Hokanson-Fasig, Hoshina, Huang, Huber, Hultqvist, Huennefeld, Hussain, In, Iovine, Ishihara, Jacobi, Japaridze, Jeong, Jero, Jones, Kalaczynski, Kang, Kappes, Kappesser, Karg, Karle, Katz, Kauer, Keivani, Kelley, Kheirandish, Kim, Kim, Kintscher, Kiryluk, Kittler, Klein, Koirala, Kolanoski, Koepke, Kopper, Kopper, Koschinsky, Koskinen, Kowalski, Krammer, Krings, Kroll, Krueckl, Kunwar, Neilson, Kuwabara, Kyriacou, Labare, Lanfranchi, Larson, Lauber, Leonard, Lesiak-Bzdak, Leuermann, Liu, Mariscal, Lu, Luenemann, Luszczak, Madsen, Maggi, Mahn, Mancina, Maruyama, Mase, Maunu, Meagher, Medici, Meier, Menne, Merino, Meures, Miarecki, Micallef, Momente, Montaruli, Moore, Morse, Moulai, Nahnhauer, Nakarmi, Naumann, Neer, Niederhausen, Nowicki, Nygren, Pollmann, Olivas, Murchadha, O’Sullivan, Padovani, Palczewski, Pandya, Pankova, Peiffer, Pepper, de~los Heros, Pieloth, Pinat, Plum, Price, Przybylski, Raab, Raedel,
  Rameez, Rawlins, Rea, Reimann, Relethford, Relich, Resconi, Rhode, Richman, Robertson, Rongen, Rott, Ruhe, Ryckbosch, Rysewyk, Safa, Saelzer, Sahakyan, Herrera, Sandrock, Sandroos, Santander, Sarkar, Sarkar, Satalecka, Schlunder, Schmidt, Schneider, Schoenen, Schoeneberg, Schumacher, Sclanfani, Seckel, Seunarine, Soedingrekso, Soldin, Song, Spiczak, Spiering, Stachurska, Stamatikos, Stanev, Stasik, Stettner, Steuer, Stezelberger, Stokstad, Stoessl, Strotjohann, Stuttard, Sullivan, Sutherland, Taboada, Tatar, Tenholt, Ter-Antonyan, Terliuk, Tilav, Toale, Tobin, Toennis, Toscano, Tosi, Tselengidou, Tung, Turcati, Turley, Ty, Unger, Usner, Driessche, Eijk, van Eijndhoven, Vandenbroucke, Vanheule, van Santen, Vogel, Vraeghe, Walck, Wallace, Wallraff, Wandler, Wandkowsky, Waza, Weaver, Weiss, Wendt, Werthebach, Westerhoff, Whelan, Whitehorn, Wiebe, Wiebusch, Wille, Williams, Wills, Wolf, Wood, Wood, Woschnagg, Xu, Xu, Xu, Yanez, Yodh, Yoshida and Yuan}]{txs0506}
\bibinfo{author}{{IceCube Collaboration}}, \bibinfo{author}{Aartsen, M.}, \bibinfo{author}{Ackermann, M.}, \bibinfo{author}{Adams, J.}, \bibinfo{author}{Aguilar, J.A.}, \bibinfo{author}{Ahlers, M.}, \bibinfo{author}{Ahrens, M.}, \bibinfo{author}{Samarai, I.A.}, \bibinfo{author}{Altmann, D.}, \bibinfo{author}{Andeen, K.}, \bibinfo{author}{Anderson, T.}, \bibinfo{author}{Ansseau, I.}, \bibinfo{author}{Anton, G.}, \bibinfo{author}{Argüelles, C.}, \bibinfo{author}{Arsioli, B.}, \bibinfo{author}{Auffenberg, J.}, \bibinfo{author}{Axani, S.}, \bibinfo{author}{Bagherpour, H.}, \bibinfo{author}{Bai, X.}, \bibinfo{author}{Barron, J.}, \bibinfo{author}{Barwick, S.}, \bibinfo{author}{Baum, V.}, \bibinfo{author}{Bay, R.}, \bibinfo{author}{Beatty, J.}, \bibinfo{author}{Becker, K.H.}, \bibinfo{author}{Tjus, J.B.}, \bibinfo{author}{BenZvi, S.}, \bibinfo{author}{Berley, D.}, \bibinfo{author}{Bernardini, E.}, \bibinfo{author}{Besson, D.}, \bibinfo{author}{Binder, G.}, \bibinfo{author}{Bindig, D.}, \bibinfo{author}{Blaufuss,
  E.}, \bibinfo{author}{Blot, S.}, \bibinfo{author}{Bohm, C.}, \bibinfo{author}{Boerner, M.}, \bibinfo{author}{Bos, F.}, \bibinfo{author}{Boeser, S.}, \bibinfo{author}{Botner, O.}, \bibinfo{author}{Bourbeau, E.}, \bibinfo{author}{Bourbeau, J.}, \bibinfo{author}{Bradascio, F.}, \bibinfo{author}{Braun, J.}, \bibinfo{author}{Brenzke, M.}, \bibinfo{author}{Bretz, H.P.}, \bibinfo{author}{Bron, S.}, \bibinfo{author}{Brostean-Kaiser, J.}, \bibinfo{author}{Burgman, A.}, \bibinfo{author}{Busse, R.}, \bibinfo{author}{Carver, T.}, \bibinfo{author}{Cheung, E.}, \bibinfo{author}{Chirkin, D.}, \bibinfo{author}{Christov, A.}, \bibinfo{author}{Clark, K.}, \bibinfo{author}{Classen, L.}, \bibinfo{author}{Coenders, S.}, \bibinfo{author}{Collin, G.}, \bibinfo{author}{Conrad, J.}, \bibinfo{author}{Coppin, P.}, \bibinfo{author}{Correa, P.}, \bibinfo{author}{Cowen, D.}, \bibinfo{author}{Cross, R.}, \bibinfo{author}{Dave, P.}, \bibinfo{author}{Day, M.}, \bibinfo{author}{de~André, J.P.A.M.}, \bibinfo{author}{Clercq, C.D.},
  \bibinfo{author}{Delaunay, J.}, \bibinfo{author}{Dembinski, H.}, \bibinfo{author}{DeRidder, S.}, \bibinfo{author}{Desiati, P.}, \bibinfo{author}{de~Vries, K.}, \bibinfo{author}{DeWasseige, G.}, \bibinfo{author}{DeWith, M.}, \bibinfo{author}{DeYoung, T.}, \bibinfo{author}{Díaz-Vélez, J.C.}, \bibinfo{author}{Lorenzo, V.D.}, \bibinfo{author}{Dujmovic, H.}, \bibinfo{author}{Dumm, J.}, \bibinfo{author}{Dunkman, M.}, \bibinfo{author}{Dvorak, E.}, \bibinfo{author}{Eberhardt, B.}, \bibinfo{author}{Ehrhardt, T.}, \bibinfo{author}{Eichmann, B.}, \bibinfo{author}{Eller, P.}, \bibinfo{author}{Evenson, P.}, \bibinfo{author}{Fahey, S.}, \bibinfo{author}{Fazely, A.}, \bibinfo{author}{Felde, J.}, \bibinfo{author}{Filimonov, K.}, \bibinfo{author}{Finley, C.}, \bibinfo{author}{Flis, S.}, \bibinfo{author}{Franckowiak, A.}, \bibinfo{author}{Friedman, E.}, \bibinfo{author}{Fritz, A.}, \bibinfo{author}{Gaisser, T.}, \bibinfo{author}{Gallagher, J.}, \bibinfo{author}{Gerhardt, L.}, \bibinfo{author}{Ghorbani, K.},
  \bibinfo{author}{Giommi, P.}, \bibinfo{author}{Glauch, T.}, \bibinfo{author}{Gluesenkamp, T.}, \bibinfo{author}{Goldschmidt, A.}, \bibinfo{author}{Gonzalez, J.}, \bibinfo{author}{Grant, D.}, \bibinfo{author}{Griffith, Z.}, \bibinfo{author}{Haack, C.}, \bibinfo{author}{Hallgren, A.}, \bibinfo{author}{Halzen, F.}, \bibinfo{author}{Hanson, K.}, \bibinfo{author}{Hebecker, D.}, \bibinfo{author}{Heereman, D.}, \bibinfo{author}{Helbing, K.}, \bibinfo{author}{Hellauer, R.}, \bibinfo{author}{Hickford, S.}, \bibinfo{author}{Hignight, J.}, \bibinfo{author}{Hill, G.}, \bibinfo{author}{Hoffman, K.}, \bibinfo{author}{Hoffmann, R.}, \bibinfo{author}{Hoinka, T.}, \bibinfo{author}{Hokanson-Fasig, B.}, \bibinfo{author}{Hoshina, K.}, \bibinfo{author}{Huang, F.}, \bibinfo{author}{Huber, M.}, \bibinfo{author}{Hultqvist, K.}, \bibinfo{author}{Huennefeld, M.}, \bibinfo{author}{Hussain, R.}, \bibinfo{author}{In, S.}, \bibinfo{author}{Iovine, N.}, \bibinfo{author}{Ishihara, A.}, \bibinfo{author}{Jacobi, E.},
  \bibinfo{author}{Japaridze, G.}, \bibinfo{author}{Jeong, M.}, \bibinfo{author}{Jero, K.}, \bibinfo{author}{Jones, B.}, \bibinfo{author}{Kalaczynski, P.}, \bibinfo{author}{Kang, W.}, \bibinfo{author}{Kappes, A.}, \bibinfo{author}{Kappesser, D.}, \bibinfo{author}{Karg, T.}, \bibinfo{author}{Karle, A.}, \bibinfo{author}{Katz, U.}, \bibinfo{author}{Kauer, M.}, \bibinfo{author}{Keivani, A.}, \bibinfo{author}{Kelley, J.}, \bibinfo{author}{Kheirandish, A.}, \bibinfo{author}{Kim, J.}, \bibinfo{author}{Kim, M.}, \bibinfo{author}{Kintscher, T.}, \bibinfo{author}{Kiryluk, J.}, \bibinfo{author}{Kittler, T.}, \bibinfo{author}{Klein, S.}, \bibinfo{author}{Koirala, R.}, \bibinfo{author}{Kolanoski, H.}, \bibinfo{author}{Koepke, L.}, \bibinfo{author}{Kopper, C.}, \bibinfo{author}{Kopper, S.}, \bibinfo{author}{Koschinsky, J.P.}, \bibinfo{author}{Koskinen, J.}, \bibinfo{author}{Kowalski, M.}, \bibinfo{author}{Krammer, B.}, \bibinfo{author}{Krings, K.}, \bibinfo{author}{Kroll, M.}, \bibinfo{author}{Krueckl, G.},
  \bibinfo{author}{Kunwar, S.}, \bibinfo{author}{Neilson, N.K.}, \bibinfo{author}{Kuwabara, T.}, \bibinfo{author}{Kyriacou, A.}, \bibinfo{author}{Labare, M.}, \bibinfo{author}{Lanfranchi, J.}, \bibinfo{author}{Larson, M.}, \bibinfo{author}{Lauber, F.}, \bibinfo{author}{Leonard, K.}, \bibinfo{author}{Lesiak-Bzdak, M.}, \bibinfo{author}{Leuermann, M.}, \bibinfo{author}{Liu, Q.}, \bibinfo{author}{Mariscal, C.J.L.}, \bibinfo{author}{Lu, L.}, \bibinfo{author}{Luenemann, J.}, \bibinfo{author}{Luszczak, W.}, \bibinfo{author}{Madsen, J.}, \bibinfo{author}{Maggi, G.}, \bibinfo{author}{Mahn, K.}, \bibinfo{author}{Mancina, S.}, \bibinfo{author}{Maruyama, R.}, \bibinfo{author}{Mase, K.}, \bibinfo{author}{Maunu, R.}, \bibinfo{author}{Meagher, K.}, \bibinfo{author}{Medici, M.}, \bibinfo{author}{Meier, M.}, \bibinfo{author}{Menne, T.}, \bibinfo{author}{Merino, G.}, \bibinfo{author}{Meures, T.}, \bibinfo{author}{Miarecki, S.}, \bibinfo{author}{Micallef, J.}, \bibinfo{author}{Momente, G.}, \bibinfo{author}{Montaruli, T.},
  \bibinfo{author}{Moore, R.}, \bibinfo{author}{Morse, R.}, \bibinfo{author}{Moulai, M.}, \bibinfo{author}{Nahnhauer, R.}, \bibinfo{author}{Nakarmi, P.}, \bibinfo{author}{Naumann, U.}, \bibinfo{author}{Neer, G.}, \bibinfo{author}{Niederhausen, H.}, \bibinfo{author}{Nowicki, S.}, \bibinfo{author}{Nygren, D.}, \bibinfo{author}{Pollmann, A.}, \bibinfo{author}{Olivas, A.}, \bibinfo{author}{Murchadha, A.}, \bibinfo{author}{O’Sullivan, E.}, \bibinfo{author}{Padovani, P.}, \bibinfo{author}{Palczewski, T.}, \bibinfo{author}{Pandya, H.}, \bibinfo{author}{Pankova, D.}, \bibinfo{author}{Peiffer, P.}, \bibinfo{author}{Pepper, J.}, \bibinfo{author}{de~los Heros, C.P.}, \bibinfo{author}{Pieloth, D.}, \bibinfo{author}{Pinat, E.}, \bibinfo{author}{Plum, M.}, \bibinfo{author}{Price, B.}, \bibinfo{author}{Przybylski, G.}, \bibinfo{author}{Raab, C.}, \bibinfo{author}{Raedel, L.}, \bibinfo{author}{Rameez, M.}, \bibinfo{author}{Rawlins, K.}, \bibinfo{author}{Rea, I.C.}, \bibinfo{author}{Reimann, R.},
  \bibinfo{author}{Relethford, B.}, \bibinfo{author}{Relich, M.}, \bibinfo{author}{Resconi, E.}, \bibinfo{author}{Rhode, W.}, \bibinfo{author}{Richman, M.}, \bibinfo{author}{Robertson, S.}, \bibinfo{author}{Rongen, M.}, \bibinfo{author}{Rott, C.}, \bibinfo{author}{Ruhe, T.}, \bibinfo{author}{Ryckbosch, D.}, \bibinfo{author}{Rysewyk, D.}, \bibinfo{author}{Safa, I.}, \bibinfo{author}{Saelzer, T.}, \bibinfo{author}{Sahakyan, N.}, \bibinfo{author}{Herrera, S.S.}, \bibinfo{author}{Sandrock, A.}, \bibinfo{author}{Sandroos, J.}, \bibinfo{author}{Santander, M.}, \bibinfo{author}{Sarkar, S.}, \bibinfo{author}{Sarkar, S.}, \bibinfo{author}{Satalecka, K.}, \bibinfo{author}{Schlunder, P.}, \bibinfo{author}{Schmidt, T.}, \bibinfo{author}{Schneider, A.}, \bibinfo{author}{Schoenen, S.}, \bibinfo{author}{Schoeneberg, S.}, \bibinfo{author}{Schumacher, L.}, \bibinfo{author}{Sclanfani, S.}, \bibinfo{author}{Seckel, D.}, \bibinfo{author}{Seunarine, S.}, \bibinfo{author}{Soedingrekso, J.}, \bibinfo{author}{Soldin, D.},
  \bibinfo{author}{Song, M.}, \bibinfo{author}{Spiczak, G.}, \bibinfo{author}{Spiering, C.}, \bibinfo{author}{Stachurska, J.}, \bibinfo{author}{Stamatikos, M.}, \bibinfo{author}{Stanev, T.}, \bibinfo{author}{Stasik, A.}, \bibinfo{author}{Stettner, J.}, \bibinfo{author}{Steuer, A.}, \bibinfo{author}{Stezelberger, T.}, \bibinfo{author}{Stokstad, R.}, \bibinfo{author}{Stoessl, A.}, \bibinfo{author}{Strotjohann, N.L.}, \bibinfo{author}{Stuttard, T.}, \bibinfo{author}{Sullivan, G.}, \bibinfo{author}{Sutherland, M.}, \bibinfo{author}{Taboada, I.}, \bibinfo{author}{Tatar, J.}, \bibinfo{author}{Tenholt, F.}, \bibinfo{author}{Ter-Antonyan, S.}, \bibinfo{author}{Terliuk, A.}, \bibinfo{author}{Tilav, S.}, \bibinfo{author}{Toale, P.}, \bibinfo{author}{Tobin, M.}, \bibinfo{author}{Toennis, C.}, \bibinfo{author}{Toscano, S.}, \bibinfo{author}{Tosi, D.}, \bibinfo{author}{Tselengidou, M.}, \bibinfo{author}{Tung, C.}, \bibinfo{author}{Turcati, A.}, \bibinfo{author}{Turley, C.}, \bibinfo{author}{Ty, B.},
  \bibinfo{author}{Unger, L.}, \bibinfo{author}{Usner, M.}, \bibinfo{author}{Driessche, W.V.}, \bibinfo{author}{Eijk, D.V.}, \bibinfo{author}{van Eijndhoven, N.}, \bibinfo{author}{Vandenbroucke, J.}, \bibinfo{author}{Vanheule, S.}, \bibinfo{author}{van Santen, J.}, \bibinfo{author}{Vogel, E.}, \bibinfo{author}{Vraeghe, M.}, \bibinfo{author}{Walck, C.}, \bibinfo{author}{Wallace, A.}, \bibinfo{author}{Wallraff, M.}, \bibinfo{author}{Wandler, F.}, \bibinfo{author}{Wandkowsky, N.}, \bibinfo{author}{Waza, A.}, \bibinfo{author}{Weaver, C.}, \bibinfo{author}{Weiss, M.}, \bibinfo{author}{Wendt, C.}, \bibinfo{author}{Werthebach, J.}, \bibinfo{author}{Westerhoff, S.}, \bibinfo{author}{Whelan, B.}, \bibinfo{author}{Whitehorn, N.}, \bibinfo{author}{Wiebe, K.}, \bibinfo{author}{Wiebusch, C.}, \bibinfo{author}{Wille, L.}, \bibinfo{author}{Williams, D.}, \bibinfo{author}{Wills, L.}, \bibinfo{author}{Wolf, M.}, \bibinfo{author}{Wood, J.}, \bibinfo{author}{Wood, T.}, \bibinfo{author}{Woschnagg, K.}, \bibinfo{author}{Xu, D.},
  \bibinfo{author}{Xu, X.}, \bibinfo{author}{Xu, Y.}, \bibinfo{author}{Yanez, J.P.}, \bibinfo{author}{Yodh, G.}, \bibinfo{author}{Yoshida, S.}, \bibinfo{author}{Yuan, T.}, \bibinfo{year}{2018}.
\newblock \bibinfo{title}{Neutrino emission from the direction of the blazar txs 0506+056 prior to the icecube-170922a alert}.
\newblock \bibinfo{journal}{Science} \bibinfo{volume}{361}, \bibinfo{pages}{147--151}.
\newblock \URLprefix \url{https://www.science.org/doi/abs/10.1126/science.aat2890}, \DOIprefix\doi{10.1126/science.aat2890}, \href{http://arxiv.org/abs/https://www.science.org/doi/pdf/10.1126/science.aat2890}{{\tt arXiv:https://www.science.org/doi/pdf/10.1126/science.aat2890}}.
\bibitem[{Kildea et~al.(2007)Kildea, Atkins, Badran, Blaylock, Bond, Bradbury, Buckley, Carter-Lewis, Celik, Chow, Cui, Cogan, Daniel, de~la Calle~Perez, Dowdall, Duke, Falcone, Fegan, Fegan, Finley, Fortson, Gall, Gillanders, Grube, Gutierrez, Hall, Hall, Holder, Horan, Hughes, Jordan, Jung, Kenny, Kertzman, Knapp, Konopelko, Kosack, Krawczynski, Krennrich, Lang, LeBohec, Lloyd-Evans, Millis, Moriarty, Nagai, Ogden, Ong, Perkins, Petry, Pizlo, Pohl, Quinn, Quinn, Rebillot, Rose, Schroedter, Sembroski, Smith, Syson, Toner, Valcarcel, Vassiliev, Wakely, Weekes, White and Observatory}]{Kildea2007}
\bibinfo{author}{Kildea, J.}, \bibinfo{author}{Atkins, R.}, \bibinfo{author}{Badran, H.}, \bibinfo{author}{Blaylock, G.}, \bibinfo{author}{Bond, I.}, \bibinfo{author}{Bradbury, S.}, \bibinfo{author}{Buckley, J.}, \bibinfo{author}{Carter-Lewis, D.}, \bibinfo{author}{Celik, O.}, \bibinfo{author}{Chow, Y.}, \bibinfo{author}{Cui, W.}, \bibinfo{author}{Cogan, P.}, \bibinfo{author}{Daniel, M.}, \bibinfo{author}{de~la Calle~Perez, I.}, \bibinfo{author}{Dowdall, C.}, \bibinfo{author}{Duke, C.}, \bibinfo{author}{Falcone, A.}, \bibinfo{author}{Fegan, D.}, \bibinfo{author}{Fegan, S.}, \bibinfo{author}{Finley, J.}, \bibinfo{author}{Fortson, L.}, \bibinfo{author}{Gall, D.}, \bibinfo{author}{Gillanders, G.}, \bibinfo{author}{Grube, J.}, \bibinfo{author}{Gutierrez, K.}, \bibinfo{author}{Hall, J.}, \bibinfo{author}{Hall, T.}, \bibinfo{author}{Holder, J.}, \bibinfo{author}{Horan, D.}, \bibinfo{author}{Hughes, S.}, \bibinfo{author}{Jordan, M.}, \bibinfo{author}{Jung, I.}, \bibinfo{author}{Kenny, G.}, \bibinfo{author}{Kertzman,
  M.}, \bibinfo{author}{Knapp, J.}, \bibinfo{author}{Konopelko, A.}, \bibinfo{author}{Kosack, K.}, \bibinfo{author}{Krawczynski, H.}, \bibinfo{author}{Krennrich, F.}, \bibinfo{author}{Lang, M.}, \bibinfo{author}{LeBohec, S.}, \bibinfo{author}{Lloyd-Evans, J.}, \bibinfo{author}{Millis, J.}, \bibinfo{author}{Moriarty, P.}, \bibinfo{author}{Nagai, T.}, \bibinfo{author}{Ogden, P.}, \bibinfo{author}{Ong, R.}, \bibinfo{author}{Perkins, J.}, \bibinfo{author}{Petry, D.}, \bibinfo{author}{Pizlo, F.}, \bibinfo{author}{Pohl, M.}, \bibinfo{author}{Quinn, J.}, \bibinfo{author}{Quinn, M.}, \bibinfo{author}{Rebillot, P.}, \bibinfo{author}{Rose, H.}, \bibinfo{author}{Schroedter, M.}, \bibinfo{author}{Sembroski, G.}, \bibinfo{author}{Smith, A.}, \bibinfo{author}{Syson, A.}, \bibinfo{author}{Toner, J.}, \bibinfo{author}{Valcarcel, L.}, \bibinfo{author}{Vassiliev, V.}, \bibinfo{author}{Wakely, S.}, \bibinfo{author}{Weekes, T.}, \bibinfo{author}{White, R.}, \bibinfo{author}{Observatory, F.L.W.}, \bibinfo{year}{2007}.
\newblock \bibinfo{title}{The whipple observatory 10 m gamma-ray telescope, 1997-2006}.
\newblock \bibinfo{journal}{Astroparticle Physics} \bibinfo{volume}{28}, \bibinfo{pages}{182--195}.
\newblock \URLprefix \url{www.elsevier.com/locate/astropart}, \DOIprefix\doi{10.1016/j.astropartphys.2007.05.004}.
\bibitem[{Meier and Clark(2023)}]{IceCube_meier2023searchextremelyhighenergy}
\bibinfo{author}{Meier, M.}, \bibinfo{author}{Clark, B.}, \bibinfo{year}{2023}.
\newblock \bibinfo{title}{Search for extremely high energy neutrinos with icecube}.
\newblock \URLprefix \url{https://arxiv.org/abs/2308.07656}, \href{http://arxiv.org/abs/2308.07656}{{\tt arXiv:2308.07656}}.
\bibitem[{Álvarez Muñiz et~al.(2019)Álvarez Muñiz, Alves~Batista, Balagopal~V., Bolmont, Bustamante, Carvalho, Charrier, Cognard, Decoene, Denton, De~Jong, De~Vries, Engel, Fang, Finley, Gabici, Gou, Gu, Guépin, Hu, Huang, Kotera, Le~Coz, Lenain, Lü, Martineau-Huynh, Mostafá, Mottez, Murase, Niess, Oikonomou, Pierog, Qian, Qin, Ran, Renault-Tinacci, Roth, Schröder, Schüssler, Tasse, Timmermans, Tueros, Wu, Zarka, Zech, Zhang, Zhang, Zhang, Zheng and Zilles}]{GRAND_lvarez_Mu_iz_2019}
\bibinfo{author}{Álvarez Muñiz, J.}, \bibinfo{author}{Alves~Batista, R.}, \bibinfo{author}{Balagopal~V., A.}, \bibinfo{author}{Bolmont, J.}, \bibinfo{author}{Bustamante, M.}, \bibinfo{author}{Carvalho, W.}, \bibinfo{author}{Charrier, D.}, \bibinfo{author}{Cognard, I.}, \bibinfo{author}{Decoene, V.}, \bibinfo{author}{Denton, P.B.}, \bibinfo{author}{De~Jong, S.}, \bibinfo{author}{De~Vries, K.D.}, \bibinfo{author}{Engel, R.}, \bibinfo{author}{Fang, K.}, \bibinfo{author}{Finley, C.}, \bibinfo{author}{Gabici, S.}, \bibinfo{author}{Gou, Q.}, \bibinfo{author}{Gu, J.}, \bibinfo{author}{Guépin, C.}, \bibinfo{author}{Hu, H.}, \bibinfo{author}{Huang, Y.}, \bibinfo{author}{Kotera, K.}, \bibinfo{author}{Le~Coz, S.}, \bibinfo{author}{Lenain, J.P.}, \bibinfo{author}{Lü, G.}, \bibinfo{author}{Martineau-Huynh, O.}, \bibinfo{author}{Mostafá, M.}, \bibinfo{author}{Mottez, F.}, \bibinfo{author}{Murase, K.}, \bibinfo{author}{Niess, V.}, \bibinfo{author}{Oikonomou, F.}, \bibinfo{author}{Pierog, T.}, \bibinfo{author}{Qian,
  X.}, \bibinfo{author}{Qin, B.}, \bibinfo{author}{Ran, D.}, \bibinfo{author}{Renault-Tinacci, N.}, \bibinfo{author}{Roth, M.}, \bibinfo{author}{Schröder, F.G.}, \bibinfo{author}{Schüssler, F.}, \bibinfo{author}{Tasse, C.}, \bibinfo{author}{Timmermans, C.}, \bibinfo{author}{Tueros, M.}, \bibinfo{author}{Wu, X.}, \bibinfo{author}{Zarka, P.}, \bibinfo{author}{Zech, A.}, \bibinfo{author}{Zhang, B.T.}, \bibinfo{author}{Zhang, J.}, \bibinfo{author}{Zhang, Y.}, \bibinfo{author}{Zheng, Q.}, \bibinfo{author}{Zilles, A.}, \bibinfo{year}{2019}.
\newblock \bibinfo{title}{The giant radio array for neutrino detection (grand): Science and design}.
\newblock \bibinfo{journal}{Science China Physics, Mechanics \& Astronomy} \bibinfo{volume}{63}.
\newblock \URLprefix \url{http://dx.doi.org/10.1007/s11433-018-9385-7}, \DOIprefix\doi{10.1007/s11433-018-9385-7}.
\bibitem[{Neronov et~al.(2017)Neronov, Semikoz, Anchordoqui, Adams and Olinto}]{chant_Neronov_2017}
\bibinfo{author}{Neronov, A.}, \bibinfo{author}{Semikoz, D.V.}, \bibinfo{author}{Anchordoqui, L.A.}, \bibinfo{author}{Adams, J.H.}, \bibinfo{author}{Olinto, A.V.}, \bibinfo{year}{2017}.
\newblock \bibinfo{title}{Sensitivity of a proposed space-based cherenkov astrophysical-neutrino telescope}.
\newblock \bibinfo{journal}{Physical Review D} \bibinfo{volume}{95}.
\newblock \URLprefix \url{http://dx.doi.org/10.1103/PhysRevD.95.023004}, \DOIprefix\doi{10.1103/physrevd.95.023004}.
\bibitem[{Olinto et~al.(2021)Olinto, Krizmanic, Adams, Aloisio, Anchordoqui, Anzalone, Bagheri, Barghini, Battisti, Bergman, Bertaina, Bertone, Bisconti, Bustamante, Cafagna, Caruso, Casolino, Černý, Christl, Cummings, De~Mitri, Diesing, Engel, Eser, Fang, Fenu, Filippatos, Gazda, Guepin, Haungs, Hays, Judd, Klimov, Kungel, Kuznetsov, Mackovjak, Mandát, Marcelli, McEnery, Medina-Tanco, Merenda, Meyer, Mitchell, Miyamoto, Nachtman, Neronov, Oikonomou, Onel, Osteria, Otte, Parizot, Paul, Pech, Perkins, Picozza, Piotrowski, Plebaniak, Prévôt, Reardon, Reno, Ricci, Romero~Matamala, Sarazin, Schovánek, Scotti, Shinozaki, Soriano, Stecker, Takizawa, Ulrich, Unger, Venters, Wiencke, Winn, Young and Zotov}]{poemma_Olinto_2021}
\bibinfo{author}{Olinto, A.}, \bibinfo{author}{Krizmanic, J.}, \bibinfo{author}{Adams, J.}, \bibinfo{author}{Aloisio, R.}, \bibinfo{author}{Anchordoqui, L.}, \bibinfo{author}{Anzalone, A.}, \bibinfo{author}{Bagheri, M.}, \bibinfo{author}{Barghini, D.}, \bibinfo{author}{Battisti, M.}, \bibinfo{author}{Bergman, D.}, \bibinfo{author}{Bertaina, M.}, \bibinfo{author}{Bertone, P.}, \bibinfo{author}{Bisconti, F.}, \bibinfo{author}{Bustamante, M.}, \bibinfo{author}{Cafagna, F.}, \bibinfo{author}{Caruso, R.}, \bibinfo{author}{Casolino, M.}, \bibinfo{author}{Černý, K.}, \bibinfo{author}{Christl, M.}, \bibinfo{author}{Cummings, A.}, \bibinfo{author}{De~Mitri, I.}, \bibinfo{author}{Diesing, R.}, \bibinfo{author}{Engel, R.}, \bibinfo{author}{Eser, J.}, \bibinfo{author}{Fang, K.}, \bibinfo{author}{Fenu, F.}, \bibinfo{author}{Filippatos, G.}, \bibinfo{author}{Gazda, E.}, \bibinfo{author}{Guepin, C.}, \bibinfo{author}{Haungs, A.}, \bibinfo{author}{Hays, E.}, \bibinfo{author}{Judd, E.}, \bibinfo{author}{Klimov, P.},
  \bibinfo{author}{Kungel, V.}, \bibinfo{author}{Kuznetsov, E.}, \bibinfo{author}{Mackovjak, S.}, \bibinfo{author}{Mandát, D.}, \bibinfo{author}{Marcelli, L.}, \bibinfo{author}{McEnery, J.}, \bibinfo{author}{Medina-Tanco, G.}, \bibinfo{author}{Merenda, K.D.}, \bibinfo{author}{Meyer, S.}, \bibinfo{author}{Mitchell, J.}, \bibinfo{author}{Miyamoto, H.}, \bibinfo{author}{Nachtman, J.}, \bibinfo{author}{Neronov, A.}, \bibinfo{author}{Oikonomou, F.}, \bibinfo{author}{Onel, Y.}, \bibinfo{author}{Osteria, G.}, \bibinfo{author}{Otte, A.}, \bibinfo{author}{Parizot, E.}, \bibinfo{author}{Paul, T.}, \bibinfo{author}{Pech, M.}, \bibinfo{author}{Perkins, J.}, \bibinfo{author}{Picozza, P.}, \bibinfo{author}{Piotrowski, L.}, \bibinfo{author}{Plebaniak, Z.}, \bibinfo{author}{Prévôt, G.}, \bibinfo{author}{Reardon, P.}, \bibinfo{author}{Reno, M.}, \bibinfo{author}{Ricci, M.}, \bibinfo{author}{Romero~Matamala, O.}, \bibinfo{author}{Sarazin, F.}, \bibinfo{author}{Schovánek, P.}, \bibinfo{author}{Scotti, V.},
  \bibinfo{author}{Shinozaki, K.}, \bibinfo{author}{Soriano, J.}, \bibinfo{author}{Stecker, F.}, \bibinfo{author}{Takizawa, Y.}, \bibinfo{author}{Ulrich, R.}, \bibinfo{author}{Unger, M.}, \bibinfo{author}{Venters, T.}, \bibinfo{author}{Wiencke, L.}, \bibinfo{author}{Winn, D.}, \bibinfo{author}{Young, R.}, \bibinfo{author}{Zotov, M.}, \bibinfo{year}{2021}.
\newblock \bibinfo{title}{The poemma (probe of extreme multi-messenger astrophysics) observatory}.
\newblock \bibinfo{journal}{Journal of Cosmology and Astroparticle Physics} \bibinfo{volume}{2021}, \bibinfo{pages}{007}.
\newblock \URLprefix \url{http://dx.doi.org/10.1088/1475-7516/2021/06/007}, \DOIprefix\doi{10.1088/1475-7516/2021/06/007}.
\bibitem[{Prohira et~al.(2020)Prohira, de~Vries, Allison, Beatty, Besson, Connolly, van Eijndhoven, Hast, Kuo, Latif, Meures, Nam, Nozdrina, Ralston, Riesen, Sbrocco, Torres and Wissel}]{RTE_Prohira_2020}
\bibinfo{author}{Prohira, S.}, \bibinfo{author}{de~Vries, K.}, \bibinfo{author}{Allison, P.}, \bibinfo{author}{Beatty, J.}, \bibinfo{author}{Besson, D.}, \bibinfo{author}{Connolly, A.}, \bibinfo{author}{van Eijndhoven, N.}, \bibinfo{author}{Hast, C.}, \bibinfo{author}{Kuo, C.Y.}, \bibinfo{author}{Latif, U.}, \bibinfo{author}{Meures, T.}, \bibinfo{author}{Nam, J.}, \bibinfo{author}{Nozdrina, A.}, \bibinfo{author}{Ralston, J.}, \bibinfo{author}{Riesen, Z.}, \bibinfo{author}{Sbrocco, C.}, \bibinfo{author}{Torres, J.}, \bibinfo{author}{Wissel, S.}, \bibinfo{year}{2020}.
\newblock \bibinfo{title}{Observation of radar echoes from high-energy particle cascades}.
\newblock \bibinfo{journal}{Physical Review Letters} \bibinfo{volume}{124}.
\newblock \URLprefix \url{http://dx.doi.org/10.1103/PhysRevLett.124.091101}, \DOIprefix\doi{10.1103/physrevlett.124.091101}.
\bibitem[{Southall et~al.(2023)Southall, Deaconu, Decoene, Oberla, Zeolla, Alvarez-Muñiz, Cummings, Curtis-Ginsberg, Hendrick, Hughes, Krebs, Ludwig, Mulrey, Prohira, Carvalho, Rodriguez, Romero-Wolf, Schoorlemmer, Vieregg, Wissel and Zas}]{BEACON_Southall_2023}
\bibinfo{author}{Southall, D.}, \bibinfo{author}{Deaconu, C.}, \bibinfo{author}{Decoene, V.}, \bibinfo{author}{Oberla, E.}, \bibinfo{author}{Zeolla, A.}, \bibinfo{author}{Alvarez-Muñiz, J.}, \bibinfo{author}{Cummings, A.}, \bibinfo{author}{Curtis-Ginsberg, Z.}, \bibinfo{author}{Hendrick, A.}, \bibinfo{author}{Hughes, K.}, \bibinfo{author}{Krebs, R.}, \bibinfo{author}{Ludwig, A.}, \bibinfo{author}{Mulrey, K.}, \bibinfo{author}{Prohira, S.}, \bibinfo{author}{Carvalho, W.R.d.}, \bibinfo{author}{Rodriguez, A.}, \bibinfo{author}{Romero-Wolf, A.}, \bibinfo{author}{Schoorlemmer, H.}, \bibinfo{author}{Vieregg, A.G.}, \bibinfo{author}{Wissel, S.A.}, \bibinfo{author}{Zas, E.}, \bibinfo{year}{2023}.
\newblock \bibinfo{title}{Design and initial performance of the prototype for the beacon instrument for detection of ultrahigh energy particles}.
\newblock \bibinfo{journal}{Nuclear Instruments and Methods in Physics Research Section A: Accelerators, Spectrometers, Detectors and Associated Equipment} \bibinfo{volume}{1048}, \bibinfo{pages}{167889}.
\newblock \URLprefix \url{http://dx.doi.org/10.1016/j.nima.2022.167889}, \DOIprefix\doi{10.1016/j.nima.2022.167889}.
\bibitem[{Thompson(2023)}]{tambo_thompson2023tambosearchingtauneutrinos}
\bibinfo{author}{Thompson, W.G.}, \bibinfo{year}{2023}.
\newblock \bibinfo{title}{Tambo: Searching for tau neutrinos in the peruvian andes}.
\newblock \URLprefix \url{https://arxiv.org/abs/2308.09753}, \href{http://arxiv.org/abs/2308.09753}{{\tt arXiv:2308.09753}}.
\bibitem[{Wang et~al.(2021)Wang, Lin, Otte, Doro, Gazda, Taboada, Brown and Bagheri}]{Trinity_Wang_2021}
\bibinfo{author}{Wang, A.}, \bibinfo{author}{Lin, C.}, \bibinfo{author}{Otte, N.}, \bibinfo{author}{Doro, M.}, \bibinfo{author}{Gazda, E.}, \bibinfo{author}{Taboada, I.}, \bibinfo{author}{Brown, A.}, \bibinfo{author}{Bagheri, M.}, \bibinfo{year}{2021}.
\newblock \bibinfo{title}{Trinity’s sensitivity to isotropic and point-source neutrinos}, in: \bibinfo{booktitle}{Proceedings of 37th International Cosmic Ray Conference — PoS(ICRC2021)}, \bibinfo{publisher}{Sissa Medialab}. p. \bibinfo{pages}{1234}.
\newblock \URLprefix \url{http://dx.doi.org/10.22323/1.395.1234}, \DOIprefix\doi{10.22323/1.395.1234}.
\bibitem[{Weekes et~al.(1989)Weekes, Cawley, Fegan, Gibbs, Hillas, Kowk, Lamb, Lewis, Macomb, Porter, Reynolds and Vacanti}]{whipplecrab_nuimeprn12618}
\bibinfo{author}{Weekes, T.}, \bibinfo{author}{Cawley, M.F.}, \bibinfo{author}{Fegan, D.}, \bibinfo{author}{Gibbs, K.}, \bibinfo{author}{Hillas, A.}, \bibinfo{author}{Kowk, P.}, \bibinfo{author}{Lamb, R.}, \bibinfo{author}{Lewis, D.}, \bibinfo{author}{Macomb, D.}, \bibinfo{author}{Porter, N.}, \bibinfo{author}{Reynolds, P.}, \bibinfo{author}{Vacanti, G.}, \bibinfo{year}{1989}.
\newblock \bibinfo{title}{Observation of tev gamma-rays from the crab nebula using the atmospheric cherenkov imaging technique}.
\newblock \bibinfo{journal}{Astrophysical Journal} \bibinfo{volume}{342}, \bibinfo{pages}{379--395}.
\newblock \URLprefix \url{https://mural.maynoothuniversity.ie/12618/}.

\end{thebibliography}






\end{document}